\documentstyle[12pt,psfig]{article}
\title{The Non-Commutative Geometry of the Quantum Hall Effect}
\author{J. Bellissard\thanks {e-mail: jeanbel@siberia.ups-tlse.fr} , A. 
van Elst
\thanks {e-mail: andreas@tolosa.ups-tlse.fr} , H. Schulz-
Baldes\thanks {e-mail: hermann@tolosa.ups-tlse.fr} ,  \\
Universit\'e Paul Sabatier, Toulouse, France \thanks {URA 505, CNRS 
and Laboratoire de Physique Quantique, 
118, Route de Narbonne, 31062-Toulouse Cedex, France}}

\newtheorem{theo}{Theorem}
\newtheorem{defini}{Definition}
\newtheorem{proposi}{Proposition}
\newtheorem{lemma}{Lemma}
\newtheorem{coro}{Corollary}

\newcommand{\CS}{$C^{\ast}$-algebra }
\newcommand{\CSs}{$C^{\ast}$-algebras}

\newcommand{\be}{\begin{equation}}
\newcommand{\ee}{\end{equation}}
\newcommand{\beqn}{\begin{eqnarray}}
\newcommand{\eeqn}{\end{eqnarray}}
\newcommand{\bed}{\begin{description}}
\newcommand{\ed}{\end{description}}
\newcommand{\CH}{{\bf Ch}}
\newcommand{\Ch}{{\rm Ch}}
\newcommand{\TR}{{\rm Tr}}
\newcommand{\TV}{{\cal T}}
\newcommand{\Tg}{{\rm T}}
\newcommand{\ST}{{\rm Tr}_{S}}
\newcommand{\TD}{{\rm Tr}_{{\mbox{\tiny  Dix}}}}
\newcommand{\Aa}{{\cal A}}
\newcommand{\Bb}{{\cal B}}
\newcommand{\Cc}{{\cal C}}
\newcommand{\Dd}{{\cal D}}
\newcommand{\Ee}{{\cal E}}

\newcommand{\Hh}{{\cal H}}
\newcommand{\Kk}{{\cal K}}
\newcommand{\Ll}{{\cal L}}
\newcommand{\Nn}{{\cal N}}
\newcommand{\Oo}{{\cal O}}
\newcommand{\Pp}{{\cal P}}
\newcommand{\Ss}{{\cal S}}
\newcommand{\Vv}{{\cal V}}
\newcommand{\Ww}{{\cal W}}
\newcommand{\AO}{{\cal A}_0}
\newcommand{\BB}{{\bf B}}
\newcommand{\CC}{{\bf C}}
\newcommand{\EE}{{\bf E}}

\newcommand{\NN}{{\bf N}}
\newcommand{\PP}{{\bf P}}

\newcommand{\RR}{{\bf R}}

\newcommand{\ZZ}{{\bf Z}}

\newcommand{\ZA}{{\cal Z}(\Delta)}
\newcommand{\pro}{{\bf P}(\omega)}

\newcommand{\EV}{\vec{\cal E}}
\newcommand{\Hha}{\hat{{\cal H}}}
\newcommand{\jV}{\vec{j}}
\newcommand{\JV}{\vec{J}}
\newcommand{\kV}{\vec{k}}
\newcommand{\XV}{\vec{X}}
\newcommand{\naV}{\vec{\nabla}}
\newcommand{\cop}{\hat{\kappa}}

\newcommand{\ov}{\overline}

\begin{document}

\maketitle

\begin{abstract}
We give an overview of the Integer Quantum Hall Effect. We propose a 
mathematical framework using Non-Commutative Geometry as 
defined by A. Connes. Within this framework, it is proved that the Hall 
conductivity is quantized and that plateaux occur when the Fermi 
energy varies in a region of localized states.


This article will appear in the October 94 special issue of the Journal of
Mathematical Physics.
\end{abstract}


\tableofcontents

\vspace{1cm}

\section{Introduction}

In 1880, E.H. Hall \cite{Ha} undertook the classical experiment which 
led to the so-called Hall effect. A century later, von Klitzing and his 
co-workers  \cite{KDP} showed that the Hall conductivity was quantized 
at very low temperatures as an integer multiple of the universal 
constant $e^{2}/h$. Here $e$ is the electron charge whereas $h$ is 
Planck's constant. This is the  Integer Quantum Hall Effect (IQHE). For 
this discovery, which led to a new accurate measurement of the fine 
structure constant and a new definition of the standard of resistance 
\cite{ResH}, von Klitzing was awarded the Nobel price in 1985.  

On the other hand, during the seventies, A. Connes \cite{Co85,Co90} 
extended most of the tools of  differential 
geometry to non-commutative \CSs, thus creating a new branch of mathematics 
called {\em Non-Commutative Geometry}. The main new result obtained in this 
field was the definition of cyclic cohomology and the proof of an index 
theorem for elliptic operators on a foliated manifold. For this work and also 
his contribution to the study of von Neumann algebras, Connes was awarded the 
Fields Medal in 1982. He recently extended this theory to what is now called 
{\em Quantum Calculus} \cite{Co93}. 

After the works by Laughlin \cite{Laug} and especially by Kohmoto, den Nijs, 
Nightingale and Thouless \cite{TKN} (called $TKN_{2}$ below), it became 
clear that the quantization of the Hall conductance at low temperature 
had a geometric origin. The universality of this effect had then an 
explanation. Moreover, as proposed by Prange \cite{Prange,JP} and Thouless 
\cite{Thou2}, the plateaux of the Hall conductance which appear
while changing the magnetic field or the charge-carrier density, are due to 
localization. Neither the original Laughlin paper nor the $TKN_{2}$ one 
however could give a description of both properties in the same model. 
Developing a mathematical framework able to reconcile topological and 
localization properties at once was a challenging problem. Attempts were made 
by Avron et al. \cite{ASS1} who exhibited quantization but were not able to 
prove that these quantum numbers were insensitive to disorder. In 1986, 
H. Kunz \cite{Ku} went further on and managed to prove this for disorder 
small enough to avoid filling the gaps between Landau levels. 

But in \cite{B85,Be87,Be85}, one of us proposed to use Non-Commutative Geometry 
to extend the $TKN_{2}$ argument to the case of arbitrary magnetic field and 
disordered crystal. It turned out that the condition under which plateaux occur 
was precisely the finiteness of the localization length near the Fermi level. 
This work was rephrased later on by Avron et al. \cite{ASS} in terms of charge 
transport and relative index, filling the remaining gap between experimental 
observations, theoretical intuition and mathematical frame. 

Our aim in this work is to review these various contributions in a 
synthetic and detailed way. We will use this opportunity to give proofs 
that are missing or scattered in the literature. In addition, we will discuss
the effect of disorder from two complementary aspects.

On the one hand, we will develop our point of view on localization produced
by quenched disorder. This is crucial for understanding the IQHE.
We review various localization criteria and formulate them in terms of
Non-Commutative Geometry. With the Dixmier trace, A. Connes introduced
a remarkable technique into Quantum Calculus. In our context, it allows us to 
give the precise condition under which the Hall conductance is quantized; this
condition is shown to be a localization condition.

On the other hand, we also propose a model for electronic transport giving
rise to the so-called ``relaxation time approximation'' and allowing to derive
a Kubo formula for the conductivity. This approach allows to describe the 
effect of time-dependent disorder in a phenomenological way. This latter 
has quite different consequences from those of the quenched disorder such as a 
non-zero finite direct conductivity. Even though this approach is not original 
in its principle, the non-commutative framework allows us to treat the case of 
aperiodic crystals and magnetic fields when Bloch theory fails. Therefore, 
strictly speaking, our Kubo formula is new. We also show, without proofs, how 
to justify the linear response theory within this framework, leaving the formal 
proofs for a future work. The advantage of this approach is to give  
control of the various approximations that have to be made to fit the ideal 
result with experiments. For this reason, we discuss the effects of temperature,
of non-linear terms in the electric field, of the finite size of samples and
finally those of collisions and disorder. In particular, we argue that the
discrepancy $\delta \sigma_{H}$ between the measured Hall conductivity and the
ideal one, given by a Chern number, is dominated by the collision terms.
In the center of a plateau, we get the rough estimate

\begin{equation}
\label{eq-discrepancy}
\frac{\delta \sigma_{H}}{\sigma_{H}}
	\leq \mbox{\rm const.} \;
		\nu \frac{e}{h}
			\frac{\lambda ^{2}}{\mu_{c}}
\mbox{ , }
\end{equation}

\noindent where  $\nu$ is the filling factor, $\lambda$ is the localization 
length and $\mu_{c}$ is the charge-carrier mobility. $e/h$ is a universal constant, 
$\nu$ about unity and the localization length
typically of the order of the magnetic length. Inserting measured
values for the mobility, one obtains $10^{-4}$ for the right-hand side
expression. This estimate does not take into account the Mott
conductivity. However, it shows why both a large quenched 
disorder (in order to have small localization lengths) and a large mobility 
(namely a low collision rate) are necessary in order to get accurate 
measurements. Such a compromise is realized in heterojunctions and to
less extend in MOSFETs.
The estimate (\ref{eq-discrepancy}) also permits to understand intuitively 
why the plateaux in the Fractional Quantum Hall Effect (FQHE) are less precise, 
since the localization length of Laughlin quasiparticles is probably larger than
that of electrons at integer plateaux, and their mobility probably lower. 

No attempt will be made however to extend our non-commutative approach to the 
FQHE. We will only give some insight and a short review of works that we feel 
relevant in view  of a mathematically complete description of the FQHE.

This rest of the article is organized as follows. In Chapter~\ref{chap-IQHE} we 
give the conventional explanations of the IQHE. In particular, we discuss the
Laughlin argument, the topological aspect introduced by $TKN_{2}$ and the 
effects of localization in a qualitative way. Chapter~\ref{chap-NCG} is devoted 
to the mathematical framework needed for Non-Commutative Geometry. In particular
we describe how to overcome the difficulty of not having Bloch's theorem for 
aperiodic media. We then show that the Brillouin zone still exists as a 
non-commutative manifold. We also give the main steps of our strategy leading to
a complete mathematical description of the IQHE. In Chapter~\ref{chap-kubo}
we discuss transport theory leading to Kubo's formula. We show that in the IQHE 
idealization, the Hall conductance is a non-commutative Chern number. We also 
relate this Chern number to a Fredholm index which leads to the quantization of 
the Hall conductance. Through the notion of relative index we show in which 
sense this approach is a rigorous version of the Laughlin argument. 
Chapter~\ref{chap-loc} is devoted to localization theory. We give various 
criteria and define various localization lengths which are commonly used in the 
literature. We also show how to express these notions in the non-commutative 
language. This part allows us to explain on a rigorous basis the occurrence of 
plateaux of the Hall conductance.
Finally, we show that such criteria are in fact satisfied in models such as the
Anderson model. In Chapter~\ref{chap-phys} we give some consequences of this
theory for practical models. In particular we show that low-lying states do
not contribute to the IQHE. We also discuss the open question where the jumps of
the Hall conductance occur. The last Chapter~\ref{chap-FQHE} is a short review 
of available results on the FQHE.

\vspace{.3cm}
\noindent {\em Acknowledgements.} This paper has benefited from many contacts
during the last ten years. It is almost impossible to give the list of all
colleagues who contributed to these discussions. One of us (J.B.) gives his
special thanks to Y. Avron, R. Seiler, B. Simon, and, of course, to A. Connes
for his outstanding contribution, continuous support and warm encouragements.
We would also like to thank T. Ziman and Y. Tan for their help during the
writing of this paper.  J.B. had very helpful discussions with B.I. Shklovskii
about the Mott conductiviy. Two of us (J.B. and H.S.B.) would like to thank 
the FORTH and the University of Crete for providing support during the final
edition of this paper.

\vspace{.5cm}

\section{IQHE: experiments and theories}
\label{chap-IQHE}

\vspace{.2cm}

\subsection{The classical Hall effect}
\label{sec-CHE}
 
Let us consider a very flat conductor placed in a constant uniform magnetic 
field in the $z$ direction perpendicular to the plane $Oxy$ of the plate 
(see Fig.1). If we force a constant current in the $x$ direction, the electron 
fluid will be submitted to the Lorentz force perpendicular to the current and 
the magnetic field. Hall realized that the electron fluid is incompressible so 
that the Lorentz force must produce a pressure, namely a potential difference 
perpendicular to the current.

Let $\jV $ be the current density, $\vec{\Bb}$ the magnetic field and $\EV$ the 
Hall electric field. In a stationary state, the electric forces acting on the 
charges are opposite to the Lorentz forces. This leads to the equation

\begin{equation}
\label{Hall1}
nq\EV +\jV  \times \vec{\Bb} =0
\mbox{ , }
\end{equation}

\noindent where $n$ is the charge-carrier density and $q$ is the charge of the 
carriers. Since the magnetic field $\vec{\Bb}$ is perpendicular to both $\jV $ 
and $\EV $,  solving (\ref{Hall1}) for $\jV $ gives 

$$
\jV =\frac{nq}{\Bb^{2}}\vec{\Bb} \times \EV 
	  = \mbox{\boldmath $\sigma$}\EV 
\mbox{ , }
$$

\noindent where $\Bb$ is the modulus of the magnetic field and 
{\boldmath $\sigma$} is the conductivity tensor. The anti-diagonal components
of the tensor are the only non-vanishing ones and can be written as 
$\pm \sigma_{H} \delta$, where $\delta $ is the plate width and $\sigma_{H}$ is 
called the Hall conductance. Thus

$$
\sigma_{H}=\frac{qn\delta }{\Bb}\mbox{ . }
$$

\noindent We remark that the sign of $\sigma_{H}$ depends upon the sign of the 
carrier charge. In particular, the orientation of the Hall field will change 
when passing from electrons to holes. Both possibilities were already observed 
by Hall using various metals. This observation is commonly used nowadays to 
determine which kind of particles carries the current. 

\begin{figure}
\includegraphics{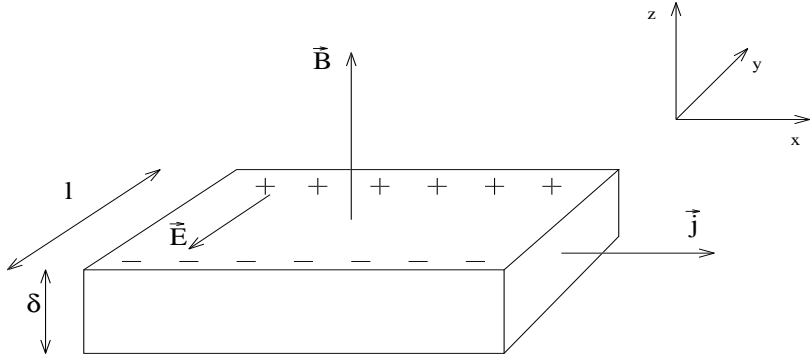}
\vglue 5.5truecm
\caption{{\sl The classical Hall effect: 
the sample is a thin metallic plate of width $\delta$. 
The magnetic field $\vec{\Bb}$ is uniform and perpendicular to the plate. 
The current density $\jV $ parallel to the $x$-axis is  stationary. 
The magnetic field pushes the charges as indicated creating the 
electric field $\EV $ along the $y$ direction. 
The Hall voltage is measured between opposite sides along 
the $y$-axis}}
\end{figure}

\noindent Let $\ell $ be the plate width in the $y$ direction (see  Fig.1). The 
current intensity inside the plate is then given by $I= j\delta \ell$ where $j$ 
is the modulus of $\jV $. The potential difference created by the Hall field is 
$V_{H}=-\ell \EV .\vec{u} $ if $\vec{u}$ is the unit vector along the $y$ axis. 
Using (\ref{Hall1}) we find:

\begin{equation}
\label{Hall2}
V_{H}=\frac{\Bb I}{nq\delta }=\frac{I}{\sigma_{H}}
\mbox{ . }
\end{equation}

\noindent In particular, for a given current intensity $I$, the thinner the 
plate the higher the potential difference. For example, for a good conductor 
like gold at room temperature, the charge carrier density is of order of 
$6\times 10^{28}m^{-3}$ (see \cite{MA} chap. 1). Thus, for a magnetic field of 
$1T$, a current intensity of $1A$ and a potential difference of $1mV$ the plate
width is about $1\mu m$. These numbers explain why the effect was so difficult 
to observe. It forced Hall to use very thin gold leaves in the beginning. In 
modern devices, much thinner ``plates'' with thickness of about $100 \AA $ are 
produced in inversion layers between two semi-conductors. 

In view of (\ref{Hall2}), the Hall conductance has the dimension of the inverse 
of a resistance. Since the product $n\delta $ is the number of charge carriers 
per unit area, the dimensionless ratio 

$$
\nu = \frac{n\delta h}{\Bb e} 
\mbox{ , }
$$

\noindent called the {\em filling factor\/}, represents the fraction of a 
Landau level filled by conduction electrons of the thin plate. In terms of this 
parameter, we obtain for a free electron gas:

\begin{equation}
\label{Hall3}
\sigma_{H}= \frac{\nu}{R_{H}}
\mbox{ , }
\hspace{2cm}
R_{H}=\frac{h}{e^{2}}
\mbox{ , }
\end{equation}

\noindent where $R_{H}$ is called the Hall resistance. It is a universal 
constant with value $R_{H}= 25812.80 \Omega$. $R_H$  can be measured directly 
with an accuracy better than $10^{-8}$ in QHE experiments. Since January 1990, 
this is the new standard of resistance at the national bureau of standards 
\cite{ResH}. 

\vspace{.2cm}

\subsection{The Quantum Hall Effect}
\label{sec-QHE}

\begin{figure}
\includegraphics{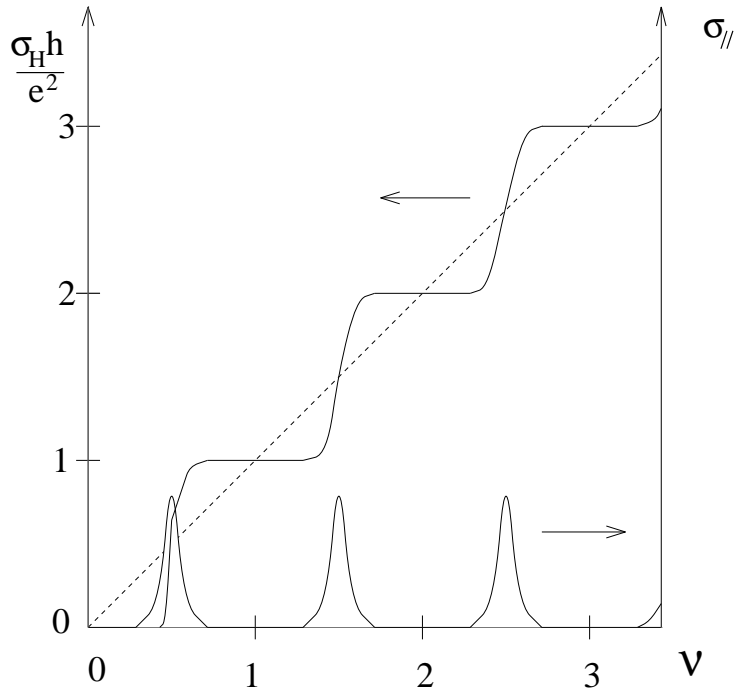}
\vglue 9.5truecm
\caption{{\sl Schematic representation of the experimental observations in the
IQHE. The Hall conductivity $\sigma_H$
is drawn in units of $e^{2}/h$ versus filling 
factor $\nu$. The dashed line shows the Hall conductivity of the Landau 
Hamiltonian without disorder.
The direct conductivity $\sigma_{//}$ is shown in arbitrary units.}}
\end{figure}

Let us concentrate upon the dependence of the Hall conductance (in units of 
$e^{2}/h$) on the filling factor $\nu$. In the classical Hall effect, these two 
quantities are just equal (eq.\ref{Hall3}). Lowering the temperature below $1K$ 
leads to the observation of plateaux for integer values of the Hall conductance 
(see Fig.2). In von Klitzing's experiment \cite{KDP}, the variation of $\nu$ was
obtained by changing the charge carrier density, whereas in later experiments
one preferred varying the magnetic field. The accuracy of the Hall conductance on
the plateaux is better than $10^{-8}$. For values of the filling factor 
corresponding to the plateaux, the direct conductivity $\sigma_{//}$, 
namely the conductivity along the current density axis, vanishes.
 These two observations are actually the most important 
ones. The main problems to be explained are the following: 

\bed
\item[(i)]  Why do the plateaux appear exactly at integer values ? 

\item[(ii)]  How do the plateaux appear ?

\item[(iii)] Why are these plateaux related to the vanishing of the direct 
conductivity ? 
\ed

\noindent To observe the QHE, physicists have used conduction electrons trapped in the 
vicinity of an interface between two semiconductors. The local potential 
difference between the two sides produces a bending of the local Fermi level. 
Near the interface, this Fermi level meets with the valence band creating states
liable to participate in the conductivity. This bending occurs on a distance of 
the order of $100 \AA$ from the interface, so that the charge carriers are 
effectively concentrated within such a thin strip. In addition, by changing the 
potential difference between the two sides, the so-called gate voltage, one can 
control the charge-carrier density. 

The samples used in QHE experiments belong to two different categories. The 
first one is called MOSFET \cite{AnAo}, 
for metal-oxide silicon field effect 
transistors. The interface separates  doped silicon from silicon oxide. This 
device was common in the beginning of the eighties and was the one used in von 
Klitzing's experiment. However, the electron mobility is relatively low because 
the control of the flatness of the interface is difficult. 

The samples of the other category are heterojunctions. The interface separates 
$GaAs$ from an alloy of $Al_xGa_{1-x}As$. This kind of device nowadays makes 
available interfaces almost without any defects. Moreover, electrons therein 
have a high mobility. These devices are most commonly used in modern quantum 
Hall experiments.

In both kinds of samples, there are many sources of defects producing 
microscopic disorder. The first comes from the doping ions. Even though they are
usually far from the interface (about $1000 \AA$), the long range Coulomb 
potential they produce is strong enough to influence the charges on the 
interface. It is not possible to control the position of these ions in the 
crystal. The second source of defects is the roughness of the interface. This is
an important effect in MOSFET's, much less in heterojunctions. In the latter the
accuracy is better than one atomic layer in every $1000 \AA $ along the  
interface \cite{PG}. Finally, long range density modulation of the compounds may
produce visible effects. This is the case especially for heterojunctions where 
the aluminium concentration may vary by a few percent on a scale of $1\mu m$ 
\cite{GG}. 

It is important to notice that the observation of plateaux supposes 
several conditions. 

\bed
\item[(i)] The effect is more easily seen if the electron fluid is concentrated 
in such a thin region that it can be considered as two-dimensional. In fact, 
owing to the trapping effect of the potential interface, the motion 
perpendicular to the interface is quantized. For good samples in high magnetic 
fields, the energy difference between two corresponding eigenvalues is big 
compared to $k_{B}T$ (where $T$ is the temperature and $k_{B}$ the Boltzmann 
constant), so that only the lowest such level has to be considered. Hence the 
problem becomes effectively two-dimensional. 

\item[(ii)] The plateaux disappear beyond a temperature of a few Kelvin. More 
exact, the inelastic relaxation time has to be large enough; otherwise 
corrections will be needed in formul{\ae} calculating the current; this will 
lead to the destruction of the plateaux. This is the reason why the IQHE is 
seen more easily in heterojunctions than in MOSFETs: the electron mobility in 
the former is higher than in the latter.

\item[(iii)] We will see that some quenched disorder producing only elastic
scattering is necessary for the appearance of the plateaux (see
Section~\ref{sec-loc}). In practice, the disorder that occurs is strong enough 
to produce a filling  of the gaps between Landau levels \cite{GG,SWW}. 

\item[(iv)] Clearly the sample size must be big enough as to allow the use of the 
infinite volume limit. Mesoscopic systems exhibit conductance fluctuations from 
sample to sample 
which may partially distroy the effect. 
Finite volume effects however have been shown to decrease exponentially fast 
with the sample size.

\item[(v)] Finally the electric field needed to produce the current has to be
small.  If it is too high, non-linear phenomena may distroy the plateaux 
\cite{KIYYWK}.
\ed

Provided the previous conditions are satisfied, the quantum Hall effect is a 
universal phenomenon. It is quite independent of the specific shape of the 
sample. Hall plateaux have also been observed in microwave experiments where the
topology of the sample is trivial \cite{KMWS}. Note, however, that the centers 
of the plateaux need not be located near integer values of the filling factor. 
It depends upon which kind of doping ion is used \cite{GHP}. The values of the 
Hall conductance, however, on the plateaux are independent of the nature of the 
used sample.

\vspace{.2cm}

\subsection{The Hall effect for the free Fermi gas}
\label{sec-freeFG}

Let us first consider a very simple model in which the charge carriers are 
spinless, free, two-dimensional fermions with charge $q$. Our aim is to show 
that no quantization of the Hall conductance is observed in this case. 

Since the particles are independent, the quantum motion is described by the 
one-particle Hamiltonian. Let $\vec{A}=(A_{1},A_{2})$ be the vector potential 
given by 

$$ \partial_{1}A_{2}-\partial_{2}A_{1}=\Bb  \mbox{ , } $$

\noindent where $\Bb $ is the modulus of the magnetic field. The energy operator
is then given by the Landau Hamiltonian \cite{Lan}

\begin{equation}
\label{Landau}
H_{L}= \frac{(\vec{P}-q\vec{A})^{2}}{2m_{\ast }}
\mbox{ , }
\end{equation}

\noindent where $\vec{P}$ is the 2D momentum operator and $m_{\ast }$ is the 
effective mass of the particle. This operator is not translation invariant 
owing to the symmetry breaking produced by the vector potential. However, if one
replaces the usual representation of the translation group by the so-called 
{\em magnetic translations\/} \cite{Zak}, the Landau Hamiltonian becomes 
translation invariant. Let us introduce the quasimomentum operators 
$\vec{K} = (\vec{P}-q\vec{A}) / \hbar$; they fulfill the canonical commutation 
relations

$$
[K_{1},K_{2}]=\imath \frac{q\Bb }{\hbar}
\mbox{ . }
$$

\noindent Here $q\Bb /\hbar$ plays the r\^ole of an effective Planck constant. We 
see that the Landau Hamiltonian describes a harmonic oscillator so that its 
spectrum is given by the Landau levels, namely

$$E_{n}=\hbar \omega _{c} (n+\frac{1}{2})
\mbox{ , }
$$

\noindent where $\omega _{c}=q\Bb /m_{\ast }$ is the cyclotron frequency and 
$n\in \NN$. Each Landau level is infinitely degenerate owing to translation 
invariance. The degeneracy per unit area is finite however and given by $q\Bb /h$. 
Perturbing this operator will give rise to a band spectrum.

Let us now compute the current. We assume the system to be in a thermodynamical 
equilibrium. In order to allow non-zero current, we describe the system in the 
grand-canonical ensemble; in practice such a fluid is open. Since the particles 
are independent fermions, the thermal averaged density per unit volume of a 
one-particle translation invariant extensive observable ${\cal O}$ at 
temperature $T$ and chemical potential $\mu$ is given by

$$
<{\cal O}>_{T,\mu}= 
  \lim_{\Lambda \uparrow \RR^{2}}
    \frac{1}{| \Lambda |}
      Tr_{\Lambda} 
        (f_{T,\mu}(H_{L}) {\cal O} )
\mbox{ , }
$$

\noindent where $f_{T,\mu}(E) =(1+e^{\beta (E-\mu)})^{-1}$ is the 
Fermi-distribution function and $\beta =1/k_{B}T$. In the limit above, 
$\Lambda $ denotes a square box centered at the origin, and $|\Lambda |$ is its 
area. The current is represented by the operator

$$
\JV   = 
  \frac{\imath  q}{\hbar}
    [H_{L},\XV ]
\mbox{ , }
$$

\noindent where $\XV $ is the position operator. It is easy to check that this 
current operator commutes with the magnetic translations. Obviously the thermal 
average of the current vanishes since no current can flow without an external
source of energy. To make it non-zero, we switch on an electric field $\EV $ at 
time $t=0$. 
For simplicity, we assume this field to be uniform and time-independent. After 
the field has been switched on, the time evolution is given by

$$
\frac{d\JV }{dt}= 
  \frac{\imath }{\hbar}
    [H_{L,\EV },\JV ]
\mbox{ , }
\hspace{1cm}
\mbox{\rm with}
\hspace{1cm}
H_{L,\EV }= H_{L}-q\EV \XV 
\mbox{ . }
$$

\noindent The solution of this equation is elementary. Using complexified 
variables, that is $M=M_{1}+\imath M_{2}$ whenever $\vec{M}=(M_{1},M_{2})$ is a 
2D vector, we find

$$
J(t)= -\imath q 
   \frac{\Ee}{\Bb } + 
     e^{-\imath \omega _{c} t} J_{0}
\mbox{ , }
$$

\noindent where $J_{0}$ is some initial datum. This solution consists of a 
time-independent part and of an oscillating part with the cyclotron frequency. 
The time average is just the constant part; it is the system's response to the
applied electric field. Taking the thermal average of this constant part, we 
have the observed current, namely, if $j=j_{1}+\imath j_{2}$, 

$$
j= 
 \lim_{t\rightarrow \infty}
   \int_{0}^{t}
     \frac{ds}{t}
       <J(s)>_{T,\mu}
= -\imath \frac{q{\cal E}}{\Bb }
    \lim_{\Lambda \uparrow \RR^{2}}
      \frac{1}{| \Lambda |}
        Tr_{\Lambda} 
          (f_{T,\mu}(H_{L}))
= -\imath qn \frac{\Ee}{\Bb }
\mbox{ . }
$$

\noindent This is nothing but the classical formula~(\ref{Hall1}), written in 
complex notation. 

Therefore we see that the classical Hall formula still holds in quantum
mechanics for the free fermion gas at all temperatures. There is no way to see 
any trace of quantization of the Hall conductance, neither is there any kind of 
plateaux of the Hall conductance! 

\vspace{.2cm}

\subsection{The r\^ole of localization}
\label{sec-loc}

The main experimental property we have not yet considered  is the vanishing of 
the direct conductivity when the filling factor corresponds to the plateaux. 
In the previous argument we ignored all effects leading to a finite non-zero 
direct conductivity. Among these effects there are several sources of 
dissipation such as phonon scattering or photoemission. At very low temperature,
these sources usually have very limited influence and impurity scattering 
dominates. This is why we are led to consider non-dissipative effects like 
Anderson localization in disordered systems. As explained in 
Section~\ref{sec-QHE}, several kinds of defects can influence the electron 
motion. These defects are usually distributed in a random way so that the forces
they create on the charge carriers are actually represented by a random 
potential. In many cases one considers these defects as isolated and of small 
influence, so that a first-order perturbation theory based upon one-electron 
scattering on one impurity already gives a good account of the observed effects.
For 2D systems however, it turns out that the low-density limit for impurities 
does not give the relevant contribution. We will see in Section~\ref{sec-homSO} 
how to define properly a potential representing the effects of a high density of
random scatterers. For the moment let us stay at an intuitive level. 

As explained by P.W. Anderson \cite{An}, the occurrence of a random potential in 
a one-particle Hamiltonian may lead to the quantum localization of particles. 
More precisely, the quantal wave representing these particles reflects on the 
potential bumps producing interferences. The Bragg condition is necessary for 
building a constructive interference pattern throughout the crystal. This 
requires some regularity of the crystal, such as periodicity or 
quasiperiodicity, in order to allow the wave to propagate. If the potential 
creating these bumps exhibits some randomness, the probability for the Bragg 
condition to be satisfied everywhere in the crystal eventually vanishes. 
Therefore, the wave function will vanish at infinity leading to the trapping of 
particles in the local minima of the potential. This is Anderson localization. 

It has been argued \cite{AALX} that one-dimensional systems of non-interacting 
particles in a disordered potential exhibit localization at any strength of 
disorder. Mathematicians have proved such a claim under relatively mild 
conditions on the randomness of the potential \cite{GoMo,KuSo,CaLa,FiPa,CFKS}. 
They have also proved localization in any dimension for strong disorder 
\cite{FrSp,MSFS,DSL,SW,MSGB,AM}.
A finite-size scaling argument, proposed by Thouless \cite{Thou}, has been used 
in \cite{AALX} to show that two-dimensional systems are also localized at any 
disorder unless there are spin-orbit couplings \cite{Hilcani}. The same argument
also shows that in higher dimensions, localization  disappears at low disorder. 
We remark that the notion of localization we are using does not exclude 
divergence of the localization length at an isolated energy.  These results were
supplemented by many numerical calculations \cite{PiSa,Kramers,Kramers2,HuKr}.

\begin{figure}
\label{DOS}
\includegraphics{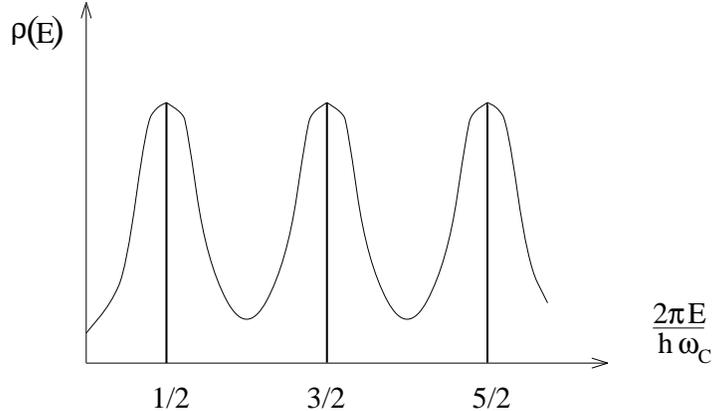}
\vglue 5.5truecm
\caption{{\sl The density of states for the Landau Hamiltonian with a 
random potential.}}
\end{figure}

On the other hand, the occurrence of a random potential will create new states 
with energies in the gaps between Landau levels. This can be measured by the 
{\em density of states\/} (DOS). To define it, let ${\cal N}(E)$ be the 
{\em integrated density of states\/} (IDS), namely the number of eigenstates of 
the Hamiltonian per unit volume below the energy $E$:

\begin{equation}
\label{IDOS}
{\cal N}(E)=\lim_{\Lambda \rightarrow \infty} \frac{1}{|\Lambda|}
\#\{ \mbox{eigenvalues of } H|_\Lambda \le E \},
\end{equation}
 
\noindent where degenerate eigenvalues are counted with their multiplicity. 
This is a non-decreasing function of $E$. Therefore its derivative 
$\rho (E)=d{\cal N}(E)/dE$ is well defined as a Stieljes-Lebesgue positive 
measure and is called the DOS. Under mild conditions on the distribution of the 
random potential, it is possible to show \cite{Klein,CFKS,FiPa} that the DOS is 
a smooth function (see Figure~\ref{DOS}). If the potential strength exceeds the 
energy difference $\hbar \omega _{c}$ between two consecutive Landau levels, the
gaps will be entirely filled. This is actually the physical situation in most 
samples used up to now. In heterojunctions, for instance, the minimum of the 
density of states may represent as much as 30\% of its maximum
\cite{GG}, although in modern samples it is usually very small
\cite{SWW}.

Recall that the chemical potential at zero temperature is called the {\em Fermi 
level } $E_{F}$. Since the charge carriers are spinless fermions, their density 
at $T=0$ is given by

\begin{equation}
\label{eq-DOS}
n=
  \int_{-\infty}^{E_{F}}
    dE \;\rho (E)=
     {\cal N}(E_{F})
\mbox{ . }
\end{equation}

\noindent The absence of spectral gaps means that the IDS is monotone 
increasing, so that changing continuously the particle density is equivalent to 
changing continuously the Fermi level. More generally, if a magnetic field is 
switched on, changing continuously the filling factor is equivalent to changing 
continuously the Fermi level. On the other hand, while the Fermi level crosses a
region of localized states, the direct conductivity must vanish whereas the Hall
conductivity cannot change. This last fact is not immediately obvious, but more 
justification will be given later on. Changing the filling factor therefore will
force the Fermi level to change continuously within this region of localized 
states while the Hall conductance will stay constant. This is the main mechanism
leading to the existence of plateaux.

In contrast, if the spectrum had no localized states, the Hall conductance would
change while changing the filling factor, as long as the Fermi level would move 
within the spectrum. Moreover, if a spectral gap occurred between two 
bands, let us say between energies $E_{-}$ and $E_{+}$, the IDS 
would equal a constant $n_{0}$ on that gap and changing the filling factor from 
$n_{0}-\epsilon$ to $n_{0}+\epsilon$ would cause the Fermi level to jump 
discontinuously from $E_{-}$ to $E_{+}$ with the value $(E_{-}+E_{+})/2$ 
whenever $n=n_{0}$. In this case, once again, no plateaux could be observed. 
This is why there is no quantized Hall effect in the free fermion theory.

One of the consequences of this argument is that between two different plateaux,
there must be an energy for which the localization length diverges 
\cite{GoMo,Ku}. Even though there should be no extended states in 2D disordered 
systems, the localization length need not be constant in energy. As it happens, 
if the impurities are electrically neutral in the average, the localization 
length diverges exactly at the Landau levels. Therefore, the influence of 
impurities decreases near the band centers, and other sources of interactions, 
like the Coulomb potential between charge carriers, may become dominant. This is
actually the basic observation leading to the understanding of the FQHE. 

We have argued that the direct conductivity should vanish whenever the Fermi 
level lies in a region of localized states. One may wonder how it is then 
possible to have a non-zero Hall conductance. The answer is that the Hall 
current is actually carried by edge states. This has been seen in numerical 
calculations \cite{SKmcK,JRT}
and there are the first theoretical hints 
\cite{Hat}. The experimental situation is not as clear as it looks from theory. 
This is probably due to the existence of zones in which states are localized 
surrounded by filamentary regions in which the Hall current can actually flow. 

\vspace{.2cm}

\subsection{The Laughlin argument}
\label{sec-IQHcond}

The first attempt to explain the integer quantization of the Hall conductance
in units of $e^2/h$ was proposed by Laughlin \cite{Laug}. Laughlin originally
chose a cylinder geometry for his argument. His justification for this was that
the effect seemed to be universal and should therefore be independent of the 
choice of the geometry. Here, we present 
a Laughlin argument in the plane in form of a singular gauge transformation. In
Section~\ref{sec-QuaRelIndex}, this presentation and an observation of Avron,
Seiler and Simon \cite{ASS} will allow us to discuss the links between the
Laughlin argument and the Chern-character approach presented in this article.

\begin{figure}
\label{LaArg}
\includegraphics{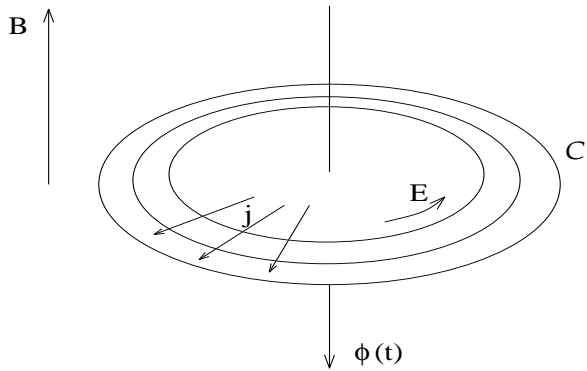}
\vglue 5.5truecm
\caption{{\sl The chosen geometry for the Laughlin Gedanken experiment.}}
\end{figure}

Let us consider a free gas of non-interacting electrons in the two-dimensional
plane subjected to an exterior magnetic field perpendicular to the plane. We
choose an origin and then pass an infinitely thin flux tube through it. A radial
electromotive force is created by means of a slowly varying flux $\phi(t)$
(see Fig \ref{LaArg}). In polar coordinates, the vector potential with symmetric
gauge for the constant field is 
$A(r,\theta)=(-\frac{B}{2}r \mbox{sin}\theta - \frac{1}{2\pi r} \mbox{sin}\theta \; \phi(t),
               \frac{B}{2}r \mbox{cos}\theta + \frac{1}{2\pi r} \mbox{cos}\theta \; \phi(t))$; 
the Hamiltonian then reads:

\begin{equation}
\label{polarcoordinates}
H_\Bb  = \frac{1}{2m_*}\left\{
                       - \hbar^2 \partial_{r}^{2} 
                       - \hbar^2 \frac{1}{r} \partial_r 
                       + \left(
                             \frac{\hbar}{\imath} \frac{1}{r} \partial_{\theta} 
                           + \frac{er\Bb }{2} 
                           + \frac{ e\phi(t)}{2\pi r}
                         \right)^2
                    \right\}.
\end{equation}

Recall that the electron charge is $-e$. We now assume that the time evolution
of the flux through the cylinder is so slow that the adiabatic approximation
describes the evolution of the states. The eigenstates may then be computed
explicitly:

\begin{equation}
\label{extended}
\psi_{n,m}(z,\theta;t) =
    C_{n,m,\phi} \; e^{-\imath m \theta}
    \left(
       \frac{r}{\ell_\Bb }
    \right)^{m+\frac{2\pi e\phi(t)}{h}} 
    e^{-\frac{r^2}{4 \ell_{\Bb}^2}} \; 
    L_n^{m+\frac{2\pi e \phi(t)}{h}}(\frac{r^2}{\ell_{\Bb}^2}),
\end{equation}

\noindent where $\ell_\Bb =\sqrt{2\hbar/q\Bb }$ is the magnetic length; $n\in \NN$ is 
the principal quantum number corresponding to the energy 
$E_n=\hbar\omega_c(n+1/2)$ and $m\in \ZZ$ is the orbital quantum number so that 
the angular momentum is $m\hbar$; $C_{n,m,\phi}$ are easily computable 
normalization constants and the polynomials that appear are the Laguerre 
polynomials given by:

$$
L_n^{\alpha}(x) = \frac{1}{n!} e^x x^{-\alpha}
                  \partial_x^n (e^{-x}x^{n+\alpha})
\qquad n\in\NN,\;\alpha\in\RR. 
$$

As the flux $\phi(t) = \frac{h t}{e \tau}$ is varied from $t=0$ to $t=\tau$, one
flux quantum is forced through the tube and a state $\psi_{n,m}$ evolves to a
state $\psi_{n,m+1}$ up to a phase factor $e^{-i\theta}$. Note that this 
reflects the fact that the singular gauge transformation introduces an extra
angular momentum $-\hbar$ into the system. Now we assume that the filling factor
is an integer $N$. The net effect of the above process is that the state with
lowest angular momentum of each Landau level is transported to infinity. Let us
fix a large circle ${\cal C}$ around the origin of radius $R$. The current
density during the process is then (approximatively) given by 
$j = \frac{1}{2\pi R} \frac{-Ne}{\tau}$; the strength of the electric field on
the circle is ${\cal E} = -\partial_t (\phi(t)/(2\pi R)) = -\hbar/(\tau e R)$; 
therefore, the Hall conductivity is calculated as $\sigma_H=j/{\cal E}=Ne^2/h$.

What did we learn from this argument ? Until now, not so much in fact. For had 
we taken another non-integer filling factor $\nu$ and supposed that the electron
density were uniform, we would obtain $\sigma_H=\nu e^{2}/h$; that is the 
classical result already calculated in Section~\ref{sec-loc}.

Using ideas of Prange, Halperin and Joynt \cite{Prange,JP,Halp} however, the
picture may be completed to furnish a qualitative understanding of the IQHE in the
following way: if the sample has impurities, these transform some of the 
extended states (\ref{extended}) into localized states; moreover, the explicit 
form of the remaining extended states changes. Now the localized states will not
change as a flux quantum is forced through the flux tube, and it can be argued 
that the remaining extended states of one Landau level still carry the same 
current, that is one charge $e$ through the circle ${\cal C}$. As some of the
localized states caused by the impurities are situated below the energy of the 
Landau levels and others above, we come to the desired qualitative explanation 
with help of the argument explained in Section~\ref{sec-loc}.

There are two ingredients in this Gedanken-experiment which ought to be
emphasized. The first is the gauge invariance which produces the periodicity
of the Hamiltonian with respect to the varying flux, the period being the flux
quantum. The second is that the Hall effect corresponds to a charge transport of
one unit for each filled Landau level. The conductance quantization is therefore
likely to be connected to the charge quantization.

\vspace{.2cm}

\subsection{The Chern-Kubo relationship}
\label{sec-ChKu}

The geometrical origin of the Hall conductance quantization was revealed by
$TKN_2$ \cite{TKN} and Avron et al. \cite{ASS1}. $TKN_2$ considered an electron 
gas submitted to a uniform magnetic field on a 2D square lattice in the 
tight-binding approximation. In the Landau gauge the Hamiltonian reads:

\begin{equation}
\label{eq-Har1}
\begin{array}{lcl}
(H\psi )(m_{1},m_{2}) & = & \psi (m_{1}-1,m_{2})+ \psi (m_{1}+1,m_{2}) 
                            \nonumber \\
                      &   & \\
                      & + & e^{-2\pi \alpha m_{1}} \psi (m_{1},m_{2}-1)
                        +   e^{2\pi \alpha m_{1}}  \psi (m_{1},m_{2}+1)  
\mbox{ , }
\end{array}
\end{equation}

\noindent where $\alpha =\phi /\phi_{0}$ is the ratio of the magnetic flux 
$\phi$ in the unit cell to the flux quantum $\phi_{0}=h/e$. Using the 
translation invariance along the $y$-axis, the solution of the eigenvalue 
equation can be written in the form

$$
\psi (m_{1},m_{2})= 
    e^{-\imath k_{2}m_{2}}
	 \varphi (m_{1})
\mbox{ , }
$$

\noindent with

\begin{equation}
\label{eq-Har2}
 \varphi (m_{1}-1)+
  \varphi (m_{1}+1)+
   2\cos{(k_{2}-2\pi \alpha m_{1})}\varphi (m_{1})
   =E(k_{2})\varphi (m_{1})
\mbox{ . }
\end{equation}

\noindent 
This is Harper's equation \cite{Har}. It can also be obtained by adding a weak
periodic potential to the Landau Hamiltonian (\ref{Landau}) and then projecting
onto one Landau level. Now $TKN_{2}$ assumed that the ratio $\alpha$ is a 
rational number of the form $p/q$ where $p$ and $q$ are relatively prime 
integers. By translation invariance of period $q$, the solution of 
eq.~(\ref{eq-Har2}) can also be written in Bloch's form

$$
\varphi (m_{1})= 
    e^{-\imath k_{1}m_{1}}
	 \xi (m_{1})
\mbox{ , }
\hspace{2cm}
\xi (m_{1}+q)=\xi (m_{1})
\mbox{ , }
$$

\noindent so that Harper's equation now becomes 

$$
\begin{array}{lcl}
(H(k_{1},k_{2})\xi )(n) & := &
    e^{\imath k_{1}}\xi (n-1)+
	 e^{-\imath k_{1}}\xi (n+1)+
	  2\cos{(k_{2}-2\pi \frac{p}{q}n)}\xi (n) \nonumber \\
	  &   &  \\
	  & = &
	   E(k_{1},k_{2})\xi (n)
\mbox{ . }
\end{array}
$$

\noindent This is the secular equation for the $q\times q$ hermitian matrix 
$H(k_{1},k_{2})$. The energy spectrum is given by energy bands corresponding
to the eigenvalues $\epsilon_{l}(k_{1},k_{2})$, $1\leq l\leq q$. Let 
$P_{l}(k_{1},k_{2})$ be the corresponding eigenprojection. We will assume for 
simplicity that all bands are well separated: for the Harper equation, all gaps 
but the central one are open \cite{BeSi,vMou,CEY}.

It is quite clear that $H(k_{1},k_{2})$ is a trigonometric polynomial in
$(k_{1},k_{2})$, implying that the eigenprojections and the eigenvalues are 
smooth periodic functions of $(k_{1},k_{2})$. Moreover, introducing the two 
unitary $q\times q$ matrices defined by 

$$
(u_{1}\xi )(n)= \xi (n-1)
\mbox{ , }
\hspace{1cm}
(u_{2}\xi )(n)= e^{-2\imath \pi \frac{p}{q}n}\xi (n)
\mbox{ , }
$$

\noindent we derive the covariance relation

\begin{equation}
\label{eq-cov}
u_{1}H(k_{1},k_{2})u_{1}^{-1}= H(k_{1},k_{2}+2\pi \frac{p}{q})
\mbox{ , }
\hspace{2cm}
u_{2}H(k_{1},k_{2})u_{2}^{-1}= H(k_{1}-2\pi \frac{p}{q},k_{2})
\mbox{ . }
\end{equation}

\noindent These relations show that the eigenvalues and the trace of any 
function of $H$ and its derivatives are actually $2\pi /q$-periodic in 
$(k_{1},k_{2})$.

The Hall conductance can now be calculated by Kubo's formula \cite{Kubo}.
It gives the current as the velocity-velocity correlation function. The formula 
may be deduced from classical arguments using the Boltzmann equation 
\cite{Kubo}, but a deduction closer to quantum mechanics using linear response 
theory will be presented in Chapter~\ref{chap-kubo}. The result derived by 
$TKN_{2}$ is then the following: at zero temperature, if the Fermi level belongs
to a gap of the Harper Hamiltonian of eq.~(\ref{eq-Har1}), the transverse 
conductivity is given by the Kubo-Chern relation (see \cite{ASS1})

$$
\sigma _{H}= 
 \frac{e^{2}}{h} {\rm Ch}(P_{F})
\mbox{ , }
$$

\noindent where

\begin{equation}
\label{eq-KuCh1}
\Ch (P_{F})=
  \int_{\Tg ^{2}} 
   \frac{d^{2}k}{4 \pi^2} \left\{ 2\pi\imath\frac{1}{q}
    \TR (
	 P_{F}(k_{1},k_{2})
	  [\partial_{1} P_{F}(k_{1},k_{2}),\partial_{2} P_{F}(k_{1},k_{2})]
	  )\right\}
\mbox{ . }
\end{equation}

\noindent Here $\partial_{a} =\partial /\partial k_{a},a=1,2$ and $P_{F}$
is the eigenprojection on energies smaller than the Fermi level. The integrand 
is a complex 2-form on the torus; its de~Rham-cohomology class is called the 
first {\em Chern class} or the first {\em Chern character}
of the projection $P_F$. 
We shall also refer to the integral over this form as Chern character, because
the integral will retain meaning in the non-commutative context; the integral 
will also be called Chern number.

For further informations on Chern classes and characters
we refer to \cite{LM,Hir}. The Chern 
character of a projection is a topological invariant. To see this, we remark that 
the data of the family $k=(k_{1},k_{2})\in \Tg ^{2}\mapsto P_{F}(k)$ of 
projections defines a complex fiber bundle $\Pp _{F}$ over the $2$-torus in a 
natural way. More precisely, $\Pp _{F}$ is the set of pairs 
$(k,\xi )\in \Tg ^{2}\times \CC ^{q}$ such that $P_{F}(k)\xi = \xi$. Since the 
Fermi level lies in a gap, $P_{F}$ is smooth and its dimension is constant. 
Therefore $\Pp _{F}$ is a smooth vector bundle over the $2$-torus. It can be 
shown that $\Ch (P_{F})$ depends only upon the equivalence class of $\Pp _{F}$ 
provided the equivalence is isomorphism of vector bundles modulo adding trivial 
bundles. In particular, it is a homotopy invariant quantity. This implies that 
adding a small perturbation to the Harper Hamiltonian will not change the value 
of $\Ch (P_{F})$, at least as long as the Fermi level does not cross the 
spectrum while turning on the perturbation! This is the argument that was needed
to explain the robustness of the Hall conductance. 

In addition to being robust, the Chern number $\Ch (P_{F})$ is actually an 
integer. To see this explicitly, we will show (see Section~\ref{sec-NCChern}) 
that the Chern character $Ch$ is additive with respect to the direct sum of
two orthogonal projections $P$ and $Q$, that is 
$\Ch (P\oplus Q)= \Ch (P)+\Ch (Q)$. Even though this is not obvious from 
eq.~(\ref{eq-KuCh1}), all cross-terms in the left hand side vanish. It is thus 
sufficient to show that if $P$ is a one dimensional smooth projection over 
$\Tg^{2}$, then its Chern character is an integer. This can be shown as 
follows. 

Let $\xi (k)$ be a unit vector in the vector space $\CC ^{q}$ such that 
$P(k)\xi (k)=\xi (k),\, \forall k$. It is easy to show that 

\begin{equation}
\label{eq-Chern1}
\Ch (P)=
\int_{\Tg ^{2}} 
   \frac{d^{2}k}{2\imath \pi}
    \TR (
	 P(k)
	  [\partial_{1} P(k),\partial_{2} P(k)]
	  )
	  =\frac{1}{\pi}
	    \int_{0}^{2\pi} dk_{1}
		 \int_{0}^{2\pi} dk_{2}
          \Im m<\partial_{1}\xi (k)|\partial_{2}\xi (k)>
\mbox{ , }
\end{equation}

\noindent where $\Im m$ is the imaginary part. Using Stokes formula, this double
integral can be written as 

\begin{equation}
\label{eq-Chern2}
\Ch (P)=
 \int_{0}^{2\pi}
  \frac{dk_{1}}{2\imath \pi}
   <\xi (k)|\partial_{1}\xi (k)>|_{k_{2}=0}^{k_{2}=2\pi}
   -
    \int_{0}^{2\pi}
     \frac{dk_{2}}{2\imath \pi}
       <\xi (k)|\partial_{2}\xi (k)>|_{k_{1}=0}^{k_{1}=2\pi}
\mbox{ . }
\end{equation}

\noindent Even though $P$ is doubly periodic in $k$, $\xi $ need not be 
periodic. The obstruction to the periodicity of $\xi$ is precisely the 
non-vanishing of the Chern character. However, we can find two real functions 
$\theta_{1}$ and $\theta_{2}$ such that 

\begin{equation}
\label{eq-per}
\xi (k_{1},2\pi )=e^{\imath \theta_{1}(k_{1})}\xi (k_{1},0)
\mbox{ , }
\hspace{2cm}
\xi (2\pi ,k_{2})=e^{\imath \theta_{2}(k_{2})}\xi (0,k_{2})
\mbox{ , }
\end{equation}

\noindent provided $0\leq k_{1}\leq 2\pi$ and $0\leq k_{2}\leq 2\pi$ 
respectively. The reason is that after one period the new vector $\xi$ defines 
the same subspace as the old one, so they are proportional. Since $\xi (k)$ is a
unit vector for any $k$, the proportionality factor must be a phase. Writing 
$\xi (2\pi ,2\pi )$ in two ways we get 
$\theta_{1}(2\pi ) - \theta_{1}(0)  = 
 \theta_{2}(2\pi )-  \theta_{2}(0) \pmod{2\pi}$. 
Replacing in eq.~(\ref{eq-Chern2}) leads to

$$
\Ch (P)=
\frac{(\theta_{1}(2\pi )-\theta_{1}(0))
  -(\theta_{2}(2\pi )-\theta_{2}(0))}{2\pi}
   \in \ZZ 
\mbox{ . }
$$

\noindent Thanks to this argument, we see that the Chern character of a smooth
projection valued function over the $2$-torus is an integer.

We remark, however, that the expression used in eq.~(\ref{eq-Chern1}) is not 
exactly the same as the one used in eq.~(\ref{eq-KuCh1}): in the latter, we have
divided by $q$. However, the covariance relation eq.~(\ref{eq-cov}) shows that 
the integrand in eq.~(\ref{eq-KuCh1}) is $2\pi/q$-periodic in $k$ so that it is 
sufficient to integrate over the square $[0,2\pi/q]^{\times 2}$ and multiply the
result by $q^{2}$. On the other hand, the periodicity condition (\ref{eq-per}) 
has to be modified to take the covariance into account. Now the argument goes 
along the same line and gives rise to an integer as well. 

\vspace{.2cm}

This series of arguments shows that indeed the Hall conductance (in units of 
$e^{2}/h$)is a very robust integer, as long as the Fermi level remains in a 
spectral gap. However, the previous argument does not work for any kind of 
perturbation. It is valid provided the perturbation of the Hamiltonian is given 
by a matrix valued function $\delta H(k)$ satisfying the same covariance 
condition (\ref{eq-cov}). Otherwise, the previous construction does not tell us 
anything about the homotopy invariance of the Hall conductance. This is exactly 
the limitation of the $TKN_{2}$ result. More precisely, no disorder is allowed 
here and, in addition, only a rational magnetic field can be treated in this 
way, a quite unphysical constraint. For indeed, in most experiments the 
variation of the filling factor is obtained through changing continuously the 
magnetic field. Even though rational numbers are dense in the real line, the 
previous expression (\ref{eq-KuCh1}) depends in an explicit way on the 
denominator $q$ of the fraction so that we have no way to check whether the 
integer obtained is robust when the magnetic field is changed.

The drastic condition of having no disorder is also disastrous: the periodicity 
of the Hamiltonian excludes localized states, and the argument in 
Section~\ref{sec-loc} shows that no plateaux can be observed in this case. 
Therefore we have not yet reached the goal. 

Nevertheless, the main result of $TKN_{2}$ is the recognition that the Hall
conductance can be interpreted as a Chern character. This is a key fact in
understanding why it is quantized and robust as well. Moreover, the topology
underlying this Chern character is that of the quasimomentum space, namely
the periodic {\em Brillouin zone} which is a $2$-torus. From this point of view,
the Laughlin argument in its original form may be misleading since it gives the
impression of emphasizing the topology of the sample in real space, which has 
nothing to do  with the IQHE. 

The solution to the $TKN_{2}$ limitation is given by the Non-Commutative
Geometry of the Brillouin zone where both of the above restrictions can be 
dropped. We will see that the Kubo-Chern relation still holds, that the Chern 
character is still a topological invariant and that it is an integer. But we 
will also discover something new, characteristic of Non-Commutative Geometry, 
namely that localization can be expressed in a very clear way in this context 
and that the conclusion will be valid as long as the Fermi level lies in a 
region of {\em localized states}.

\vspace{.5cm}

\section{The Non-Commutative Geometry of the IQHE}
\label{chap-NCG}

{\bf The four-traces way}

\noindent In this chapter the strategy and the main steps of the non-commutative
approach are given. In particular we will introduce the four different traces
that are technically needed to express the complete results of this theory. The 
first one is the usual trace on matrices or on trace-class operators. The second
one, introduced in the Section~\ref{NCBzone} below, is the {\em trace per unit 
volume} which permits to compute the Hall conductance by the Kubo formula. The 
third one is the {\em graded trace} or {\em supertrace} introduced in 
Section~\ref{sec-HallChern}. This is the first technical tool proposed by 
A. Connes \cite{Co85} to define the cyclic cohomology and constitutes the first
important step in proving quantization of the Hall conductance \cite{Be85}. The 
last one is the {\em Dixmier trace} defined by Dixmier in 1964 \cite{Dix} and of
which the importance for Quantum Differential Calculus was emphasized by 
A. Connes \cite{Co87,Co90,Co93}. It will be defined in Section~\ref{sec-Dixmier}
but we already explain in Section~\ref{sec-LocHallChern} below how to use it in 
order to relate localization properties to the validity of the previous results.

\vspace{.2cm}

\subsection{The non-commutative Brillouin zone} 
\label{NCBzone}

In a $D$-dimensional perfect crystal, the description of observables is provided
by Bloch's theory. Owing to the existence of a translation group symmetry, each 
observable of interest commutes with this group. It is a standard result in 
solid state physics that such an observable is a matrix valued periodic function
of the quasi-momenta $k$. The matrix indices usually label both the energy bands
and the ions in the unit cell of the crystal whenever it is not a Bravais 
lattice. The period group in the quasi-momentum space is the reciprocal lattice.
 
If the crystal suffers disorder or if a magnetic field is turned on, this 
description fails. In most situations however, physicists have found ways to 
overcome this difficulty. For instance, impurities are treated as isolated 
objects interacting with the electron Fermi sea using perturbation theory. 
Magnetic fields in 3-dimensional real crystals are usually so small that a 
quasi-classical approximation gives a very good account of the physical 
properties.

Still the conceptual problem of dealing with the breaking of translation 
symmetry remains. It has not been considered seriously until new physical
results forced physicists to face it. One important example in the past was 
Anderson localization due to a random potential \cite{An,AALX,FiPa,CaLa,CFKS}. 
Even though important progress has been made, the solution is still in a rather 
rough shape and numerical results are still the only justification of many 
intuitive ideas in this field. When arriving at the QHE, the r\^ole of 
localization became so important that there was no way to avoid it.

The main difficulty in such cases is that translates of the one-electron
Hamiltonian no longer commute with the Hamiltonian itself. Nevertheless, the
crystal under study is still macroscopically homogeneous in space so that its 
electronic properties are translation invariant. Thus choosing one among the
translates of the Hamiltonian is completely arbitrary. In other words, the 
one-electron Hamiltonian is only known up to translation. Our first proposal to 
deal with this arbitrariness is to consider all the translates at once. More 
precisely, our basic object is the observable algebra $\Aa$ generated by the 
family of all translates of the one-electron Hamiltonian in the space $G$, where
$G$ is either $\RR^D$ or $\ZZ^D$.

It is remarkable that the algebra $\Aa$ exists in the periodic case as well
and coincides then with the algebra of matrix-valued, periodic, continuous 
functions of the quasi-momenta \cite{Be91}. In this case, due to the periodicity
with respect to the reciprocal lattice, it is sufficient to consider 
quasi-momenta in the first Brillouin zone $\BB$. Then $\BB$ is topologically a 
$D$-dimensional torus. Therefore the matrix elements of a typical observable are
continuous functions over $\BB$.

Proceeding by analogy, we will consider the algebra $\Aa$ in the non-periodic
case as the non-commutative analogue of the set of continuous functions over a
virtual object called the ``non-commutative Brillouin zone'' \cite{B85,Be91}.
This is, however, ineffective as long as we have not defined rules for calculus.
In the periodic case physical formul{\ae} require two kinds of calculus 
operations, namely integrating over the Brillouin zone and differentiating with 
respect to quasi-momenta.

If $A$ is an observable in a periodic crystal, its average over quasi-momenta
is given by 

$$
<A>=
  \int_{\BB}
   \frac{d^{D}k}{|\BB |}
    \TR (A(k))
\mbox{ . }
$$

\noindent Here $|\BB |$ represents the volume of the Brillouin zone in momentum 
space. It turns out that such an average coincides exactly with the ``trace per 
unit volume'', namely we have

\begin{equation}
\label{tracevol}
<A>=
  \lim_{\Lambda \rightarrow \infty}
   \frac{1}{|\Lambda |}
    \TR _{\Lambda }(A_{\Lambda})
					=\TV (A)
\mbox{ , }
\end{equation}

\noindent where $\Lambda$ is a square centered at the origin, $|\Lambda |$ is
its volume, $\TR _{\Lambda }$ and $A_{\Lambda}$ are the restrictions of the
trace and the observable $A$ respectively to $\Lambda$. 

\noindent In a non-periodic, but homogeneous crystal it is still possible to 
define the trace per unit volume $\TV (A)$ of an observable $A$. 
Formula~(\ref{tracevol}) however becomes ambiguous in this case, because the
limit need not exist. We will show in Section~\ref{sec-ObsCal} how to define
$\TV$ in general. It will satisfy the following properties:

\bed
\item[(i)] $\TV$ is linear, like the integral;

\item[(ii)] $\TV$ is positive, namely if $A$ is a positive observable, 
$\TV(A)\geq 0$;

\item[(iii)] $\TV$ is a trace, namely, even though observables may not commute 
with each other, we still have $\TV (AB)=\TV (BA)$. 
\ed

Differentiating with respect to quasi-momenta can be defined along the same
line. In the periodic case, we remark that the derivative
$\partial_{j}A=\partial A/\partial k_{j}$ is also given by 

\begin{equation}
\label{diff}
\partial_{j}A=
	-\imath [X_{j},A]
\mbox{ , }
\end{equation}

\noindent where $\vec{X}=(X_{1},\ldots ,X_{D})$ is the position operator. We 
will use the formula~(\ref{diff}) as a definition of the derivative in the 
non-periodic case. Owing to the properties of the the commutator it satisfies:

\bed
\item[(i)] $\partial_{j}$ is linear;

\item[(ii)] $\partial_{j}$ obeys the Leibniz rule 
$\partial_{j}(AB) = (\partial_{j}A)B + A(\partial_{j}B)$;

\item[(iii)] $\partial_{j}$ commutes with the adjoint operation, namely 
$\partial_{j}(A^{\ast}) = (\partial_{j}A)^{\ast}$.
\ed

\noindent Such a linear map on the observable algebra is called a
``$^{\ast}$-derivation''. In our case it is moreover possible to exponentiate 
it, for, if $\vec{\theta}=(\theta_{1},\ldots ,\theta_{D})$, we find

$$
e^{\vec{\theta}\cdot \naV}(A)=
	e^{-\imath \vec{\theta} \cdot\vec{X}} A
	e^{\imath \vec{\theta}\cdot \vec{X}}
\mbox{ , }
$$

\noindent where 
$\vec{\theta} \cdot\vec{X} = (\theta_{1}X_{1}+\cdots +\theta_{D}X_{D})$
and 
$\vec{\theta}\cdot\naV  = 
 (\theta_{1}\partial_{1} + \cdots+\theta_{D}\partial_{D})$. 
Equipped with the trace per unit volume and with these rules for 
differentiations, $\Aa$ becomes a ``non-commutative manifold'', namely the 
{\em non-commutative Brillouin zone}.

\vspace{.2cm}

\subsection{Hall conductance and non-commutative Chern character}
\label{sec-HallChern}

Using the dictionary created in the previous section between the periodic
crystals and the non-periodic ones and in view of eq.~(\ref{eq-KuCh1}), one is 
led to the following formula for the Hall conductance at zero temperature

\begin{equation}
\label{eq-KuCh2}
\sigma_{H}=
	\frac{e^{2}}{h}\Ch (P_{F})
		\mbox{ , }
\hspace{2cm}
\Ch (P_{F})=
	\frac{1}{2\imath \pi}
		\TV (P_{F} [ \partial_{1}P_{F},\partial_{2}P_{F} ] )
\mbox{ , }
\end{equation}

\noindent where $P_{F}$ is the eigenprojection of the Hamiltonian on energies
smaller than or equal to the Fermi level $E_{F}$. We will justify this
formula in Section~(\ref{sec-Kubo}) below within the framework of the relaxation
time approximation. However this formula is only valid if:

\bed
\item[(i)] the electron gas is two-dimensional (so $D=2$);

\item[(ii)] the temperature is zero;

\item[(iii)] the thermodynamic limit is reached;

\item[(iv)] the electric field is vanishingly small;

\item[(v)] the collision time is infinite;

\item[(vi)] electron-electron interactions are ignored. 
\ed

\noindent We will discuss in Section~(\ref{sec-correction}) the various 
corrections to that formula whenever these conditions are not strictly
satisfied.

The main result of this paper is that {\em the non-commutative Chern character 
$\Ch (P_{F})$ of the Fermi projection is an integer provided the Fermi level
belongs to a gap of extended states} (see Theorem~\ref{theo-main}). 
Moreover, we will show that {\em it is a continuous function of the Fermi level
as long as this latter lies in a region of localized states} 
(see Theorem~\ref{theo-main}). This last result implies that changing the 
filling factor creates plateaux corresponding to having the Fermi level in an 
interval of localized states (see Proposition~\ref{theo-continuity}). In 
addition, whenever the Hall conductance jumps from one integer to another one, 
the localization length must diverge somewhere in between (see 
Corollary~(\ref{divergence})). 

In order to get this result we will follow a strategy introduced by A. Connes
in this context \cite{Co85}, but originally due to M. Atiyah \cite{Ati}. Before
describing it, we need some notations and definitions. We build a new Hilbert
space which is made of two copies $\Hh_{+}$ and $\Hh_{-}$ of the physical 
Hilbert space $\Hh$ of quantum states describing the electron. This is like 
working with Pauli spinors. In this doubled Hilbert space $\Hha =
\Hh_{+} \oplus \Hh_{-}$, the grading operator $\hat{G}$ and the ``Hilbert 
transform'' $F$ are defined as follows:

\begin{equation}
\label{eq-GandF}
\hat{G} = \left( 
	    \begin{array}{cc}
		+1 &  0 \\
		 0 & -1 
	    \end{array}
	  \right)
\mbox{ , }
\hspace{2cm}
F = \left( 
      \begin{array}{cc}
                    0             &  \frac{X}{|X|} \\
	 \frac{\overline{X}}{|X|} &        0 
      \end{array}
    \right)
\mbox{ , }
\end{equation}

\noindent where $X=X_{1}+\imath X_{2}$ (here the dimension is $D=2$). It is 
clear that $F$ is selfadjoint and satisfies $F^{2}={\bf 1}$. An operator $T$ 
on $\Hha$ will said to be of degree $0$ if it commutes with $\hat{G}$ and of 
degree $1$ if it anticommutes with $\hat{G}$. Then, every operator on $\Hha$ 
can be written in a unique way as the sum $T=T_{0}+T_{1}$ where $T_{j}$ has 
degree $j$. The graded commutator (or supercommutator) of two operators and 
the graded differential $dT$ are defined by

$$
			[T,T']_{S}= TT' - (-)^{deg(T)deg(T')}T'T
\mbox{ , }
\hspace{2cm}
			dT=[F,T]_{S}
\mbox{ . }
$$

\noindent Then, $d^{2}T=0$. The graded trace $\ST$ (or supertrace) is defined 
by

\begin{equation}
\label{eq-supertrace}
\ST (T)=
	\frac{1}{2}
		\TR_{\hat{\Hh}} (\hat{G}F[F,T]_{S})=
			\TR_{\Hh} (T_{++}-uT_{--}\overline{u})
\mbox{ , }
\end{equation}

\noindent where $u=X/|X|$ and $T_{++}$ and $T_{--}$ are the diagonal components
of $T$ with respect to the decomposition of $\hat{\Hh}$. It is a linear map on 
the algebra of operators such that $\ST (TT')= \ST (T'T)$. Moreover, operators 
of degree $1$ have zero trace. However, this trace is not positive. Observables
in $\Aa$ will become operators of degree $0$, namely $A \in \Aa$ will be 
represented by $\hat{A}=A_{+}\oplus A_{-}$ where $A_{\pm}$ acts as $A$ on each 
of the components $\Hh_{\pm}= \Hh$.

With this formalism, A. Connes \cite{Co85} gave a formula which was extended to
the present situation in \cite{Be85} to the following one (see
Section~\ref{sec-Connesformula1}) :

\begin{equation}
\label{eq-connes1}
\Ch (P_{F})=
	\int_{\Omega}
		d{\bf P}(\omega)
			\ST (\hat{P}_{F}d\hat{P}_{F}d\hat{P}_{F})
\mbox{ , }
\end{equation}

\noindent where the space $\Omega$ represents the disorder configurations while
the integral over $d{\bf P}$ is the average over the disorder.  We will show 
that the integrand does not depend upon the disorder configuration. Moreover, 
we will show that the right hand side of (\ref{eq-connes1}) is $\PP$-almost 
surely equal to the index of a Fredholm operator, namely

\begin{equation}
\label{eq-connes2}
\Ch (P_{F})=
	\mbox{Ind} ({P}_{F}F_{+-}\uparrow_{{P}_{F}(\Hh)})
\mbox{ . }
\end{equation}

\noindent The Chern character is therefore an integer, at least whenever it is 
well-defined.

One may wonder what is the physical meaning of this integer. Actually, an
answer was provided quite recently by Avron et al. \cite{ASS}. Considering the
definition of the graded trace in formula~(\ref{eq-supertrace}), they 
interpreted the index found for the Chern character as the relative index of the 
projections $P_{F}$ and $uP_{F}\overline{u}$. This is to say even though these 
projections are infinite-dimensional, their difference has finite rank giving 
rise to a new index called the ``relative index''. Then they argue that $u$ 
represents the action of the singular Laughlin gauge (see 
Section~\ref{sec-IQHcond}) and that this difference can be interpreted as a 
charge transported to infinity as in the original Laughlin argument 
\cite{Laug}.

\vspace{.2cm}

\subsection{Localization and plateaux of the Hall conductance}
\label{sec-LocHallChern}

In the last Section~(\ref{sec-HallChern}), we have explained the 
topological aspect of the Hall conductance quantization. However, these
formul{\ae} only hold if both sides are well defined. For instance, it is not 
clear whether the Fermi projection is differentiable. If it is not, what is the
meaning of formula~(\ref{eq-KuCh2}) ? This is the aspect that we want to 
discuss now.

First of all, using the Schwarz inequality, a sufficient condition for
formula~(\ref{eq-KuCh2}) to hold is that the Fermi projection satisfy $\TV
(|\naV P_{F}|^{2}) < \infty$ if we set $\naV =(\partial_{1},\partial_{2})$. It 
is important to remark that this condition is much weaker than demanding 
$P_{F}$ to be differentiable. This is rather a non-commutative analogue of a 
Sobolev norm (or the square of it). By definition, this expression can be 
written as

\begin{equation}
\label{sobolev}
\TV (|\naV  P_{F}|^{2}) =
	\int_{\Omega}
		d{\bf P}(\omega)
			\int_{x\in G}d^{2}x
				|<0|P_{F}|\vec{x}>|^{2}|\vec{x}|^{2}
\mbox{ . }
\end{equation}

\noindent Note that equation (\ref{sobolev}) can be defined in any dimension
$D$.
From this expression, we see that the finiteness of the Sobolev norm
is related to the finiteness of some localization length. Indeed, we will show
that only the part of the energy spectrum near the Fermi level gives a 
contribution to formula~(\ref{sobolev}) so that $P_{F}$ can be replaced by the 
projection $P_{\Delta}$ onto energies in some interval $\Delta$ close to
$E_{F}$. On the other hand, the matrix element $<0|P_{\Delta}|\vec{x}>$ gives a
measure of how far states with energies in $\Delta$ are localized. Thus, the 
Sobolev norm is some measurement of the localization length. We will develop 
this fact in Section~(\ref{sec-Sobolev}) below. One way of defining a 
localization length consists in measuring how far a wave packet goes in time. 
This gives 

$$
l^2({\Delta})=
	\lim_{T\rightarrow \infty}
		\frac{1}{T}
	           \int_{t=0}^{T} dt
			\int_{\Omega} d{\bf P}(\omega)
				<0||(\vec{X}(t)-\vec{X})|^{2}|0>
\mbox{ , }
$$

\noindent where the time evolution is governed by $HP_{\triangle}$ and where
$|A|^{2}=\sum_{i} A_{i}^{\ast}A_{i}$ if 
$A = (A_{1},\ldots,A_{D})$. In this expression, $\vec{X}$ is the position 
operator and $\vec{X}(t)$ is the evolution of $\vec{X}$ under the one-electron 
Hamiltonian after time $t$. Let now $\Nn$ be the density of states (DOS) 
already introduced in Section~\ref{sec-loc}. It can be defined equivalently as 
the positive measure $\Nn$ on the real line such that for any interval $\Delta$ 

$$\TV(P_{\Delta})=\int_{\Delta} d\Nn (E)\mbox{ . }$$

\noindent That these two definitions coincide is guaranteed by a theorem of
Shubin \cite{Be91,Sh}. We will show that if 
$l^2(\triangle)$ is finite, there is a positive 
$\Nn$-integrable function $l (E)$ over $\Delta$ such that 

$$
l^2({\Delta '})=\int_{\Delta '} d\Nn (E)l (E)^{2}\mbox{ , }
$$

\noindent for any Borel subset $\Delta '$ of $\Delta$ 
(see Theorem~\ref{theo-locpoint}). The number $l (E)$ will be called the 
``localization length'' at energy $E$. Moreover, we also find

$$\TV (|\naV  P_{\Delta '}|^{2})
			\leq \int_{\Delta '} d\Nn (E)l (E)^{2}
				\mbox{ , }$$

\noindent such that 
the mapping $E_{F}\in \Delta \mapsto P_{F}$ is continuous
with respect to the Sobolev norm (Theorem \ref{theo-continuity}). 

The previous argument shows that, whenever the localization length at the Fermi
level is finite, the Chern character of the Fermi projection is well defined. 
Now let us consider the right hand side of formula~(\ref{eq-connes1}). In a 
recent work A. Connes \cite{Co87,Co90} proposed to use another trace which was 
introduced in the sixties by J. Dixmier \cite{Dix}. This Dixmier trace $\TD$ 
will be defined in Section~(\ref{sec-Dixmier}) below. Let us simply say that it 
is a positive trace on the set of compact operators, such that any trace-class 
operator is annihilated by it whereas if $T$ is a positive operator with finite 
Dixmier trace, then $T^{1+\epsilon}$ is trace class for any $\epsilon >0$. In
\cite{Co87}, A. Connes proved a formula relating the so-called Wodzicky residue 
of a pseudodiferential operator to its Dixmier trace. Adapting this formula to 
our situation leads to the following remarkable result (see 
Theorem~\ref{theo-DixmierConnes}):

$$
\TV (|\naV  P_{F}|^{2})=
	\frac{1}{\pi}
		\TD (|d\hat{P}_{F}|^{2})
\mbox{ . }
$$

\noindent Note that this formula only holds for electrons in two dimensions. In
higher dimensions, things are more involved. Therefore, we see that as soon as 
the localization length is finite in a neighborhood of the Fermi level, the 
operator $d\hat{P}_{F}$ is square summable with respect to the Dixmier trace 
implying that its third power is trace-class. In view of the formul{\ae}
~(\ref{eq-supertrace},\ref{eq-connes1}), we see that this is exactly the 
condition for the existence of the right-hand side of eq.~(\ref{eq-connes1}). 
Moreover, the continuity of $P_{F}$ in $E_{F}$ with respect to the Sobolev norm 
implies that the index in formula~(\ref{eq-connes2}) is constant as long as the 
Fermi level stays in a region in which the localization length is finite. 
Therefore the Chern character is an integer and the Hall conductance is 
quantized and exhibits plateaux. 

In this way, the mathematical frame developed here gives rise to a complete
mathematical description of the IQHE, within the approximations that have been
described previously.

\vspace{.2cm}

\subsection{Summary of the main results}
\label{sec-summary}

Let us summarize our mathematical results in this section.

\begin{theo}
\label{theo-main}
Let $H=H^{\ast}$ be a Hamiltonian affiliated to the \CS 
$\Aa=C^{\ast}(\Omega \times G,\Bb )$ defined in Section~\ref{sec-ObsCal}. Let
$\PP$ be a $G$-invariant and ergodic probability measure on $\Omega$. Then we
have the following results:

\bed
\item[i)] {\bf (Kubo-Chern formula)} In the limit where \\
\indent a) the volume of the sample is infinite, \\
\indent b) the relaxation time is infinite and \\
\indent c) the temperature is zero, \\
the Hall conductance of an electron gas without interaction and described by
the one-particle Hamiltonian $H$ is given by the formula

$$
\sigma_H=
 \frac{e^2}{h}\CH (P_{F})=
  \frac{e^2}{h}
   2\imath \pi\;
    \TV(P_{F} [ \partial_{1} P_{F},\partial_{2} P_{F} ] )
\mbox{ , }
$$

\noindent where $P_F$ is the eigenprojection of $H$ on energies smaller than or equal
to the Fermi level $E_F$ and $\TV$ is the trace per unit volume associated to
$\PP$. 

\item[ii)] {\bf (Quantization of the Hall conductance)} If 
in addition $P_F$ belongs to
the Sobolev space $\Ss$ associated to $\Aa$, then $\CH (P_F)$ is an integer
which represents the charge transported at infinity by a Laughlin adiabatic
switching of a flux quantum.

\item[iii)] {\bf (Localization regime)} Under the same conditions as 
the ones in {\rm ii)}, the direct conductivity vanishes.

\item[iv)] {\bf (Existence of Plateaux)} Moreover, if $\Delta$ is an 
interval on which 
the localization length $l^2(\Delta)$ defined in Section~\ref{sec-LocHallChern}
is finite, then as long as the Fermi energy stays in $\Delta$ and is a continuity
point of the density of states of $H$, $\CH (P_F)$ is constant and $P_F$ belongs
to the Sobolev space $\Ss$. 
\ed
\end{theo}

\noindent The following corollary is an immediate consequence. 
It was proved in \cite{Halp,Ku}. 

\begin{coro}
\label{coro-main}
If the Hall conductance $\sigma_{H}$ jumps from one integer to another in
between the values $\nu_{1}$ and $\nu_{2}$ of the filling factor, there is an
energy between the corresponding values of the Fermi levels at which the
localization length diverges. 
\end{coro}

Strictly speaking, this theorem has been completely proved only for the case of 
a discrete lattice (tight-binding representation). Most of it
is valid for the continuum, but parts of the proofs require extra technical
tools so that the proof of this theorem is not complete in this paper.
We postpone the complete proof of it for the continuum case to a future work
\cite{SCB}.
 
As a side result, we emphasis that we have developed a non-commutative 
framework valid to justify the transport theory for aperiodic media
(see Section~\ref{sec-RTA} below). It allows us to give rigorous estimates on the 
error terms whenever the conditions of the previous theorem are not strictly
satisfied (see Section~\ref{sec-correction} below). We will not give the mathematical
proofs that these errors are 
rigorous bounds here even though they actually are. This will also be the main topic
of a future work. But we have estimated them and we show that they are 
compatible with the accuracy observed in the experiments.

Another result which is actually completely developed here due to its importance
in the Quantum Hall Effect, concerns the definition and the properties of the
localization length. We give a non-commutative expression for it and show how it is 
related to the existence of plateaux. The main results in this direction are the 
Theorems~\ref{theo-locpoint} and \ref{theo-continuity} in Section~\ref{sec-Sobolev}.
We also show that the localization
length is indeed finite in models like the Anderson model for disordered systems
for which proofs of exponential localizations are available.

The proof of Theorem~\ref{theo-main} will be divided up into a number of partial
results; there will be no explicit paragraph `{\bf Proof of 
Theorem~\ref{theo-main}}', let us therefore outline the main steps.
In Sections~\ref{sec-homSO} and \ref{sec-ObsCal},
we give a precise mathematical description of a homogeneous Schr\"odinger 
operator and its hull and we construct the observable algebra. In 
Chapter~\ref{chap-kubo} the Kubo formula for the Hall conductance is derived
and we present the (non-commutative) geometrical argument for the integer
quantization of the Hall conductance (compare Theorem~\ref{theo-indexloc}).
That the proven index theorem is precisely valid under the weak localization 
condition $P_F\in\Ss$ results from Theorem~\ref{theo-DixmierConnes}. Point
iv)
is proved in Chapter~\ref{chap-loc}. 

\vspace{0.2cm}

\subsection{Homogeneous Schr\"{o}dinger's operators}
\label{sec-homSO}

Most of the results of the next two sections have already been proved in 
\cite[Section 2]{Be91}. We will only give the main steps that are necessary in 
this paper for the purpose of the IQHE.

In earlier works \cite{B85,Be91}, one of us has introduced the notion of a
homogeneous Hamiltonian. The purpose of this notion is to describe materials
which are translation invariant at a macroscopic scale but not necessarily at
a microscopic one. In particular, it is well suited for the description of 
aperiodic materials. In such a medium, there is no natural choice of an origin 
in space. If $H$ is a Hamiltonian describing one particle in this medium, we can
replace it by any of its translates $H_{a}=U(a)HU(a)^{-1},\; a\in \RR^D$; the 
physics will be the same. This choice is entirely arbitrary, so that the 
smallest possible set of observables must contain at least the full family 
$\{ H_{a}; a\in \RR^{D}\} $; this family will be completed with respect to a 
suitable topology. We remark that $H$ need not be a bounded operator, so that 
calculations are made easier if we consider its resolvent instead. Let us define
precisely what we mean by ``homogeneity'' of the medium described by $H$. 

\begin{defini}
\label{homogeneity}
Let $\Hh$ be a separable Hilbert space. Let $G$ be a locally compact group (for 
instance $\RR ^{D}$ or $\ZZ^{D}$). Let $U$ be a unitary projective 
representation of $G$, namely for each $a\in G$, there is a unitary operator 
$U(a)$ acting on $\Hh  $ such that the family $U=\{ U(a); a\in G\} $ 
satisfies the following properties:

(i) $U(a) U(b) = U(a+b) e^{\imath \phi (a,b)}\, \forall a,b\in G$, where
$\phi (a,b)$ is some phase factor.

(ii) For each $\psi \in \Hh $, the map $a\in G\mapsto U(a)\psi \in \Hh $ is
continuous.

\noindent Then a selfadjoint operator $H$ on $\Hh $ is homogeneous with respect
to $G$ if the family $\{ R_{a}(z)=U(a)(z{\bf 1}-H)^{-1}U(a)^{-1}; a\in G\}
$ admits a compact strong closure.			

\end{defini}

\noindent {\em Remark.} A sequence $(A_{n})_{n>0}$ of bounded operators on
$\Hh $ converges strongly to the bounded operator $A$ if for every $\psi \in
\Hh $, the sequence $\{ A_{n}\psi \} $ of vectors in $\Hh $ converges in norm to
$A\psi $. The set considered in the definition has therefore a strong compact 
closure if, for given $\varepsilon >0$ and for a finite set 
$\psi_{1},\ldots ,\psi_{N}$ of vectors in $\Hh $, there is a finite set 
$a_{1},\ldots ,a_{m}$ in $G$ such that for every $a\in G$ and every 
$1\leq j\leq N$, there is $1\leq i\leq m$ such that
$\parallel (R_{a}(z)-R_{a_{i}}(z)) \psi_{j}\parallel \leq \varepsilon $. In
other words, the full family of translates of $R(z)$ is well approximated on
vectors by a finite number of these translates; this finite number then repeats 
itself infinitely many times up to infinity.

The virtue of this definition comes from the construction of the ``hull''.
Let $z$ belong to the resolvent set $\rho (H)$ of $H$ and let $H$ be 
homogeneous. We denote by $\Omega (z)$  the strong closure of the family
$\{ R_{a}(z)=U(a)(z{\bf 1}-H)^{-1}U(a)^{-1}; a\in G\}$; it is therefore a
compact space. Moreover, it is endowed with an action of the group $G$ by means 
of the representation $U$. This action defines a group of homeomorphisms of 
$\Omega(z)$. Thanks to the resolvent equation, it is quite easy to prove that if
$z'$ is another point in the resolvent set $\rho (H)$, the spaces $\Omega (z)$ 
and $\Omega (z')$ are homeomorphic \cite{Be91}. Identifying them gives rise to 
an abstract compact space $\Omega $ endowed with an action of $G$ by a group of 
homeomorphisms. If $\omega \in \Omega $ and $a\in G$, we will denote by
$T^{a}\omega $ the result of the action of $a$ on $\omega $, and by
$R_{\omega}(z)$ the representative of $\omega $ in $\Omega (z)$. Then we have

$$
U(a)R_{\omega}(z)U(a)^{\ast}=R_{T^{a}\omega}(z) \mbox{ , }
$$

\begin{equation}
\label{reseq}
R_{\omega}(z')-R_{\omega}(z)=(z-z')R_{\omega}(z')R_{\omega}(z)
\mbox{ . }
\end{equation}

\noindent In addition, $z\mapsto R_{\omega}(z)$ is norm-holomorphic on
$\rho (H)$ for every $\omega \in \Omega$, and the application 
$\omega \mapsto R_{\omega}(z)$ is strongly continuous.

\begin{defini}
\label{hull}
Let $H$ be an operator, homogeneous with respect to the representation $U$ of 
the locally compact group $G$ on the Hilbert space $\Hh $. Then the hull of $H$ 
is the dynamical system $(\Omega ,G,T)$ where $\Omega$ is the compact space 
given by the strong closure of the family 
$\{ R_{a}(z)=U(a)(z{\bf 1}-H)^{-1}U(a)^{-1}; a\in G\} $, and $G$ acts on 
$\Omega $ through $T$.
\end{defini}

\noindent In general, the equation~(\ref{reseq}) is not sufficient to ensure
that $R_{\omega}(z)$ is the resolvent of some self-adjoint operator
$H_{\omega}$, for indeed, one may have $R_{\omega}(z)=0$ if no additional
assumption is demanded. A sufficient condition is that $H$ be given by
$H_{0}+V$ where: (i) $H_{0}$ is self-adjoint and $G$-invariant, (ii) $V$ is
relatively bounded with respect to $H_{0}$, i.e. $\parallel
(z-H_{0})^{-1}V\parallel < \infty $, (iii) $\lim_{|z|\mapsto \infty}
\parallel (z-H_{0})^{-1}V\parallel =0$. 
Then, $R_{\omega}(z)=\{ {\bf 1}-(z{\bf 1}-H_{0})^{-1}V_{\omega}\} ^{-1}
(z{\bf 1}-H_{0})^{-1}$ where $(z{\bf 1}-H_{0})^{-1}V_{\omega}$ is defined by
the strong limit of $(z{\bf 1}-H_{0})^{-1}V_{a_{i}}$, which obviously exists. 
So $R_{\omega}(z)$ is the resolvent of $H_{0}+V_{\omega}$. 

If $H$ is a Schr\"{o}dinger operator, the situation becomes simpler. Let us
consider the case of a particle in $\RR ^{2}$ with mass $m$ and
charge $q$, submitted to a bounded potential $V$ and a uniform magnetic
field $\Bb $ with vector potential $\vec{A}$. We will describe the vector potential
in the symmetric gauge, namely $\vec{A}=(-\Bb x_{2}/2,\Bb x_{1}/2)$. The
Schr\"{o}dinger operator is given by 

\begin{equation}
\label{eq-hamilmagn}
H=
	\frac{1}{2m}
		\sum_{j=1,2} (P_{j}-qA_{j})^{2} +V=H_{L}+V
\mbox{ . }
\end{equation}

\noindent The unperturbed part $H_{L}$ is translation invariant,
provided one uses magnetic translations \cite{Zak} defined by (if $\vec{a} \in
\RR^{2},\, \psi \in L^{2}(\RR ^{2})$ )

\begin{equation}
\label{eq-magntrans}
U(\vec{a}) \psi (\vec{x}) = 
	exp\left\{ 
	  \frac{-\imath q\Bb }{2\hbar } \vec{x} \wedge \vec{a}
		\right\} 
			\psi (\vec{x}-\vec{a})
\mbox{ , }
\end{equation}

\noindent where $\vec{x} \wedge \vec{a} = x_1 a_2 - x_2 a_1$.
It is easy to check that the operators $U(a)$ form a projective unitary
representation of the translation group $\RR ^{2}$. The main result in this
case is given by 

\begin{theo}
\label{theo-hull}
Let $H$ be given by eq.~(\ref{eq-hamilmagn}) with $V$ a
measurable essentially bounded function on $\RR ^{2}$. Then

(i) $H$ is homogeneous with respect to the magnetic
translations (\ref{eq-magntrans});

(ii) the hull $\omega$ of $H$ is homeomorphic to the hull of $V$,
 namely the weak closure
of the family $\{ V(.-\vec{a}); \vec{a}\in \RR^{2}\}$ in
$L_{\RR}^{\infty}(\RR^{2})$;

(iii) there is a Borelian function $v$ on $\Omega$ such that, if we denote
by $V_{\omega}$ the bounded function representing the point $\omega \in \Omega$,
then $V_{\omega}(\vec{x}) = v(T^{-\vec{x}}\omega)$ for almost every 
$\vec{x} \in \RR ^{2}$ and all
$\omega \in \Omega $. If in addition $V$ is uniformly continuous and bounded,
then $v$ is continuous.
\end{theo} 

\noindent The proof of this theorem can be found in \cite[Section 2.4]{Be91}. 

In many cases, it is actually better to work in the tight-binding
representation. The reason is that only electrons with energies near the Fermi
level contribute to the current. One usually defines an effective
Hamiltonian by reducing the Schr\"{o}dinger operator to the interval of
energies of interest \cite{B85}. The Hamiltonian can then be described as a 
matrix $H(x,x')$ indexed by sites in the lattice $\ZZ^{2}$ acting on elements 
of the Hilbert space $\ell^{2}(\ZZ^{2})$ as follows

\begin{equation}
\label{eq-discrhamil}
H\psi (\vec{x})=
	\sum_{\vec{x'} \in \ZZ^{2}}
		H(\vec{x},\vec{x}')
			exp\left\{ \imath \pi
			\frac{\phi}{\phi_{0}}
				 \vec{x} \wedge \vec{x}' \right\} 
					\psi (\vec{x'})
\mbox{ , }
\hspace{1cm}
	\psi \in \ell^{2}(\ZZ^{2})
\mbox{ , }
\end{equation}

\noindent where $\phi$ is the flux in the unit cell, whereas $\phi_{0}=h/q$ is
the flux quantum. In most examples, one can find a sequence $f$ such
that $\sum_{\vec{a} \in \ZZ^{2}} f(\vec{a}) < \infty$ and 
$|H(\vec{x},\vec{x}')|\leq f(\vec{x} - \vec{x}')$. In
particular, $H$ is bounded and there is no longer a 
need to consider its resolvent.
 Let now $U$ be the unitary projective representation of the translation group
$\ZZ^{2}$ given  by  

\begin{equation}
\label{discrrep}
U(\vec{a})\psi (\vec{x})=
	exp\left\{ \imath \pi
			\frac{\phi}{\phi_{0}}
				 \vec{x} \wedge \vec{a} \right\} 
					\psi (\vec{x} - \vec{a})
\mbox{ , }
\hspace{1cm}
	\psi \in \ell^{2}(\ZZ^{2})
\mbox{ . }
\end{equation}

\begin{theo}
\label{theo-discrhull}

Let $H$ be given by eq.~(\ref{eq-discrhamil}). Then $H$ is homogeneous with
respect to the projective representation $U$ (eq.~(\ref{discrrep})) of the
translation group. Moreover, if $\Omega$ is the hull of $H$, there is a
continuous function $\hat{h}$ on $\Omega \times \ZZ^{2}$, vanishing at infinity,
such that 
$H_{\omega}(\vec{x},\vec{x}') = \hat{h}(T^{-\vec{x}}\omega, \vec{x}'-\vec{x})$,  
for every pair $(\vec{x},\vec{x}')\in \ZZ^{2}$ and $\omega \in \Omega$.
\end{theo}

\noindent Again the proof can be found in \cite[Section 2.4]{Be91}.

\vspace{.2cm}

\subsection{Observables and calculus}
\label{sec-ObsCal}

In the previous section we have constructed the hull of a Hamiltonian. 
It is a compact space that represents the degree of
aperiodicity of the crystalline forces acting on the charge carriers. For
disordered systems, the hull is nothing but the space of disorder
configurations. In
principle, the algebra of observables should be constructed from the
Hamiltonian $H$ by taking all functions of $H$ and its translates. However, we
proceed in a different way giving a more explicit construction which may give a
bigger algebra in general, but will be easier to use. 

Let $\Omega$ be a compact topological space equipped with a
$\RR^{2}$-action by a group $\{ T^{a}; a\in \RR^{2}\}$ of homeomorphisms. Given
a uniform magnetic field $\Bb $, we can associate to this dynamical system a
$C^{\ast}$-algebra $C^{\ast}(\Omega \times \RR^{2},\Bb )$ as follows. We
first consider the topological vector space $\Cc_{\kappa}(\Omega \times 
\RR^{2})$ of continuous functions with compact support on $\Omega \times  
\RR^{2}$. It is endowed with the following structure of a $^{\ast}$-algebra by

\begin{eqnarray}
\label{eq-staralg}
AB (\omega, \vec{x}) & = &
 \int_{\RR^{2}}d^{2}y
		A(\omega, \vec{y}) B(T^{-\vec{y}} \omega, \vec{x}-\vec{y}) 
			exp\left\{ 
			      \frac{\imath q\Bb }{2\hbar }
			      \vec{x} \wedge \vec{y} 
			   \right\} 
				\mbox{ , }\nonumber \\
 		&    &		\\
A^{\ast}(\omega, \vec{x}) & = & \overline{A(T^{-\vec{x}}\omega, -\vec{x})}
		\mbox{ , }	\nonumber
\end{eqnarray}

\noindent where $A,B\in \Cc_{\kappa}(\Omega \times  \RR^{2})$, 
$\omega \in \Omega$ and $x \in \RR^{2}$. For $\omega \in \Omega$, this 
$^{\ast}$-algebra is represented on $L^{2}(\RR^{2})$  by 

\begin{equation}
\label{eq-repreg}
\pi_{\omega}(A)\psi (\vec{x})=
	\int_{\RR^{2}}d^{2}y A(T^{-\vec{x}} \omega, \vec{y} - \vec{x})
		exp\left\{ \frac{\imath q\Bb }{2\hbar} 
		\vec{y} \wedge \vec{x} \right\} \psi (\vec{y}) 
\mbox{ , }
\hspace{1cm}
\psi \in L^{2}(\RR^{2})
\mbox{ , }
\end{equation}

\noindent namely, $\pi_{\omega}$ is linear, $\pi_{\omega}(AB)
= \pi_{\omega}(A)\pi_{\omega}(B)$ and $\pi_{\omega}(A)^{\ast}=\pi_{\omega}
(A^{\ast})$. In addition, $\pi_{\omega}(A)$ is a bounded operator for
$\parallel \pi_{\omega}(A)\parallel \leq \parallel f\parallel _{\infty ,1}$ 
where

$$
\parallel A\parallel _{\infty ,1}=
\max 
	\left\{ 
	\sup_{\omega \in \Omega}
		\int_{\RR^{2}}d^{2}y |A(\omega, \vec{y})|,\: 
			\sup_{\omega \in \Omega}
				\int_{\RR^{2}}d^{2}y |A^{\ast}(\omega,\vec{y})|
					\right\}
\mbox{ . }
$$

\noindent This defines a norm which satisfies 
$\parallel AB\parallel _{\infty ,1}\leq 
\parallel A\parallel _{\infty ,1}\parallel B\parallel _{\infty ,1}, \:
\parallel A\parallel _{\infty ,1} = \parallel A^{\ast}\parallel _{\infty ,1}$.
Since $A$ has compact support, its $(\infty ,1)$-norm is finite. We will denote
by $L^{\infty ,1}(\Omega \times \RR^{2};\Bb )$ the completion of
$\Cc_{\kappa}(\Omega \times \RR^{2})$ under this norm. We then remark that 
these representations are related by the covariance condition 

$$
U(\vec{a})\pi_{\omega}(A)U(\vec{a})^{-1} =\pi_{T^{\vec{a}}\omega}(A)
				\mbox{ . }
$$

\noindent Now we set 

$$
\parallel A\parallel =
	\sup_{\omega \in \Omega}
		\parallel \pi_{\omega}(A)\parallel 
			\mbox{ , }
$$

\noindent which defines a $C^{\ast}$-norm on $L^{\infty ,1}(\Omega \times
\RR^{2};\Bb )$. This permits us to define $C^{\ast}(\Omega \times \RR^{2},\Bb )$ as
the completion of $\Cc_{\kappa}(\Omega \times \RR^{2})$ or of $L^{\infty
,1}(\Omega \times \RR^{2};\Bb )$ under this norm. Clearly, the 
representations $\pi_\omega$ can be continuously extended to this 
$C^\ast$-algebra. We remark that this algebra has no unit. 

In particular, for the Hamiltonian~(\ref{eq-hamilmagn}) we get
\cite[Section 2.5]{Be91}:

\begin{theo}
\label{theo-scrocstar}

Let $H$ be given by eq.~(\ref{eq-hamilmagn}) where $\vec{A}$ is the vector
potential of a uniform magnetic field in the symmetric gauge and
let  V  be in $L^{\infty}(\RR^{2})$. We denote by $\Omega$ its hull. Then for
each $z$ in the resolvent set of $H$ and for every $\vec{x} \in \RR^{2}$, there
is an element $R(z;\vec{x})\in C^{\ast}(\Omega \times \RR^{2},\Bb )$, such that 
for each $\omega \in \Omega$, 
$\pi_{\omega}(R(z;\vec{x}))=(z{\bf 1}-H_{T^{-\vec{x}}\omega})^{-1}$. 

\end{theo}

\noindent In the discrete case (see
eq.~(\ref{eq-discrhamil},\ref{theo-discrhull}), the same construction holds
provided we replace $\RR^{2}$ by $\ZZ^{2}$ and the integral over $\RR^{2}$ by
a sum over $\ZZ^{2}$, namely

\begin{theo}
\label{theo-scrocstar2}

Let $H$ be given by eq.~(\ref{eq-discrhamil}) where $\phi$ is the magnetic flux
through the unit cell and $\phi_{0}=h/q$ is the flux quantum. We denote by
$\Omega$ its hull as in Theorem~\ref{theo-discrhull}. Then the 
function $\hat{h}$ appearing in Theorem~\ref{theo-discrhull} 
belongs to $C^{\ast}(\Omega \times \ZZ^{2},\Bb )$ and for each $\omega  \in
\Omega$, $\pi_{\omega}(\hat{h})=H_{\omega }$. 

\end{theo}

\noindent We remark that in the discrete case, the function 
${\bf 1}(\omega, \vec{x}) = \delta_{\vec{x}, \vec{0}}$ is a unit of 
$C^{\ast}(\Omega \times \ZZ^{2},\Bb )$; this is
the main difference between the continuous and the discrete case. In the 
non-commutative terminology, the discrete case corresponds to a compact 
non-commutative manifold, whereas the continuous case corresponds to a locally
compact, but not compact, non-commutative manifold. 
In the sequel, many results will hold for both of the constructed \CSs. We 
therefore introduce the notation $\Aa$ for both of them. 
The algebras of functions with compact support which are at the base of the
construction are denoted by $\AO$. Moreover, let $G$ be the physical space; it
is either $\RR^2$ or $\ZZ^2$. 
 
If the Hamiltonian $H$ is unbounded, it does not belong
to the $C^{\ast}$-algebra $\Aa$. However, we have seen
in Theorem~\ref{theo-scrocstar} that its resolvent belongs to $\Aa$. As a
consequence, $f(H)$ belongs to $\Aa$ for every continuous function $f$ on the
real line vanishing at infinity. This leads to the following definition.

\begin{defini}
\label{def-affiliation}
Let $\Aa$ be the $C^{\ast}$-algebra $C^{\ast}(\Omega \times G,\Bb )$ defined
above and let $\Hh $ be the Hilbert space $L^{2}(G)$ endowed with the
projective representation $U$ defined in
eq.(~\ref{eq-magntrans},~\ref{discrrep}). We will say that a selfadjoint
homogeneous operator $H$ is affiliated to $\Aa$ whenever its hull can be
embedded in the dynamical system $(\Omega ,G,T)$ and if for each $z$ in its
resolvent set, there is an element $R(z)\in \Aa$ such that

$$(z{\bf 1}-H_{\omega})^{-1} =\pi_{\omega} (R(z)),$$

\noindent for all $\omega \in \Omega$.
\end{defini}

\noindent By `embedded' we mean that the hull is a $T$-invariant subset of
$\Omega$.

The two rules of calculus, namely integration and derivations, are now
constructed as follows. First of all, let $\PP $ be a $G$-invariant
probability measure on $\Omega$.
For most of the results of this paper,
the choice of such a measure is irrelevant. We will discuss its physical
relevance in Section~\ref{sec-Index} and
Section~\ref{sec-pointsp2}. A trace on $\Aa$ is constructed as follows. If
$A\in\AO$ we set:

$$
\TV_\PP (A)=\int_{\Omega} d\PP (\omega ) A(\omega ,\vec{0})
				\mbox{ . }
$$

\noindent This defines a positive trace (see Section~\ref{NCBzone}). It
is faithful (namely $\TV_{\PP}(AA^{\ast})=0$ implies $ A=0$) whenever the
support of $\PP$ is $\Omega$. The trace is normalized if $G=\ZZ^{2}$ (namely
$\TV_{\PP} ({\bf 1})=1$) and is unbounded if $G=\RR^{2}$. It actually coincides with
the trace per unit area. More precisely we have \cite{B85,Be91}:

\begin{proposi}
Let $A$ belong to $\AO$. Then for $\PP$-almost all
$\omega$'s 

\begin{equation}
\label{eq-tracepervol}
\TV_\PP (A)=
	\lim_{\Lambda \uparrow G}
		\frac{1}{|\Lambda |}
			\TR_{\Lambda }(\pi_{\omega }(A))
				\mbox{ , }
\end{equation}

\noindent where $\Lambda$ denotes a sequence of squares in $G$ centered at the
origin and covering $G$ and $\TR_{\Lambda }$ is the restriction to $\Lambda$
of the usual trace. 
\end{proposi}

\noindent In the sequel, we will drop the subscript $\PP$. 

Given a selfadjoint element $H\in \Aa$, we define its DOS as the
positive measure $\Nn$ on the real line such that for any continuous function
$f$ with compact support on $\RR$ 

$$
\TV (f(H))=\int_{\RR} d\Nn (E) f(E)
\mbox{ . }
$$

\noindent In view of (\ref{eq-tracepervol}), this definition agrees with the
definition given in eq.~(\ref{eq-DOS}). 

If $p\geq 1$, we denote by $L^{p}(\Aa,\TV)$ the completion of $\AO$ 
under the norm

$$
\parallel A\parallel _{L^{p}}=
	\left( \TV (\{ AA^{\ast}\} ^{p/2} ) \right) ^{2/p} 
				\mbox{ . }
$$

\noindent In particular, one can check that the space $L^{2}(\Aa,\TV)$ is a
Hilbert space (GNS construction) identical to $L^{2}(\Omega \times G,d\PP\otimes
dg)$. 
The map $\phi \in L^{2}(\Aa,\TV)\mapsto A\phi\in L^{2}(\Aa,\TV)$ for $A\in\Aa$
defines a
representation $\pi_{GNS}$ of $\Aa$. The weak closure $L^{\infty}(\Aa,\TV)
 =\pi_{GNS}(\Aa)"$ is a von Neumann algebra. By construction, 
the trace $\TV$ extends to a trace on this algebra. We remark that if $H$ is a
self-adjoint element of $\Aa$, its eigenprojections are in general elements of
the von Neumann algebra $L^{\infty}(\Aa,\TV)$. 

Let us give another
characterization of the von Neumann algebra $L^{\infty}(\Aa,\TV)$ which can be
found in \cite{Co78}. Let $\Ww$ be the set of weakly measurable families
$\omega \in \Omega \mapsto A_{\omega }$ of bounded operators
on $L^{2}(G)$ which are covariant

$$
U(a)A_{\omega}U(a)^{-1}= A_{T^{a}\omega }
				\mbox{ , }
\hspace{1cm}
a\in G,\, \omega \in \Omega 
		\mbox{ . }
$$

\noindent and $\PP$-essentially bounded. This latter means that the norm of
$A_{\omega}$ is bounded in $\omega$ except possibly on a subset of zero
probability with respect to $\PP$. We endow $\Ww$ with the norm

$$
\parallel A\parallel_{\PP}= 
	\PP-{\rm ess}\sup_{\omega} 
		\parallel A_{\omega }\parallel_{\Ll( L^{2}(G))}
			\mbox{ , }
\hspace{1cm}
A\in \Ww
		\mbox{ . }
$$

\noindent Sum, product and adjoint of elements of $\Ww$ are defined
pointwise in the obvious way. Then Connes \cite{Co78} proved that $\Ww$ is
a von Neumann algebra, namely a $C^{\ast}$-algebra with predual \cite{Sak}, and
that

\begin{theo}
\label{the-Wstar}
$L^{\infty}(\Aa,\TV)$ is canonically isomorphic to the von Neumann algebra $\Ww$
of $\PP$-es\-sen\-tial\-ly bounded measurable covariant families of operators on 
$L^{2}(G)$. 
\end{theo}

\noindent Actually, this isomorphism is obvious if we realize that the Hilbert
space $L^{2}(\Aa ,\TV )$ of the $GNS$-representation of $\Aa$ with respect to
the trace $\TV$ can be written as the direct $\PP$-integral of $L^{2}(G)$. We
will not give details. A consequence of this result is that the family $\{
\pi_{\omega};\omega \in \Omega \} $ of representations of $\Aa$ extends to a
(faithful) family of (weakly measurable) representations of $\Ww$. In
particular,
any spectral projection $P$ of a Hamiltonian $H\in \Aa$ can be seen as a
covariant $\PP$-essentially bounded family $P_{\omega}$ of projections,
where $P_{\omega}$ is an eigenprojection of $H_{\omega}$.

The differential structure is obtained through the derivations defined on $\AO$
by 

\begin{equation}
\label{eq-deri}
\partial_{j}A(\omega,\vec{x})=
	\imath x_{j} A(\omega,\vec{x})
				\mbox{ . }
\end{equation}

\noindent It is a $^{\ast}$-derivation in the sense given in
Section~\ref{NCBzone}. By exponentiation it defines a two-parameter group of
$^{\ast}$-automorphisms given by 

$$
\rho_{\vec{k}} (A)(\omega,\vec{x})=
	e^{\imath \vec{k}\cdot  \vec{x}} A(\omega,\vec{x})
				\mbox{ , }
$$

\noindent where $\vec{k} \in \RR^{2}$ and 
$\vec{k} \cdot \vec{x} = k_{1}x_{1} + k_{2}x_{2}$. We notice that in the 
discrete case, $\vec{k}$ is defined modulo $2\pi \ZZ^{2}$. Introducing on
$L^{2}(G)$ the position operator $\vec{X}=(X_{1},X_{2})$, namely the 
multiplication by $\vec{x}$, we get 

\begin{equation}
\label{eq-derirepres}
\pi_{\omega}(\rho_{\vec{k}} (A))=
    e^{-\imath \vec{k}\cdot \vec{X}} \pi_{\omega}(A)  e^{\imath k\cdot \vec{X}}
				\mbox{ , }
\hspace{.7cm}
\pi_{\omega}(\partial_{j}A)=
	-\imath  [ X_{j},\pi_{\omega}(A) ]
				\mbox{ . }
\end{equation}

\noindent We remark that in the case of periodic media, this derivation is just
differentiation in quasi-momentum space.
We will denote by $\Cc^{N}(\Aa )$ the set of elements $A\in \Aa$
for which the map $\vec{k}\in \RR^{2}\mapsto \rho_{\vec{k}}(A)\in \Aa$ is 
$N$-times
continuously differentiable. If $N$ is an integer, this is equivalent to say
that $\parallel \partial_{1}^{a}\partial_{2}^{b}A\parallel <\infty$ for any pair
$a,b$ of integers such that $a+b\leq N$. In much the same way non-commutative
Sobolev spaces can be defined. For the purpose of this work we will use the
Sobolev space $\Ss=\Hh ^{2}(\Aa ,\TV )$, namely the Hilbert space obtained by
completion of $\AO$ under the Hilbert norm given by the inner product

$$
<A|B>_{\Hh^{2}}= 
	\TV (A^{\ast}B)+
		\TV (\naV  A^{\ast}\cdot \naV  B)
\mbox{ , }
\hspace{1cm}
A,B \in \AO 
				\mbox{ , }
$$

\noindent where $\naV =(\partial_{1},\partial_{2})$ is the 
non-commutative gradient operator.

We will finish this section by giving a technical result which will be used
later on. 

\begin{proposi}
\label{prop-FeynHell}
Let $H$ be a selfadjoint element in $\Cc^{1}(\Aa )$ where
$\Aa=C^{\ast}(\Omega \times \ZZ^2 ;\Bb )$ and let $\Nn$ be its density of states. 
Then for any function $f\in \L^{1}(\RR ,d\Nn )$ we have:

$$\TV (f(H)\partial_{j}H) =0.$$

\end{proposi}

\noindent {\bf Sketch of the proof.} By density, it is enough to prove this
result for a smooth function $f$ on $\RR$ with compact support. Then one can
write 

$$f(H)=
			\int_{\RR} dt\;
				e^{-\imath t H} \tilde{f} (t)
\mbox{ , }
$$

\noindent where $\tilde{f}$ is the Fourier transform of $f$. Since $H$ is
bounded, this integral converges in norm. Classical results on Fourier
transform  show that $\tilde{f}$ is a smooth rapidly-decreasing function over 
$\RR$. Thus, it is sufficient to show the result for $f(E)=e^{-\imath tE}$. In 
this case, Duhamel's formula gives \cite{Duhamel}: 

$$\naV e^{-\imath tH}=
			-\imath \int_{0}^{t} ds\;
				e^{-\imath (t-s) H}\naV  H e^{-\imath s H}
\mbox{ . }
$$

\noindent Taking the trace, the left hand side vanishes, since the trace is
$\rho_{\vec{k}}$-invariant. The right hand side is given by 
$\TV (e^{-\imath t H} \naV  H)$. The proof is then complete. 
\hfill $\Box$

\vspace{0.2cm}

\noindent In the continuum case, a similar result holds, but it is technically
more complicated. We will restrict ourselves to the case of Schr\"{o}dinger 
operators.

\begin{proposi}
\label{prop-FeynHell2}
Let $H$ be given by eq.~(\ref{eq-hamilmagn}) with $V\in L^{\infty}(\RR ^{2})$.
We denote by $\Omega$ the hull of $V$. Then $\naV  H$ is a well-defined
selfadjoint operator bounded from above by $(H+b)^{1/2}$ for some positive $b$.
Moreover, if $\Nn$ denotes the DOS of $H$, for every $\Nn$-measurable function
$f$ such that the map $E \mapsto Ef(E)$ is in $L^{1}(\RR,d\Nn)$, we have 

$$\TV (f(H)\naV H) =0.$$

\end{proposi}

\noindent {\bf Sketch of the proof.} We use the results of \cite[Section
2.5]{Be91} to conclude that $e^{-tH}$ admits an integral kernel $F(t;\omega
,\vec{x})$ for $\Re e(t)>0$ because $H$ is bounded from below. This kernel is
jointly continuous in $\omega ,\vec{x}$ and holomorphic in $t$ in this domain.
Moreover, it decays rapidly at infinity in $\vec{x}$ (uniformly on compact 
subsets for the other variables). Then we use the same argument as in
Proposition~\ref{prop-FeynHell} provided we replace $e^{-\imath t H}$ by
$e^{-(\epsilon +\imath t) H}$ for any $\epsilon >0$. 
\hfill $\Box$

\vspace{.5cm}

\section{The Kubo-Chern formula}
\label{chap-kubo}

This chapter is devoted to the Kubo formula and its relation to the Chern
character. The first three sections are devoted to transport theory and are not
treated on a completely rigorous footing. This is because technical proofs
would require too much details and not shed more light on the physics. 

\vspace{.2cm}

\subsection{The relaxation time approximation}
\label{sec-RTA}

The theory of transport is an essential tool in dealing with electronic
properties of solids. 
There are numerous theoretical approaches with complexity varying 
from the Drude-Sommerfeld theory to the N-body
framework.
Whatever the starting point, the Greenwood-Kubo formul{\ae} for the
transport coefficients, such as the electric or thermal conductivity,
are the main consequences. They are widely used and
accepted because of their correspondence with experimental results. Still their
derivation from first principles is questionable. One does not
really know the precise domain of validity of the linear response
approximation. The occurrence of many new devices in solid state
physics liable to test these ideas, such as mesoscopic systems, has
raised this question again.

It is not the purpose of this work to investigate that problem. However, we
have seen in Section~\ref{sec-CHE} that in the classical Hall
effect, the relation between the Hall current and the Hall voltage is
linear although there is no dissipation mechanism. For this reason, one
might expect that the derivation of Kubo's formula for the Hall
conductance in a quantum system should not require a many particle
theory. This is actually not true. We will see in this section that
such a point of view is very na\"{\i}ve and gives rise to paradoxa.
Without dissipation, a quantum theory leads either to a vanishing
conductivity or to an infinite one in most physically sound
situations. 

Moreover the great accuracy of the IQHE has been used in metrology for the
definition of a new standard of resistance \cite{ResH}. It is thus
necessary to derive a formula allowing the control of deviations from
the ideal QHE. We have already indicated in
Section~\ref{sec-HallChern} what the physical conditions are under
which the IQHE is exact, namely under which it can be stated as a
theorem within a well-defined mathematical framework. 

This is why we propose a one-particle model including
collision effects, such as interaction with phonons or other
electrons, realizing thereby the so-called {\em relaxation time
approximation} (RTA). We derive a Kubo formula for the
conductivity tensor which allows the linear response
approach to be justified 
and gives control on the order of magnitude of the deviation
from the ideal theory. The construction of this model is based upon
the phenomenological approach which can be found in standard books in
solid state physics such as \cite[Chap.1,2,12,13]{MA}.

\vspace{.2cm}

We consider the electron fluid of our system as a gas of independent
fermions. Neglecting interactions between electrons is actually very rough and
in many cases completely wrong. 
Nevertheless the Landau theory of Fermi liquids shows that
such an approximation is quite acceptable if the particles are
actually quasi-particles ``dressed'' by the interactions
\cite{NozPi}. In particular, their mass need not necessarily be the electron mass.
Moreover their lifetime is finite, but it goes to infinity as their energy gets
closer to the Fermi level. Quasi-particles
carrying current are therefore stable at zero temperature. 
In the sequel, we continue to talk of electrons having in mind that we are
actually treating quasi-particles.

The advantage of this independent-electron approximation is that one can avoid
using second quantization and can restrict oneself to a one-particle 
description. The constraint given by Pauli's principle is then represented by 
the use of the Fermi-Dirac distribution function. 

For this reason, our starting point will be the one-particle Hamiltonian $H$ of
the form already described in Section~\ref{sec-homSO}. It includes kinetic
energy of the electrons as well as  whatever time-independent forces there are 
acting on
them in the crystal. These latter include the periodic potential created by
the ions and the aperiodic corrections due to impurities, defects and
other kinds of deformations. This Hamiltonian goes beyond the
band theory since it may include Anderson localization for instance. We will
assume that $H$ is affiliated to the observable algebra $\Aa =C^{\ast}(\Omega
,G,\Bb )$ where $G=\RR^{2}$ or $\ZZ^{2}$ and $\Omega$ is some compact
space (see Section~\ref{sec-ObsCal}). In the grand canonical ensemble
the thermal equilibrium 
at inverse temperature $\beta =1/k_{B}T$ (where $k_{B}$ is the Boltzmann
constant and $T$ is the temperature) and chemical potential $\mu$ is described 
by means of
the Fermi-Dirac distribution.
In the algebraic language it means that, if $A\in \Aa$ is a one-particle 
extensive observable,  its thermal average per volume is given by

\begin{equation}
\label{eq-average}
<A>_{\beta ,\mu}=
	\TV (Af_{\beta ,\mu}(H))
\mbox{ , }
\hspace{.7cm}
\mbox{\rm with}
\hspace{.7cm}
f_{\beta ,\mu}(H)=(1+e^{\beta (H-\mu)})^{-1}
\mbox{ . }
\end{equation}

\noindent One important example for an observable is the current operator

$$
\JV _{\omega}=q \frac{d\XV}{dt} =
	\frac{\imath q}{\hbar}
		[ {H}_{\omega},\XV  ] 
\mbox{ . }
$$

\noindent Using the differential structure (\ref{eq-deri}), this can be
written as 

$$
\JV _{\omega}=
	\frac{q}{\hbar}(\naV H)_{\omega}
\mbox{ . }
$$

\noindent It is physically obvious that the average current vanishes since
the velocity distribution is usually symmetric under the change of its
sign. This is actually what happens within our framework because of the
Propositions~\ref{prop-FeynHell} and \ref{prop-FeynHell2} and the
formula~(\ref{eq-average}) above. Producing a non-vanishing 
average current requires
imposing an external force such as an electromagnetic field. Let us consider the
simplest case in which we superimpose a constant uniform electric field $\EV$ on
our system for time $t \ge 0$. Then the evolution of an observable is no longer
given by $H_{\omega}$ but rather by $H_{\omega ,\EV}=
H_{\omega}-q\EV\cdot \XV $. We notice that while $H$ is affiliated to
$\Aa$, this is certainly not true for this Hamiltonian since the position
operator is not homogeneous. However, the evolution under this operator
leaves $\Aa$ invariant. For indeed, whenever $A$ is smooth in $\Aa$,
Heisenberg's equation reads

\begin{equation}
\label{eq-Heisen}
\frac{dA_{\omega}}{dt}=
 \frac{\imath}{\hbar}
		[ H_{\omega},A_{\omega} ] +
			\frac{q\EV}{\hbar}\cdot
				(\naV  A)_{\omega}
\mbox{ . }
\end{equation}

\noindent If in particular $H$ belongs to $\Aa$, the right hand side
of~(\ref{eq-Heisen}) stays in $\Aa$ so that we expect this equation
to be integrable within $\Aa$. This can be proved by use of the
Trotter product formula \cite{Kato}. We denote by
$\eta_{t}$ the evolution given by $H$, namely

\begin{equation}
\label{eq-freeevolu}
\eta_{t}(A)=
	e^{\imath \frac{tH}{\hbar}}Ae^{-\imath \frac{tH}{\hbar}}
\mbox{ . }
\end{equation}

\noindent Since $H$ is affiliated to $\Aa$, this evolution leaves $\Aa$
invariant and defines a one-parameter group of *-automorphisms of $\Aa$.
Because of formula~(\ref{eq-derirepres}), we also have

$$
\pi_{\omega}
	\left( 
		\rho_{\frac{qt}{\hbar}\EV}(A)
			\right) =
				e^{-\imath \frac{qt\EV \cdot\XV }{\hbar}}
				\pi_{\omega}(A)
					e^{\imath \frac{qt\EV\cdot \XV }{\hbar}}
\mbox{ . }
$$

\noindent This last evolution also leaves $\Aa$ invariant. By means of the
Trotter product formula we find

$$
e^{\imath \frac{t}{\hbar}H_{\omega ,\EV}}
	\pi_{\omega}(A)
		e^{-\imath \frac{t}{\hbar}H_{\omega ,\EV}}
			= s\mbox{\small -}\lim_{N \to \infty}
				\pi_{\omega }
					\left( 
						(\eta_{\frac{t}{N}}
						\rho_{\frac{qt\EV }{N\hbar}})^{N}
							(A)
							\right)
\mbox{ . }
$$

\noindent Here, s-lim is the limit in the strong operator topology.
So this defines a one-parameter group of automorphisms
$\eta_{t}^{\EV}$ of the von Neumann algebra $\Ww$. 
It represents the evolution of the observables after
the electric field has been turned on. We will not investigate here
whether this group of automorphisms leaves $\Aa$ itself
invariant. Let us only notice that, whenever $H$ is
bounded, a Dyson expansion shows that $\eta_{t}^{\EV}$ is an
automorphism of $\Aa$. Moreover, we only have to work in the
Hilbert space $L^{2}(\Aa ,\TV )$ so that the previous result will
be sufficient if $H$ is unbounded. The new current at time $t\geq 0$
is then formally given by 

$$
\JV (t)=\eta_{t}^{\EV} (\JV )
\mbox{ . }
$$

\noindent Since the Hamiltonian $H_{\omega ,\EV}$ no longer
commutes with $H_{\omega}$, the thermal average $\jV (t) =
<\JV (t)>_{\beta ,\mu}$ of the current will no longer vanish in
general. The macroscopic response we expect from a constant uniform electric 
field is a constant and uniform current, but the microscopic quantal forces
lead to fluctuations of $\jV (t)$ in time. To extract the response, it
is therefore necessary to consider the time average of this current
which is actually what one measures in experiment. For indeed, the
typical relaxation time of an apparatus measuring the current is of
the order of $1ms$ to $1\mu s$ (unless very short times are needed), and this
has to
be compared with the typical collision time of the order of $10^{-13}s$
for the best conductors. Thus, we set

\begin{equation}
\label{eq-meancurrent}
  \jV _{\beta ,\mu, \EV}=
			\lim_{T \to \infty}
				\frac{1}{T}
					\int_{0}^{T} dt \;\jV (t)=
			\lim_{\delta \downarrow 0}
				\delta 
					\int_{0}^{\infty} dt \;e^{-t\delta }\jV (t)
\mbox{ . }
\end{equation}

\noindent However, our model does not take collisions into account. Indeed we 
have:

\begin{proposi}
\label{prop-vanishcurrent}
If the Hamiltonian $H$ is bounded in $\Aa$, the projection of
the time and thermal averaged current $\jV _{\beta ,\mu, \EV}$
along the direction of the electric field $\EV$ vanishes.

\end{proposi}

\noindent {\bf Proof.} Let us compute $\EV \cdot\JV (t)$. Using the
Heisenberg equation it is easy to see that 

$$ \EV \cdot\JV (t) =
				\frac{d H(t)}{dt}
					\mbox{ , }
\hspace{1cm}
	H(t) = \eta_{t}^{\EV} (H)
					\mbox{ . }$$

\noindent Taking the time average gives us

$$	 \frac{1}{T}
					\int_{0}^{T} dt \;
						\EV \cdot\JV (t) =
							\frac{H(T)-H}{T}
								\mbox{ . }$$

\noindent Since $H$ is bounded in norm and since $\parallel
H(t)\parallel =\parallel H\parallel$, the right hand side vanishes
as $T \to \infty$.
\hfill  $\Box $

\vspace{0.2cm}

\noindent This result is easy to understand in the
one-band approximation. In this case, the electric field
produces a time shift of the quasi-momentum, namely $\kV \in \BB
\mapsto \kV (t)=\kV -q\EV t/\hbar$. Taking the time average
is therefore equivalent to averaging over quasi-momenta, and this is
exactly zero. We point out that this result applies also
to models with disorder of the form given in
(\ref{eq-discrhamil}). The Anderson model is a special example of
such Hamiltonians. More generally, if we accept that
transitions to bands far from the Fermi level are
essentially forbidden, the effective Hamiltonian is always
bounded and the Proposition~\ref{prop-vanishcurrent}
then leads to the vanishing of the current component parallel to $\EV$.

For these reasons, the presented
 approach is definitely too na\"{\i}ve. As we already pointed out,
collisions occur after time periods very short compared 
to the time
over which we average the current. These collisions produce a loss
of memory in the current evolution and are actually responsible for
the non-vanishing of the time average. Everything
looks like as if time evolution has to be considered over short intervals only.
We propose to take collision effects into account by means of the
following model: the time evolution is described by the new
time-dependent Hamiltonian $H_{coll}(t)=H-q\EV\cdot \XV+W_{coll}(t)$
where 

$$
W_{coll}(t)=
	\sum_{n\in \ZZ}
		W_{n} \delta (t-t_{n})
\mbox{ . }
$$

\noindent In this expression, the $t_{n}$'s represent collision times.
They are labeled such that $\ldots, t_{-1}< t_{0}= 0 <t_{1}
<\ldots <t_{n}<t_{n+1}<\ldots$.
 Because these times occur randomly, we will assume
that the time delays $\tau _{n}=t_{n}-t_{n-1}$ are independent
identically distributed random variables distributed according to
Poisson's law with mean value $\EE (\tau_{n})=\tau $ 

The $W_{n}$'s are the collision operators. Their main effect is to
produce a loss of memory during the time evolution of the current.
This process should enforce thermal equilibrium. In particular, it
should not modify the Fermi-Dirac distribution. The only way to
respect this constraint is to force the $W_{n}$'s to commute with the
Hamiltonian $H$, but to be random otherwise. More precisely, we will
assume that the $W_{n}$'s are independent identically distributed
random operators, commuting with the Hamiltonian $H$. Their distribution is
supposed to be
symmetric under the change of sign $W_{n}\leftrightarrow -W_{n}$. Let
then  $\cop$ be the operator acting on $\Aa$ as 

$$
\cop (A)=
	\EE (
		e^{\frac{\imath W_{n}}{\hbar}}
			A
				e^{-\frac{\imath W_{n}}{\hbar}}
					)
\mbox{ . }
$$

\noindent It can be extended to a bounded operator on
$L^{2}(\Aa ,\TV )$. Moreover, because of the change-of sign symmetry,
it is selfadjoint. Let then $\Aa_{H}$ be the closed subspace
of $L^{2}(\Aa ,\TV )$ generated by bounded functions of $H$. We will
then assume that there is $0\leq \kappa <1$ such that

$$
	\parallel 
		\cop (A)
			\parallel 
				\leq 
					\kappa \parallel A\parallel 
\mbox{ , }
\hspace{1cm}
\;\forall \; A \in (\Aa_{H})^{\perp}
$$

\noindent Since $W_{n}$ commutes with $H$ it follows that
$\cop $ leaves $\Aa_{H}$ invariant so that it forces any
operator $A\in \Aa$ along the direction of $H$. We will not give
explicit examples of such random operators because we will only
use the {\em collision efficiency operator} $\cop$ later on, so
that it solely characterizes our model. The parameter $\kappa$ is purely
phenomenological and represents an {\em efficiency coefficient} of
the collision process. The smaller $\kappa$, the more efficient are the
collisions. We see that the relaxation time, namely the time
after which there is a complete loss of memory of the initial
evolution, has to be renormalized by the efficiency coefficient. We
will see below that $\tau_{rel}=\tau /(1-\kappa )$ is a good
measure for this relaxation time. We will discuss later on how to
choose this parameter to fit with real systems.

\vspace{.2cm}

\subsection{Kubo's formula}
\label{sec-Kubo}

We now follow the strategy defined in the previous section and
compute the current average with the collisions taken 
into account. This requires to calculate the evolution
operator $S_{\xi}(t)$ between time $0$ and $t\geq 0$, where $\xi$
represents the random variables $\xi =(\tau_{n},W_{n})_{n>0}$. It
is well known that a kick term like $W_{n}\delta
(t-t_{n})$ produces a contribution $e^{\imath W_{n}/\hbar}$ in the
evolution between times $t_{n}-0$ and $t_{n}+0$, namely at the kick time 
\cite{CCIF,Gu1}. Therefore, if $n\geq 1$ and $t_{n-1}\leq t<t_{n}$, we 
find:

$$
	S_{\xi}(t)=
		e^{\imath (t-t_{n-1})H_{\EV}/\hbar }
			\prod_{j=1}^{n-1}
				e^{\imath W_{j}/\hbar }
					e^{\imath (t_{j}-t_{j-1})H_{\EV}/\hbar }
\mbox{ . }
$$

\noindent We will now set $\Ll_{H}(A)=(\imath /\hbar ) [ H,A ] $ for
$A$ in a suitable dense subalgebra of $\Aa$. This is a
*-derivation of $\Aa$ because of (\ref{eq-freeevolu}).
Therefore, it defines an anti-selfadjoint operator on $L^{2}(\Aa ,\TV
)$. Moreover, the evolution of observables is given by the
automorphism 

$$
	\eta_{\xi ,t}^{\EV}=
		e^{(t-t_{n-1})(\Ll_{H}+\frac{q}{\hbar}\EV .\naV )}
			\prod_{j=1}^{n-1}
				e^{\Ll_{W_{j}}}
					e^{(t_{j}-t_{j-1})(\Ll_{H}+\frac{q}{\hbar}\EV .\naV )}
\mbox{ , }
$$

\noindent where $\Ll_{W_j}(A)=(\imath/\hbar)[W_j,A]$.
The operator $\eta_{\xi ,t}^{\EV}$
may also be seen as a unitary on $L^{2}(\Aa
,\TV )$. In view of eq.~(\ref{eq-meancurrent}), the time average of
this evolution requires the calculation of 

$$\hat{\eta}_{\delta}^{\EV}=
			\delta \int_{0}^{\infty} dt e^{-t\delta} 
				\EE (\eta_{\xi ,t}^{\EV}),$$

\noindent where $\EE$ denotes the average over $\xi$ and $\delta
>0$. After a straightforward calculation we find:

$$
	\hat{\eta}_{\delta}^{\EV}=
		\delta 
			\left( 
				\delta +\frac{1-\cop}{\tau} -
					(\Ll_{H}+\frac{q}{\hbar}\EV .\naV )
			\right) ^{-1}
\mbox{ . }
$$

\noindent This expression is meaningful because the operator in
parentheses has a real part bounded from below by $\delta $. The average
current is then given by:

\begin{equation}
\label{eq-meancurrent2}
	\jV _{\beta ,\mu ,\EV }(\delta )=
		\frac{q}{\hbar}
			\TV 
				\left( 
					f_{\beta ,\mu}(H)
						\hat{\eta}_{\delta}^{\EV}(\naV H)
				\right)
\mbox{ . }
\end{equation}

\noindent We then remark that, for $\EV =0$, the right hand side of
(\ref{eq-meancurrent2}) vanishes because $\cop (f(H))=f(H)$ and
$\Ll_{H}(f(H))=0$ for any function $f$, whereas $\Ll_{H}$ is
anti-selfadjoint. Subtracting this expression with
$\EV =0$ and using the inner product $<A|B>=\TV (A^{\ast}B)$ in
$L^{2}(\Aa ,\TV )$, gives 

\begin{equation}
\label{eq-meancurrent3}
	\jV _{\beta ,\mu ,\EV }(\delta )=
		(\frac{q}{\hbar})^{2} 
			\sum_{i=1,2}\Ee_{i} 
				<\partial_{i}f_{\beta ,\mu}(H)|
					\frac{1}{
						\delta +\frac{1-\cop}{\tau} -
							\Ll_{H}-\frac{q}{\hbar}\EV .\naV 
									}
							\naV H>
\mbox{ . }
\end{equation}

\noindent If we assume that the non-linear term in the electric
field is negligible, we can let $\delta$ converge to zero. For
indeed, $\naV H$ and $\naV f_{\beta ,\mu}(H)$
are orthogonal to $\Aa_{H}$. Thus it is enough to consider the
restriction of the operator $\delta +\frac{1-\cop}{\tau} -
\Ll_{H}$ to the subspace orthogonal to $\Aa_{H}$. This restriction
has a real part bounded from below by $(1-\kappa )/\tau >0$. Thus it is
an invertible operator. If we do not neglect the non-linear term,
we have to investigate more thoroughly what happens as $\delta
\downarrow 0$. We will not discuss that matter here and postpone
it to a future work. Letting $\delta $ converge to
zero, we will get the Kubo formula for the conductivity tensor:

\vspace{.2cm}

\noindent{\bf Kubo's formula}
\begin{equation}
\label{eq-Kuboformula}
	\jV _{\beta ,\mu ,\EV }= {\bf \sigma} \EV
\mbox{ , }
\hspace{.3cm}
\mbox{\rm with}
\hspace{.3cm}
\sigma_{ij}=
	(\frac{q^{2}}{\hbar})
			< \partial_{j}f_{\beta ,\mu}(H) |
					\frac{1}{
					\hbar \frac{(1-\cop )}{\tau} - \hbar \Ll_{H}
					}			\partial _{i}H >
\mbox{ . }
\end{equation}

\noindent Let us remark that the quantity in the bracket is
dimensionless in $2D$: the trace $\TV$ is a trace
per unit volume (so that it has the dimension of the inverse of an
area) while the derivative $\partial_{j}$ has the dimension of a
length. In addition $\hbar \Ll_{H}$ has the dimension of an
energy. Therefore, we get ${\bf \sigma}$ dimensionless in units of
$q^{2}/h$. 

If, in addition, the system is symmetric under rotation by $\pi /2$
in space, namely if the Hamiltonian $H$ and the efficiency
operator $\cop$ are invariant under such a rotation, then the
conductivity tensor can be written in matrix form as 

$$
{\bf \sigma}=
	\left( 
		\begin{array}{cc}
			\sigma_{//} & \sigma_{H} \\
			-\sigma_{H} & \sigma_{//}
		\end{array}
	\right) 
\mbox{ , }
$$

\noindent where the Hall conductance is the off-diagonal term
$\sigma_{H}$, while $\sigma_{//}$ is the direct conductivity. 

Let us now consider the limit for which the IQHE is valid. Namely
the electric field is vanishingly small, the temperature is zero
and the relaxation time is infinite. In this limit, the Fermi
distribution becomes 

\begin{equation}
\label{eq-lowtemp}
\lim_{\beta \uparrow \infty}f_{\beta ,\mu}(H)=P_{F}
\mbox{ , }
\end{equation}

\noindent where $P_{F}$, the {\em Fermi projection}, is the
spectral projection onto energy levels lower than the Fermi
energy. Here the limit is taken with respect to the  
norm in $L^{2}(\Aa, \TV)$.
Actually, this result is correct only if the Fermi level $E_{F}$ 
is a continuity point of the DOS of $H$. Otherwise the
eigenprojection $P_{\{E_{F}\}}$ corresponding to the eigenvalue
$E_{F}$ satisfies $\TV (P_{\{E_{F}\}}) >0$ and therefore defines a
non-zero element of $L^{2}(\Aa ,\TV )$. This is what happens for instance
in the case of
 the Landau Hamiltonian whenever $E_{F}=(n+1/2)\hbar
\omega_{c}$. Moreover, we need the Fermi projection to be {\em
Sobolev differentiable}, namely it has to to satisfy $\naV P_{F}\in
L^{2}(\Aa ,\TV )$, otherwise the formal limit is meaningless. We
will see in Section~\ref{sec-Sobolev} below that such a condition
is related to the finiteness of localization length at the Fermi
level. We also need to show that the limit of the
derivative of the Fermi distribution is the derivative of its
limit. Even though we  know this to be true
for physically reasonable conditions on $H$, we will not give the
proof here but postpone it to a future work.

On the other hand, in the limit of infinite relaxation time, we are
left with the formal expression $\Ll_{H}^{-1}\partial_{i}H$ which
is meaningless in general. If we formally diagonalize the
Hamiltonian $H$, matrix elements of that expression are given by

$$<E|(\hbar \Ll_{H})^{-1}\partial_{i}H |E'>\;=\;
				\frac{<E|\partial_{i}H|E'>}{E-E'}.$$

\noindent In particular it diverges whenever $E\approx E'$ unless
the numerator vanishes for some reason. This divergency
does, however, not occur in the expression of the conductivity
tensor. The reason is that the derivative $\naV P_{F}$ has 
non-vanishing matrix elements only between energies $E$ and $E'$ such that
$E<E_{F}<E'$ or $E'<E_{F}<E$. For indeed, if $P$ is a projection
and $d$ is a derivation, we see that

$$dP=d(P^{2}),
			\hspace{.5cm}
				\Longrightarrow 
					\hspace{.5cm}
						dP = (1-P)dPP + PdP(1-P).$$

\noindent So we need to consider the operators
$P_{F}\Ll_{H}^{-1}\partial_{i}H\, (1-P_{F})$ and
$(1-P_{F})\Ll_{H}^{-1}\partial_{i}H\, P_{F}$ only. We then obtain

\begin{lemma}
\label{lem-proj}
If the Fermi level is not a discontinuity point of the DOS of
$H$, and if the Fermi projection is Sobolev differentiable, the
following formul{\ae} hold

$$P_{F}\Ll_{H}^{-1}\partial_{i}H\, (1-P_{F})=
 		-\imath\hbar P_{F}\partial_{i}P_{F}(1-P_{F}),$$

$$(1-P_{F})\Ll_{H}^{-1}\partial_{i}H\, P_{F}=
 		\imath \hbar(1-P_{F})\partial_{i}P_{F}P_{F},$$

\end{lemma}

\noindent {\bf Proof.} Let us consider the first formula only,
because the other can be treated in the same way (notice however
the change of sign). Let $B_{+}$ be the right hand side. Then 

$$ \hbar \Ll_{H} (B_{+})=
				P_{F} [ H, \partial_{i}P_{F} ] (1-P_{F}).$$

\noindent Since $H$ commutes with $P_{F}$, we find
$ [ H,\partial_{i}P_{F} ] = - [
\partial_{i}H,P_{F} ]$ (after using basic properties of
derivations). This gives immediately: 

$$ \hbar \Ll_{H} (B_{+})=
				P_{F}\partial_{i}H (1-P_{F}).$$

\noindent Since $B_{+}$ connects only energies below the
Fermi level with energies above it, and since the Fermi level is
not an eigenvalue of $H$, $\Ll_{H}$ is invertible on the
subspace of such operators, giving the result of the lemma.
\hfill  $\Box $

\begin{coro}[IQHE-Kubo Formula]
\label{coro-ChernKubo1}
If the Fermi level is not a discontinuity point of the DOS of
$H$, in the zero temperature and infinite relaxation time limit
and provided the Fermi projection is Sobolev differentiable, namely
$\naV P_{F} \in L^{2}(\Aa ,\TV )$, the conductivity tensor
is given by

$$\sigma_{ij}=
			\frac{q^{2}}{h}
				2\imath \pi\;
					\TV \left( 
						P_{F} [ \partial_{i}P_{F},\partial_{j}P_{F} ]
									\right)\, .$$

\noindent In particular the direct conductivity vanishes. 
\end{coro}

\vspace{.2cm}

\subsection{Estimating the deviations from the IQHE limit}
\label{sec-correction}

Before returning to the IQHE, we want to give some idea of the
accuracy of the IQHE-Kubo formul{\ae} given in the
Corollary~\ref{coro-ChernKubo1}. We recall that it is obtained
under the following conditions:

\bed
\item[(i)] the area of the sample is considered as infinite;

\item[(ii)] we work within the relaxation time approximation;

\item[(iii)] the electric field is vanishingly small;

\item[(iv)] the temperature is zero;

\item[(v)] the relaxation time is infinite. 
\ed

\noindent In this section we will evaluate roughly the size
of the correction terms whenever one of these conditions is
relaxed. We know that the relative error on the IQHE measurement of
the universal constant $e^{2}/h$ is of order of $2\times 10^{-8}$ at
best \cite{PG}. Therefore the neglected terms should be
smaller than this number in order that the experiment is reliable.

We will not estimate seriously the finite-size effects even
though they are in principle accessible to a mathematical
estimate within the non-commutative framework. However, it is
generally accepted that these effects  decrease to zero as
$e^{-L/r}$ where $L$ is the sample size and $r$ some typical
length. We will choose $r$ to be of the order of magnitude of the
localization length. We shall see that it diverges precisely
whenever the Hall conductance jumps from one plateau to another.
 Then as this localization length increases, there is a
critical value beyond which the Coulomb interaction between
electrons can no longer be neglected. This is the situation in which the FQHE
 occurs. To  estimate this
value, one can define $r$ to be such that the electrostatic energy of a
pair of electrons separated by a distance $r$ is of the order of
magnitude of the cyclotron energy. This gives $r\approx 1\mu m$ namely an 
overestimated large distance compared to the magnetic length. Thus as soon as
$L\geq 20 \mu m$, the finite size effects are negligible on the integer plateaux.
Fluctuations from sample to sample due to finite size and disorder will then be
negligible. Let us indicate that breakdown of the IQHE due to finite size
effects has been observed \cite{KIYYWK}.

The relaxation time approximation should actually be enough to
estimate other effects. Corrections to such an approximation
should not be effective at zero temperature, since the relaxation
time depends almost only upon the energy level. Only if transitions between
different bands are taken into account, it is necessary to go beyond this
approximation. This problem is too
difficult to be investigated here.

Let us consider the effect of a non-zero electric field. 
Non-linear effects such as bistability or hysteresis have indeed been
observed in such devices \cite{TKHS}.
If formula~(\ref{eq-meancurrent3}) is correct, 
the correction term due to a non-zero electric field is given by 

$$\delta \sigma_{i,j}=
			(\frac{q}{\hbar})^{2}
				< \partial_{j}f_{\beta ,\mu}(H) |
				\frac{1}{
				\frac{(1-\cop )}{\tau} - \Ll_{H}
				}
				\frac{q\EV .\naV }{\hbar}
				\frac{1}{
				\frac{(1-\cop )}{\tau} - \Ll_{H}
				}
				\partial _{i}H > .$$

\noindent The relative error is measured by the size of the ratio
$q\EV .\naV /\hbar \Ll_{H}$ between the electric energy
and the level separation. If we choose a level separation of the
order of the distance between Landau levels, namely $\hbar
\omega_{c}$, and let the electric energy be of the order of $e\Ee a$
(where $a$ is the atomic distance, namely $1\AA$), we find $\delta
\sigma /\sigma \approx 10^{-7}$ for an electric field of $1V/m$.
Thus non-linear effects require higher electric fields and it is
very easy to choose the electric field in such a way as to make
this correction negligible.

Non-zero temperature effects at infinte relaxation time can be
estimated as follows. Coming back to eq.~(\ref{eq-lowtemp}), using
eq.~(\ref{eq-Kuboformula}) and the Lemma~\ref{lem-proj}, we can
write the Kubo formula at finite temperature as

\begin{equation}
\label{eq-Hallcondatfintemp}
\sigma_{H}=
	\int_{-\infty}^{\infty}
		dE\, f_{\beta ,\mu}'(E)
			\frac{q^{2}}{h}
				2\imath \pi\;
					\TV \left( 
						P_{\leq E} [ \partial_{1}P_{\leq E},\partial_{2}P_{\leq E} ]
									\right)
\mbox{ , }
\end{equation}

\noindent provided we assume the relaxation time to be infinite.
Here $P_{\leq E}$ is the eigenprojection of the Hamiltonian on
energies smaller than or equal to $E$. The
spectral theorem allowed us to write (with an integration by parts):

$$ f(H)= 
				\int_{-\infty}^{\infty}
					dP_{\leq E} \; f(E)=
						-\int_{-\infty}^{\infty}
							dE\; f'(E)\; P_{\leq E}. $$

\noindent whenever $f$ is a bounded differentiable function
vanishing rapidly at $+\infty$. We will show that the integrand is
quantized and equal to $ne^{2}/h$ whenever $\hbar \omega_{c}
(n-1/2)<E<\hbar \omega_{c} (n+1/2)$ with $n=0,1,2,\ldots$.  This is
actually true for the most common devices. If they have charged
impurities the jump does not necessarily occur at the Landau level
\cite{GHP}. But this is of no importance for the present
discussion since we only need orders of magnitude for the
correction terms. Incidentally eq.~(\ref{eq-Hallcondatfintemp})
shows that the direct conductivity vanishes at finite
temperature as well (if the relaxation time is infinite). The error term
can easily be computed and is given by

\begin{equation}
\label{eq-distem}
\frac{\delta \sigma_{H}}{\sigma_{H}}=
			\sum_{n'\leq 0,n'\neq n}
				\frac{n'}{n}
					\int_{\hbar \omega (n'-1/2)}^{\hbar \omega (n'+1/2)}
						dE \frac{\beta }{4\cosh ^{2}(\beta (E-\mu)/2)},
\end{equation}

\noindent provided $\hbar \omega_{c} (n-1/2)<E_{F}<\hbar \omega_{c}
(n+1/2)$. We know that the chemical potential equals $E_{F}$ up
to small terms of order $T^{2}$ (where $T$ is the temperature).
If $E_{F}$ is located at a distance $r\hbar \omega_{c}$ from
the nearest Landau level, where $0<r\leq 1/2$, this correction is
thus of the order of $e^{-r\beta \hbar \omega_{c}}$. For a magnetic
field of $10T$ and a charge carrier mass of $.1m_{e}$, $\hbar
\omega_{c}/k_{B}\approx 140^{\circ}K$. This term is smaller than
$10^{-7}$ for temperature lower than $4^{\circ}K$. Accurate
experiments are often performed at $50mK$ and magnetic fields
can be as high as $18T$. Moreover, in heterojunctions, the
effective mass of the charge carriers is one order of
magnitude smaller. Therefore, the pure temperature effect (at infinite
relaxation time) is indeed negligible.

The infinite relaxation time approximation is in fact the most
important effect. The relevant parameter is the relaxation time
$\tau_{rel}=\tau /(1-\kappa )$; it can be estimated by means of
Drude's formula by measuring conductivities. Assuming the
relaxation time to be large, the lowest order contribution to the
Hall conductivity is given by (see eq.~(\ref{eq-Kuboformula}))

\begin{equation}
\label{eq-discrep}
\sigma_{H}\approx 
	\frac{q^{2}}{\hbar}
		\frac{1}{\tau_{rel}}
			\TV \left( 
				P_{F} [ \Ll_{H}^{-1}(\partial_{1}P_{F}),
					\partial_{2}P_{F} ] 
							\right) 
\mbox{ . }
\end{equation}

\noindent Estimating $\Ll_{H}^{-1}$ by $1/\omega_{c}$, we see that 
the error term due to
this contribution is controlled by $\varepsilon =(\tau
\omega_{c})^{-1}$. For semiconductors used in the IQHE, the
mobility $\mu_{c}$ ($c$ denotes the type of charge carriers) at
zero magnetic field and $4.2^{\circ}K$ varies from about
$10^{4}cm^{2}/Vs$ for the MOSFET to $10^{6}cm^{2}/Vs$ for the
AlGaAS or InAs-GaSb heterojunctions \cite{PG}[pp.40\& 41]. These
high-mobility devices are used for the FQHE. On the other hand,
the effective mass of charge carriers varies from $0.2m_{e}$ for
the MOSFET to $0.03m_{e}$ for InAs-GaSb heterojunctions. Since
$\tau \omega_{c}=\mu_{c} \Bb $, we find $\varepsilon \approx 0.1\sim
0.005$ for $\Bb =10T$. We remark, however, that our relaxation time 
${\tau_{rel}}$ only incorporates time-dependent disorder and may
therefore be
significantly smaller than the one calculated with Drude's formula.

It is usually estimated \cite{KDP} that the deviation of the Hall
conductance from its ideal value is linked to the direct conductivity
$\sigma_{//}$ by 

$${\delta \sigma_{H}}\approx
			\frac{\sigma_{//}}{\tau_{rel} \omega_{c}}.$$

\noindent Measurements give $min({\sigma_{//}})\leq
10^{-7} max({\sigma_{//}})\approx 10^{-7}e^2/h$ \cite{KDP} so that 
the relative error on the Hall conductance is indeed of the order of 
$10^{-8}$. 

Why is $\sigma_{//}$ so small ? Looking at the equivalent of
equation (\ref{eq-discrep}) for the direct conductivity, 
we see that one way to estimate it
consists of replacing $\Ll_{H}^{-1}$ by $1/\omega_{c}$ whereas the
remaining terms are related to the localization length
$\lambda$. But since $\TV (P_{F})=n$ is the charge carrier density,
we expect this term to be of the order of
$n\lambda^{2}/\tau_{rel}\omega_{c}$. 
The mobility $\mu_{c}$ of the charge carriers $c$ is
$\tau_{rel}\omega_{c}/\Bb $. Moreover we introduce the filling factor 
$\nu =nh/q\Bb $ leading to the very rough estimate

$$\frac{\delta \sigma_{H}}{\sigma_{H}}\leq
			{\rm const.}\nu 
				\frac{e}{h} 
					\frac{\lambda^{2}}{\mu_{c}}.$$

\noindent For the heterojunction, $\nu$ is typically between $1$ and $10$ at
most. The ratio $q/h$ is given by the electron charge $e/h$ and is thus 
universal. The localization length is always bigger than $80\AA$. The mobility
is at most equal to $2\cdot 10^{6} cm^{2}/Vs$ for the best heterojunctions. We
then find a ratio of the order of $10^{-4}$. Therefore, this estimate
is too crude to explain the high accuracy observed in experiment.
Nevertheless, it shows that collision effects
are dominant whereas localization is a necessity.

In any way, physical arguments indicate \cite{PS,PG} that because of the
small value of the density of states between Landau levels, the
leading contribution to the direct conductivity is given by
phonon-assisted hopping, at least while the Fermi energy is not too
close to a Landau level. 
Estimation (\ref{eq-distem}) only includes conductivity by thermal
acctivation, whereas (\ref{eq-discrep}) only incorporates effects due
to a finite collision rate. Here, we shall not treat the interplay of
the two phenomena, but only indicate that Mott theory leads to
$\sigma_{//}\approx\sigma_0 \;\mbox{exp}(-(T_0/T)^{1/3})$ in two
dimensions, while including Coulomb interaction would give
$\sigma_{//}\approx\sigma_0 \;\mbox{exp}(-(T_0/T)^{1/2})$
\cite{SE}; this latter is in better agreement with experiment
\cite{PS,vK}.

\subsection{Dixmier trace and Sobolev space}
\label{sec-Dixmier}

In this section, we introduce the Dixmier trace and we will prove a formula
that can be found in a similar form in a paper by A. Connes \cite{Co87}. We will
see that this tool is a key point both in proving the integrality of the Chern 
number of a projection and in the study of localization. Most of the material 
presented her can be found in \cite{Co87,Co93}, so that we 
will give no proof unless absolutely necessary.

First, let us recall that,
given a separable Hilbert space $\Hh$, 
$\Kk(\Hh)$ is the \CS
of compact operators on $\Hh$, namely the norm-closure of the set of finite rank
linear operators on $\Hh$.
The Schatten ideal $\Ll^p(\Hh)$ is defined as
the set of compact operators $T$ such that
$\sum_{n=1}^{\infty} \mu_{n}^{p} < \infty$, where the $\mu_n$'s are the eigenvalues
of $(TT^*)^{1/2}$ labeled in the decreasing order.
The following proposition summarizes the main properties of Schatten
ideals; it can be found in \cite{RS,Co90}, for
example.

\begin{proposi} 
Let $\Ll(\Hh)$ be the algebra of bounded
operators on $\Hh$ and $\TR$ the usual trace on $\Ll(\Hh)$. Then 
we have the following:

\bed
\item[(i)] $\Ll^p(\Hh)=\{ T\in\Ll(\Hh)\;|\;\TR(|T|^p)<\infty\}$. 

\item[(ii)] $\Ll^p(\Hh)$ is a two-sided ideal in $\Ll(\Hh)$.
 
\item[(iii)] $\Ll^p(\Hh)$ is a Banach space for the norm $\parallel T \parallel_{_p}= 
(\TR(|T|^p))^{\frac{1}{p}} = (\sum_n(\mu_n(T)^p))^{\frac{1}{p}}$.

\item[(iv)]  $\Ll^p(\Hh) \subset  \Ll^q(\Hh)$ for $p\leq q$.

\item[(v)] Let $p,q,s \in [1,\infty)$ with $\frac{1}{r} = 
\frac{1}{p}+\frac{1}{q}$, $S\in \Ll^p(\Hh)$ and $T\in \Ll^p(\Hh)$. Then, 
H\"older's inequality holds:
$\parallel ST\parallel_r \leq\parallel S\parallel_p \parallel T\parallel_q$.
\ed

\end{proposi}

\noindent Now we introduce the
Ma\v{c}aev ideals $\Ll^{p+}(\Hh)$ and $\Ll^{p-}(\Hh)$ and the Dixmier trace.

\begin{defini}
Let $\Hh$ be a separable Hilbert space and ${\cal K}$ be the
ideal of compact operators on $\Hh$. For $p\in [1,\infty)$, the Ma\v{c}aev ideal 
$\Ll^{p+}(\Hh) \subset {\cal K}$ is the set of compact operators $T$ of which the
characteristic values satisfy 

$$
\limsup_{N\rightarrow \infty}
 \frac{1}{\ln N} 
  \sum_{n=1}^N 
   \mu_n^p\; 
    < \;\infty
\mbox{ , }
$$

\noindent where the characteristic values are the eigenvalues $(\mu_n)$ of
$(TT^*)^{\frac{1}{2}}$ labeled in decreasing order.
$\Ll^{p-}(\Hh)$ is defined in the much same way but with the $\limsup $
equal to zero. We will also set

$$
\parallel T \parallel_{p+}=
 \sup_{N\rightarrow \infty}
  \frac{1}{\ln N} 
   \sum_{n=1}^N 
    \mu_n^p\; 
\mbox{ . }
$$

\noindent 

\end{defini}

\noindent 

\begin{theo}
 i) $\Ll^{p+}$ and $\Ll^{p-}$ are two-sided ideals in
${\cal L}(\Hh)$.

\noindent ii) For $p\in [1,\infty)$ one has: $\Ll^p \subset \Ll^{p-} \subset \Ll^{p+} 
\subset \Ll^{p+\varepsilon} \;\;\;\;\forall \varepsilon > 0$.

\noindent iii) The expression $\parallel T \parallel_{p+}$ defines a norm
on $\Ll^{p+}$, making it into a Banach space.
\end{theo} 

\noindent Next, the Dixmier trace is constructed as follows \cite{Dix}. Let
Lim be a positive linear functional on the space of bounded sequences
$l_{+}^{\infty}(\NN)$ of positive real numbers which is translation and scale 
invariant. If $\alpha \in l_{+}^{\infty}(\NN)$ converges, then
the functional Lim satisfies:

\begin{equation}
\label{limit}
\mbox{\rm Lim}(\alpha) 
 = 
  \lim_{n \rightarrow \infty} 
   \alpha_n
\mbox{ . }
\end{equation}

\noindent Scale invariance means that $\mbox{\rm Lim}(\alpha) = \mbox{\rm Lim}(\alpha_1,
\alpha_1,\alpha_2,\alpha_2,\ldots)$.
To construct Lim, Dixmier uses an invariant mean on the Euclidean group of {\bf R} 
(the existence of such means results from a theorem of von Neumann).

\begin{defini}
\label{def-Dixmiertr}
For positive $T\in \Ll^{1+}$ and a fixed scale-invariant and positive linear
functional $\mbox{\rm Lim}$ on $l_{+}^{\infty}(\NN)$ satisfying (\ref{limit}),
the Dixmier trace is defined by: 

$$
\TD(T)=
 \mbox{\rm Lim}
  (
   \frac{1}{\ln N} 
    \sum_{n=1}^N \mu_n
     )
\mbox{ . }
$$

\end{defini}

\noindent Remark that $T\in \Ll^{1+}$ if and only if  $\TD(| T|)< \infty$.
Moreover, if the sequence $(\frac{1}{\ln N} \sum_{n=1}^N \mu_n)$ 
converges, then all functionals Lim of the sequence are equal to the limit 
and the Dixmier trace is given by this limit.
From this definition, one can show that $\TD$ is a trace in the following
sense \cite{Dix}:

\begin{proposi}
\label{prop-Dixmiertr}
The functional $\TD$ defined in Definition~\ref{def-Dixmiertr} can be extended as
a linear form on $\Ll^{1+}$ such that:

\noindent (i) {\em positivity:} if $T\in \Ll^{1+}$ is a positive operator, then $\TD (T)>0$,

\noindent (ii) {\em trace property:} if $S,T\in \Ll^{1+}$ then $\TD (ST)=\TD (TS)$,

\noindent (iii) {\em unitary invariance:} if $T\in \Ll^{1+}$ and $U$ is unitary 
then $\TD (UTU^{-1})=\TD (T)$,

\noindent (iv) {\em continuity:} it is continuous with respect to the semi-norm 
$\parallel T \parallel_{1+}$. Moreover, $\TD $ vanishes on $L^{1 -}$.

\end{proposi}

\noindent Let us introduce the operator $\hat{\delta}$ acting on a 
linear operator $A$ on 
$l^2(G)$ as

$$
\hat{\delta} A=[u,A] \mbox{ , }\qquad u=\frac{X_1+\imath X_2}{|X_1+\imath X_2|} \mbox{ , }
$$

\noindent where $X_1,X_2$ are the components of the position operator.
The main result of this section is given by the following proposition.

\begin{proposi}
\label{prop-DixmierConnes}
Let $\Omega$ be a compact metrizable space on which $G$ acts by homeomorphisms. 
Let $\PP$ be a $G$-invariant ergodic probability on $\Omega$. One then denotes
by $\Aa$ the \CS of this dynamical system and by $\TV  $ the trace on $\Aa$ 
corresponding to $\PP$. Let $\AO$ be the dense subalgebra of continuous functions with 
compact support on $\Omega \times G$. 

\noindent Then for every $A\in \AO$, the following formula holds:

\begin{equation}
\label{eq-Connes2}
\TV (| \naV A| ^{2})=
 \frac{2}{\pi}
  \TD (| \hat{\delta} A_{\omega}| ^{2})
\mbox{ , }
\hspace{1cm}
 \mbox{for $\PP$-almost all $\omega$}
\mbox{ . }
\end{equation}

\noindent If $\Ss$ denotes the Sobolev space associated to $\TV  $, this formula
can be continued to elements $A\in \Ss$. In particular, if $A\in \Ss$, then 
$\hat{\delta} A_{\omega} \in \Ll^{2+}$ for $\PP$-almost all $\omega$.

\end{proposi}

\noindent In the remaining part of this section we present the proof of 
Proposition~\ref{prop-DixmierConnes} for the case of discrete physical space
$G=\ZZ^2$; the continuous case will be treated in future work \cite{SCB}.
The first step in this proof is the following lemma:

\begin{lemma} 
\label{lem-diagdixmier}
Let $T$ be a bounded operator on $\ell^2(\ZZ^D)$ such that

\bed
\item[i)] $\exists \;r$ such that $<n|T|m> =0 \;\; \forall \; |n-m|\geq r$.
\item[ii)] There exists a positive constant $C$ such that 
$|<n|T|m>|\leq C/(1+|n|^D)\;\;\forall \;m\in \ZZ^D$.
\ed

\noindent $T$ is then in the Ma\v{c}aev
ideal $\Ll^{1+}$ and, for any linear functional \mbox{\rm Lim}, its Dixmier trace can be
calculated by

$$
\TD(T)=\TD(\mbox{\rm Diag}(T))
$$

\noindent where {\rm Diag}($T$) is the diagonal matrix such that $<n|\mbox{\rm Diag}(T)|m>=\delta_{n,m}
<n|T|m>$.

\end{lemma}

\noindent {\bf Proof.} Since $T$ has only a finite number of diagonals, it can be written as a finite sum
of operators having only one non-zero diagonal. Using the additivity of the Dixmier trace
there is no loss of generality in assuming that $T$ has only one non-zero diagonal, namely that 
it acts on $\ell^2(\ZZ^D)$ as

$$T\psi (n)=
   t(n)\psi (n-a)
   \mbox{ , }
   \hspace{1cm}
    \psi \in \ell^2(\ZZ^D),\, a\in \ZZ^D
    \mbox{ , }
$$

\noindent where $t$ is a sequence on $\ZZ^D$ such that $| t(n)| 
\leq C(1+| n| )^{-D}$. It is thus enough to prove that $T\in \Ll^{1+}$
whatever the value of $a$ and that, if $a\neq 0$, its Dixmier trace 
vanishes.

It is clear that the modulus $| T |$ of $T$ is a diagonal operator dominated 
by $C R$ where $R$ is the multiplication operator by $(1+| n| )^{-D}$. 
Let us show that $R\in \Ll^{1+}$ which implies $T\in \Ll^{1+}$. 
Its eigenvalues are $1/j^{D}$ with a multiplicity $O(j^{D-1})$, therefore labeling
them in decreasing order $\mu_{1}\geq \ldots \geq \mu_{s}\geq \ldots$ with their 
multiplicity, we get:

$$\sup_{N>0}
   \frac{1}{N}
    \sum_{s=1}^{N}
     \mu_{s}
      \leq 
  \sup_{N>0}
   \frac{1}{N}
    \sum_{j=1}^{O(N^{1/D})}
     const. \frac{1}{j}
     < \infty 
\mbox{ . }
$$

\noindent Thus $R\in \Ll^{1+}$.

Let us now assume $a\neq 0$. We will show that $T$ is then unitarily equivalent to $-T$.
Since the Dixmier trace is invariant by unitary transformations 
(Proposition~\ref{prop-Dixmiertr}) it will follow that $\TD (T)=0$. We remark that for 
any $n\in \ZZ^{D}$ the subspace $\Ee _{n}=l^{2}(n+a\ZZ)$ is invariant under the action
of $T$. Clearly $\Ee _{n}$ is isomorphic to $l^{2}(\ZZ)$ and through this isomorphism
$T$ acts as $T_{n}\varphi (j)=t(n+ja)\varphi (j-1)$ on $\varphi \in l^{2}(\ZZ)$.
Let us define the unitary operator $U$
on $l^{2}(\ZZ)$ as the multiplication by $(-)^{j}$. Then one easily finds $UT_{n}U^{-1}
=-T_{n}$. Lifting $U$ to $\Ee_{n}$ gives a unitary denoted by $U_{n}$. Now $l^{2}
(\ZZ^{D})$ is the direct sum of the $\Ee_{n}$ whenever $n$ runs in a fundamental domain 
of the subgroup $a\ZZ$ acting on $\ZZ^{D}$ by translation. Taking the direct sum of 
the corresponding $U_{n}$'s gives a unitary $\hat{U}$ on $l^{2}(\ZZ^{D})$ such that 
$\hat{U}T\hat{U}^{-1}=-T$. \hfill $\Box$

\vspace{.2cm}

The next step in the proof of Proposition~\ref{prop-DixmierConnes} is the following

\begin{lemma}
\label{lem-density}
Let $\Sigma$ be a subset of $\ZZ^D$ not including the origin and with finite density
{\rm Dens}($\Sigma$), namely 

$$
\mbox{\rm Dens}(\Sigma)=
 \lim_{N\rightarrow \infty }
  \frac{1}{N^D}
   \sum_{n\in\Sigma, |n|\leq N} 1
\mbox{ , }
$$

\noindent where $|n|$ is the euclidean norm of the vector $n$. Then if $R_{\Sigma}$
is the restriction to $\Sigma$ of the operator of multiplication by $1/|n|^D$ in
$l^{2}(\ZZ^{D})$, we find

$$
\TD(R_{\Sigma})= 
 \frac{\omega_{D}}{D}
  \mbox{\rm Dens}(\Sigma) 
\mbox{ , }
$$

\noindent where $\omega_{D}$ is the area of the $D-1$ unit sphere of $\RR^{D}$.
In particular, for $D=2$, the geometrical constant in the right hand side is $\pi$.

\end{lemma}

\noindent {\bf Proof.} The eigenvalues of $R$ are $1/j^D$. 
The multiplicity $g_{j}(\Sigma)$ of such an eigenvalue is therefore given by the 
number of $n$'s in $\Sigma$ such that $|n|=j$. Let $\Sigma_{N}$ is the subset of 
$\Sigma$ of elements $n$ with $|n|\leq N$.
Since the eigenvalues of $R$ are already labeled in decreasing
order, we obtain

\begin{equation}
\label{eq-trdixersigma}
\TD(R_{\Sigma})= 
 \lim_{N\rightarrow \infty}
  \frac{1}{\ln |\Sigma_{N}|}
   \sum_{j=1}^{N} 
    \frac{g_{j}(\Sigma)}{j^D}
\mbox{ , }
\end{equation}

\noindent where $|\Sigma_{N}|$ denotes the number of points in $\Sigma_{N}$. Using the
definition of the density of a subset, we see that, as $j\rightarrow \infty$, the 
multiplicity $g_{j}(\Sigma)$ is asymptotically given by the product of the density
of $\Sigma$ by the volume between the balls of radii $j-1$ and $j$. Namely

$$
g_{j}(\Sigma)
\stackrel{j\rightarrow \infty}{\sim}
 \, \mbox{\rm Dens}(\Sigma)\, \,
  \omega_{D}\, j^{D-1}
\mbox{ . }
$$

\noindent In much the same way, we get $|\Sigma_{N}|\, \stackrel{N\rightarrow 
\infty}{\sim} \, \mbox{\rm Dens}(\Sigma)\, \Omega_{D} \, N^{D}$ if $\Omega_{D}$ is
the volume of the unit ball of $\RR^{D}$. Taking the logarithm we are left
with $D\ln N +O(1)$ in the expression of the Dixmier trace of $R_{\Sigma}$.
Plugging all these relations in eq.~(\ref{eq-trdixersigma}), we get the result.
\hfill $\Box$

\vspace{.2cm}

The last technical step in the proof of Proposition~\ref{prop-DixmierConnes} is provided
by the following lemma where the dimension is $D=2$.

\begin{lemma}
\label{lem-eqconnesdeux}
Let $f$ be a continuous non-negative function on $\Omega$ and $a\in\ZZ^2$, 
$a\neq 0$. Let $F^a_{\omega}$ the operator on $\ell^2(ZZ^2)$ defined 
by:

$$
F^{a}_{\omega} \psi (n)=
 f(T^{-n}\omega)\;
  |\frac{n}{|n|}-\frac{n-a}{|n-a|}|^2\;
   \psi(n)
\mbox{ , } 
 \qquad \psi\in \ell^2(ZZ^2)
\mbox{ . }
$$

\noindent Then $F^{a}_{\omega}\in \Ll^{1+}$ and its Dixmier trace is given by

$$
\TD(F^a_{\omega})=
 \frac{\pi}{2}\, |a|^2 
  \int d\PP (\omega) f(\omega)
   \mbox{ , } \qquad 
\mbox{for $\PP$-almost every } \omega\in \Omega
\mbox{ . }
$$

\end{lemma}

\noindent {\bf Proof.} As $|n|\rightarrow \infty$ the function 

$$\phi(n)=
   |\frac{n}{|n|}-\frac{n-a}{|n-a|}|^2
\mbox{ , }  
$$

\noindent admits the asymptotics $\phi (n)\, \stackrel{|n|\rightarrow 
\infty}{\sim}\, |a|^2 \sin^{2}{\alpha_{n}}/|n|^2$ modulo terms of order $1/|n|^3$,
where $\alpha_{n}$ is the angle between the directions of $a$ and $n$.

Let us now slice the space $\Omega$ according to the finite partition 
$\{ \Omega_{j,\delta}\}$, where $\delta >0$ is small enough and $j$ an integer 
such that $\Omega_{j,\delta}$ is the set of points $\omega$ for which
$(j-1/2)\delta \leq f(\omega)<(j+1/2)\delta$. Since $f$ is continuous
with compact support, it is bounded so that only a finite number of
$j$'s are needed here. Let then 
$\Sigma_{j,\delta}(\omega)$ be then set of $n$'s in $\ZZ^2$ such that
$T^{-n}\omega \in \Omega_{j,\delta}$. Using Birkhoff's ergodic theorem, 
for $\PP$-almost every $\omega$,  
$\Sigma_{j,\delta}(\omega)$ has a finite density given by the probability 
$\PP (\Omega_{j,\delta})$. 

We then slice $\Sigma_{j,\delta}$ into a finite subpartition 
$\{ \Sigma_{j,\delta ,r} \}$ where $\Sigma_{j,\delta ,r}$ corresponds to those 
points $n\in \Sigma_{j,\delta}$ for which $(r-1/2)\delta \leq \alpha_{n} 
< (r+1/2)\delta $. Thus modulo an error of order $O(\delta)$ we get
$f(T^{-n}\omega )\phi(n)=|a|^2\, j\delta \, \sin^{2}{(r\delta)}(1+O(\delta))
/|n|^2$ on $\Sigma_{j,\delta ,r}$. Moreover since this slicing concerns only a 
finite partition it permits to write $F^a_{\omega}$ as a finite sum namely

$$F^a_{\omega}=
   \sum_{j,r}
    F^a_{\omega}|_{\Sigma_{j,\delta ,r}}
\mbox{ . }
$$

\noindent It is thus sufficient to compute the Dixmier trace of the restriction
to $\Sigma_{j,\delta ,r}$ of $F^a_{\omega}$. But up to an error of order 
$O(\delta)$ this restriction is nothing but $|a|^2j\delta \sin^{2}{(r\delta)}
R_{\Sigma_{j,\delta ,r}}$. Using the Lemma~\ref{lem-density} we then get

$$\TD (F^a_{\omega}|_{\Sigma_{j,\delta ,r}})=
   \pi \, |a|^2 \, j\delta \sin^{2}{(r\delta)} (1+O(\delta))\,
    \mbox{\rm Dens}(\Sigma_{j,\delta ,r})
\mbox{ . }    
$$

\noindent Due to the slicing of the angles one gets 
$\mbox{\rm Dens}(\Sigma_{j,\delta ,r})= \delta /2\pi 
\mbox{\rm Dens}(\Sigma_{j,\delta})= \PP(\Omega_{j,\delta})\delta /2\pi $.
Plugging everything together, summing up over $j,r$ and letting $\delta$ 
converge to zero, the sum over $r$ gives the averaged value of $\sin^2({\alpha})$,
namely $1/2$, whereas the sum over $j$ reconstructs the integral of $f$. 
\hfill $\Box$

\vspace{.2cm}

\noindent {\bf Proof of Proposition~\ref{prop-DixmierConnes} (end).} Thanks to
Lemma~\ref{lem-diagdixmier}, it is enough to compute the diagonal elements of 
$|\hat{\delta} A_{\omega}|^2$ because $A\in \AO$ so that the number of non-zero diagonals
is finite. The diagonal elements are

$$<n||\hat{\delta} A_{\omega}|^2|n>=
   \sum_{a\in \ZZ^2}
    |A(T^{-n}\omega,a)|^2 
     |\frac{n}{|n|}-\frac{n-a}{|n-a|}|^2
\mbox{ . }     
$$

\noindent The number of terms in this sum is finite. Using 
Lemma~\ref{lem-eqconnesdeux} we find

$$\TD (|\hat{\delta} A_{\omega}|^2)=
   \frac{\pi}{2}
    \sum_{a\in \ZZ^2} |a|^2
     \int_{\Omega}d\PP (\omega)
      |A(\omega ,a)|^2 
\mbox{ , }     
$$

\noindent for $\PP$-almost all $\omega$'s. On the other hand, the definition
of the differential on $\Aa$ gives (see eq.~(\ref{eq-deri})) 
$\naV A(\omega ,a)=\imath \vec{a} A(\omega ,a)$. In particular

$$|\naV A|^2(\omega,0)=
   \sum_{a\in \ZZ^2} |a|^2 |A(\omega ,a)|^2
\mbox{ . } 
$$

\noindent To get the trace per unit volume, we just have to integrate both
sides of this equation giving the Connes formula. 

Since the left hand side of the Connes formula is the dominated by the 
Sobolev norm of $A$, one can extends this formula to $A\in \Ss$. In 
particular the finiteness of the right hand side implies that 
$\hat{\delta} A_{\omega}\in \Ll^{2+}$ $\PP$-almost surely.
\hfill $\Box$

\vspace{.2cm}

\subsection{Non-commutative Chern character}
\label{sec-NCChern}

We denote by $\Pp (\Aa)$ the set of orthogonal projections in the
\CS $\Aa$, namely the set of elements $P$ in $\Aa$ such
that $P=P^{2}=P^{\ast}$. If in addition $P$ is differentiable, we
define its Chern character as 

\begin{equation}
\label{eq-ChernP}
\Ch (P)= 2\imath \pi \;
	\TV (P [ \partial_{1}P,\partial_{2}P ] ) 
\mbox{ . }
\end{equation}

\noindent If we work on a lattice, $\TV $ is normalized such that
$\TV ({\bf 1})=1$. In the continuum case, we will normalize it in
reference to the projections onto Landau levels, namely the
eigenvalues of the Landau Hamiltonian.
(\ref{Landau}). An elementary calculation gives for
the lowest Landau level projection the following integral kernel

\begin{equation}
\label{eq-lowestLandau}
\Pi_{0}(\vec{x},\vec{y})=
	\frac{q\Bb }{h}
		e^{-
					\frac{q\Bb }{4\hbar}(\vec{x}-\vec{y})^{2}-
						\imath \frac{q\Bb }{2\hbar}
							\vec{x}\wedge\vec{y}
					}
\mbox{ . }
\end{equation}

\noindent In particular, $\Pi_{0}$ defines an element of $\Aa =
C^{\ast}(\Omega,\RR^{2},\Bb )$ for any choice of the hull $\Omega$. In much the
same way, we denote by $\Pi_{n}$ the projection on the $n^{th}$ Landau level
($n=0,1,2,\ldots$). We then deduce the following results.

\begin{lemma}
\label{lem-Chern Landau}
The trace and the Chern character of the Landau levels are given by

\begin{equation}
\label{eq-TraceChernLandau}
\TV (\Pi_{n})=
	\frac{q\Bb }{h}
\mbox{ , }
\hspace{1cm}
\Ch (\Pi_{n})=-1
\mbox{ . }
\end{equation}

\end{lemma}

\noindent {\bf Proof.} Let us prove this result for $\Pi_{0}$ first.
Its trace per unit volume is given by the space average of
$\Pi_{0}(\vec{x},\vec{x})=q\Bb /h$. This gives the first formula in
(\ref{eq-TraceChernLandau}). To compute its Chern character, we remark
that $\naV \Pi_{0}(\vec{x},\vec{y})=-\imath (\vec{x}-\vec{y})
\Pi_{0}(\vec{x},\vec{y})$. Introducing the complex
variables $x=x_{1}+\imath x_{2}$ and $y=y_{1}+\imath y_{2}$, we get

$$
\Ch (\Pi_{0})=
	\pi (\frac{q\Bb }{h})^{3}
		\int_{\CC \times \CC} d^{2}x \; d^{2}y \;
			e^{-
					\frac{q\Bb }{2\hbar}
						(|x|^{2} +|y|^{2} -x\overline{y})
						}
						(x\overline{y}-y\overline{x})
\mbox{ . }
$$

\noindent To compute this integral, we develop the exponential in
powers of $x\overline{y}$ and notice that all contributions vanish
except the one involving the term $|x|^{2}|y|^{2}$. The corresponding
integral has separated variables and can be computed explicitly.
This gives $-1$. 

For the other Landau levels, we remark that $\Pi_{n}\sim \Pi_{0}$ in the
sense of von Neumann equivalence (see below). More concretely, it is
enough to exhibit an element $U_{n}\in \Aa$ such that
$\Pi_{n}=U_{n}^{\ast}U_{n}$ and $\Pi_{0}=U_{n}U_{n}^{\ast}$. This
implies that the traces are identical and we will show their
Chern characters to be identical as well
(see Lemma~\ref{lem-cherninvariance} below).
To construct $U_{n}$, we introduce the annihilation operator 

$$
a=(P_{1}-qA_{1} +\imath P_{2}-qA_{2})/\sqrt{2\hbar q\Bb }.
$$

\noindent Then $ [ a,a^{\ast} ] ={\bf 1}$. Thus $aa^{\ast}$ is
bounded below by $1$ and is invertible. We set
$u=(aa^{\ast})^{-1/2}a$ and $U_{n}= \Pi_{0}u^{n}$. It is easy to
check that $uu^{\ast}={\bf 1}$ implying that
$\Pi_{0}=U_{n}U_{n}^{\ast}$. On the other hand, 
$U_{n}^{\ast}U_{n}= a^{\ast n}\Pi_{n}a^{n}/n!$. But it is a standard
result that we obtain the $n^{th}$ Landau level by applying the creation
operator $n$ times to the groundstate, namely
$|n>=(1/\sqrt{n!})a^{\ast n}|0>$. In particular  $a^{\ast
n}\Pi_{n}a^{n}/n!=\Pi_{n}$.

\noindent It remains to show that $U_{n}\in \Aa$. A straightforward
calculation gives $U_{n}=(n!)^{-1/2}\Pi_{0}a^{n}$. Now using the explicit
form of the matrix elements (\ref{eq-lowestLandau}) and of $a$, we get the
matrix elements of $U_{n}$ in the form of a polynomial in $x$ and $y$ times
$\Pi_{0}(\vec{x},\vec{y})$, showing that $U_{n}(\vec{0},\vec{x})$ is
absolutely summable in $\vec{x}$. This is enough to show that it belongs to
$\Aa$ (see Section~\ref{sec-ObsCal}).
\hfill $\Box$

\vspace{0.2cm}

Our next step will be the von Neumann equivalence. Namely if
$P,Q\in \Pp (\Aa)$, then $P\sim Q$ if there is $U\in \Aa$ such that 
$P=U^{\ast}U, Q=UU^{\ast}$. In particular if $P$ is trace class, it
follows that $\TV (P) = \TV (Q)$ then. 
The following results can be found in \cite{Ped}

\begin{lemma}
\label{lem-homotopy}
Let $P,Q\in \Pp (\Aa)$ be such that $\parallel P-Q\parallel <1$.
Then $P\sim Q$.
\end{lemma}

\begin{lemma}
\label{lem-countableKth}
If $\Aa$ is separable, then the set of equivalence classes of
projections in $\Aa$ is at most countable.
\end{lemma}

\begin{lemma}
\label{lem-classaddit}
Let $P,Q\in \Pp (\Aa)$ be two mutually orthogonal
projections (namely $PQ=QP=0$). Then the equivalence class of their
sum $P\oplus Q$ depends only upon the equivalence classes of $P$ and
$Q$. This defines a commutative
and associative composition law  on the set of equivalence classes, 
which we denote by $ [ P\oplus Q ]
= [ P ] + [ Q ] $
\mbox{ . }
\end{lemma}

\noindent See \cite{Bla} and \cite[Lemma 4.2.3]{Be91} for a proof.

Remark that we can add only mutually orthogonal projections in this way,
because the sum of two projections is not a projection in general. So giving
any pai $P,Q$ of projections it is not always possible to find equivalent
projections $P'\sim P$ and $Q'\sim Q$ such that $P'$ and $Q'$ be orthogonal.
In other words the sum is not everywhere defined in $\Pp (\Aa)$. In order to 
deal with this problem, we replace the
algebra $\Aa$ by its {\em stabilization}, namely the tensor product 
$\Aa \otimes \Kk$ with the algebra of compact operators on a separable Hilbert 
space, which is nothing but the smallest \CS 
containing all finite dimensional 
matrices. Then its possible to show that one can always choose pairs of
projection as orthogonal up to equivalence. By the Grothendieck method one 
builts a group out of the equivalence classes of projections of 
$\Aa \otimes \Kk$. This group is denoted
$K_{0}(\Aa)$ (see \cite{Bla} and \cite[Theorem 10]{Be91}). 

\begin{lemma}
\label{lem-smoothproj}
For any $P\in \Pp (\Aa)$ there is a differentiable
projection $P_{0}\in \Pp (\Aa)$ such that $P\sim P_{0}$.
\end{lemma}

\noindent This last result is a consequence of the fact that the
set of differential elements in $\Aa$ is norm dense. 

\begin{lemma}
\label{lem-smoothhomotopy}
For any pair $P,Q\in \Pp (\Aa)$ and any $\varepsilon$ small
enough, there are differentiable projections
$P_{\varepsilon},Q_{\varepsilon}\in \Pp (\Aa)$ and a
differentiable element $U_{\varepsilon}$ such that $\parallel
P_{\varepsilon} -P\parallel \leq \varepsilon ,\, \parallel
Q_{\varepsilon} -Q\parallel \leq \varepsilon$ and $P_{\varepsilon}
=U_{\varepsilon}U_{\varepsilon}^{\ast}$ whereas $Q_{\varepsilon}
=U_{\varepsilon}^{\ast}U_{\varepsilon}$. 
\end{lemma}

\noindent The proof is straightforward: it follows from the density
of $\Cc^{1}(\Aa )$ and from the proof of Lemma~\ref{lem-homotopy}.
We will say that $P$ and $Q$ are smoothly equivalent whenever the
element $U$ which connects them can be chosen differentiable. 

\begin{lemma}
\label{lem-cherninvariance}
For any pair $P,Q\in \Pp (\Aa)$ of smoothly equivalent projections
$\Ch (P)=\Ch (Q)$. 
\end{lemma}

\noindent The proof of this result is purely combinatorial
provided we use the cyclicity of the trace $\TV$. We will omit it
here \cite{Co85}. 

\begin{lemma}
\label{lem-additivity}
For any pair $P,Q\in \Pp (\Aa)$ of mutually orthogonal smooth
projections we have $\Ch (P\oplus Q)=\Ch (P)+\Ch
(Q)$.
\end{lemma}

\noindent The proof of this result is standard and can be found in
\cite{Co85} for instance. To summarize this set of results we have

\begin{theo}
\label{theo-additivity}
The Chern character $\Ch$ defines a group homomorphism from $K_{0}(\Aa)$ into a
countable subgroup of the real line.
\end{theo}

\noindent It remains to show that the image of this map is the set of
integers. This will be done in the following
Sections~\ref{sec-Connesformula1} and  \ref{sec-Index}.

\vspace{.2cm}

\subsection{Connes formul\ae}
\label{sec-Connesformula1}

In order to compute eventually the Chern character of a projection, we need an 
intermediate tool, namely a cyclic cocycle. This is actually the heart of
Connes work on the non-commutative extension of cohomology. 
Here, we shall actually only need a 2-cocycle $\tau_2$, which is a trilinear
form on the algebra $\AO$ defined by:

\begin{equation}
\label{cocycle}
\TV_2(A_0,A_1,A_2)=
 {2\pi \imath}\; 
  \TV 
   (A_0\partial_1 A_1\partial_2 A_2
    -A_0\partial_2 A_1\partial_1 A_2) 
\mbox{ . }
\end{equation}

By the Schwarz inequality for the trace $\tau$, we see that $\tau_2$ can be extended
to the non-commutative Sobolev space $\Ss$, which is linear subspace of the
von Neumann algebra $\Ww$.
The proof of the following lemma is algebraic and standard by now; it can be
found in \cite{Co85}.

\begin{lemma}
 $\TV_2$ is a  2-cocycle, i.e. it satisfies the following 
algebraic properties: \\
\indent i)  $\TV_2$ is cyclic: $\TV_2(A_0,A_1,A_2)=\TV_2(A_2,A_0,A_1)$ \\
\indent ii) $\TV_2$  is closed under Hochschild's boundary operator, that is:
$b$:
\begin{eqnarray*}
\lefteqn{(b\TV_2)(A_0,A_1,A_2,A_3) } \\
& \equiv & 	\TV_2(A_0A_1,A_2,A_3)-\TV_2(A_0,A_1A_2,A_3)+
			\TV_2(A_0,A_1,A_2A_3)-\TV_2(A_3A_0,A_1,A_2)\;=\;0
\mbox{ . }
\end{eqnarray*}

\end{lemma}

\noindent We will now give a
formula which permits to compute the cocycle $\tau_2$ by
means of the the physical representations. 
For this purpose, we present the formalism introduced by Connes \cite{Co85}
(we already gave some 
indications in 
Section~\ref{sec-HallChern}).  A graded Fredholm module is defined as follows. 
Let $\Hh_+$ and $\Hh_-$ be two separable Hilbert spaces;
their direct sum
$\Hha=\Hh_+ \oplus \Hh_-$ becomes a graded Hilbert space through the graduation
operator $\hat{G}$ equal to 
$\pm 1$ on $\Hh_{\pm}$. A representation 
$\hat{\pi}: \Dd \rightarrow \Ll(\Hha)$ of an algebra $\Dd$ is said to be 
trivially graded, if $[\hat{\pi}(A),\hat{G}]=0$ for all $A\in \Dd$. One says 
that $\hat{\pi}(A)$ is of degree 0; operators on $\Hha$ which anticommute with
the graduation operator $\hat{G}$ are said to be of degree 1.
Any operator on $\Hha$ can be uniquely decomposed into the sum of an 
operator of degree $0$ with an operator of degree $1$.

\begin{defini}
\label{def-fredmod}
 A Fredholm module is a family 
$(\Dd,\Hha,\hat{\pi},F)$ where
$\Dd$ is an algebra with a trivially graded representation $\hat{\pi}$ on 
a graded, separable Hilbert space $\Hha$ and where $F\in \Ll(\Hha)$ is a
selfadjoint operator such  that the following three conditions are satisfied:

$$
\mbox{i) }F \hat{G} = - \hat{G} F \qquad \mbox{ii) }F^2=1 \qquad 
\mbox{iii) } [\hat{\pi}(A),F]\in {\cal K}
\;\;\forall
A\in \Dd
\mbox{ . }
$$ 

\noindent Here $\Kk$ is the ideal of compact operators on $\Hha$.
An element $A\in\Dd$ is called p-summable (resp. p+-summable) whenever
$[\hat{\pi}(A),F]\in\Ll^p(\Hha)$ (resp. $\in \Ll ^{p+}(\Hha)$).
\end{defini}

\noindent The graded commutator of two  graded operators
$T,T'\in\Ll(\Hha)$ is defined by

$$
[T,T']_S=TT'-(-1)^{d^o(T)d^o(T')}T'T
\mbox{ . }
$$

\noindent This commutator extends to the whole algebra $\Ll(\Hha)$
 by bilinearity. The non-commutative differential is given by

$$
dT=[F,T]_S \qquad T\in \Ll(\Hha)
\mbox{ . }
$$

\noindent One can check that it obeys the graded Leibniz rule 
$d(TT')=dT\, T'+(-)^{d^o(T)}T\, dT'$ and that $d^2=0\;$.
Finally, the graded trace or  the supertrace is defined by

$$
\TR_S(T)=\frac{1}{2}\TR(\hat{G} F dT)
\mbox{ , }
$$

\noindent whenever the right-hand side is well defined.
Here, $\TR$ is the usual trace in $\Ll(\Hha)$. Remark that $\TR_S$ is
linear and satisfies: 

$$
\TR_S(TT')=(-1)^{d^o(T')d^o(T)}\TR_S(T'T)\mbox{ . }
$$

\noindent Moreover, if $d^o(T)=1$, then $\TR_S(T)=0$.
 
Now we shall consider the concrete family of Fredholm modules which will be 
of interest to us.
The algebra will be $\AO$. The Hilbert space is
$\Hha=\Hh^+\oplus \Hh^-=L^2(G) \oplus L^2(G)$ (where $G=\RR^2$ or $G=\ZZ^2$)
and for $A\in \AO$ the
representation is given by

$$
\hat{\pi}_{\omega}(A) = 
   	\hat{A}_{\omega}=
		\left(\begin{array}{cc}A_{\omega}&0 \\ 
		0&A_{\omega} \end{array}\right) \mbox{ , }
			\qquad A_{\omega}=\pi_{\omega}(A)
\mbox{ , }
$$

\noindent where $\pi_{\omega}$ is the family of representations defined in
(\ref{eq-repreg}). Next

$$
F= \left(\begin{array}{cc}0&{F^{+-}} \\ {F^{-+}}&0 \end{array} \right)
 = \left(\begin{array}{cc}0&{u} \\ {u^*}&0  \end{array}\right) \mbox{ , }
	 \qquad u=\frac{X_1+\imath X_2}{|X_1+\imath X_2|}
\mbox{ , }
$$

\noindent here $X_1,X_2$ are the two components of the position operator on
$L^2(G)$.

The following result holds for $G=\ZZ^2$ or $\RR^2$. However, we will prove
it only in the discrete case and leave the continuous case for a forthcoming 
work \cite{SCB}.

\begin{theo}[First Connes formula]
\label{theo-DixmierConnes}
The Fredholm module $(\AO,\Hha ,\hat{\pi}_{\omega}, F)$
defined above
is $2+$-summable for $\PP$-almost all $\omega$'s. Moreover for every
$A\in \AO$, the following formula holds:

\begin{equation}
\label{eq-Connes3}
\TV (|\naV A|^{2})=
 \frac{1}{\pi}
  \TD (|dA_{\omega}|^{2})
\mbox{ , }
\hspace{1cm}
 \mbox{for $\PP$-almost all $\omega$}
\mbox{ . }
\end{equation}

\noindent If $\Ss$ denotes the Sobolev space associated to $\TV  $, this formula
can be continued to elements $A\in \Ss$. In particular, if $A\in \Ss$, then 
$dA_{\omega} \in \Ll^{2+}$ for $\PP$-almost all $\omega$.

\end{theo}

\noindent This result is an elementary extension of 
Proposition~\ref{prop-DixmierConnes}. Its proof is left to the reader.

\vspace{.2cm}

The following formula links the cocycle $\tau_2$ defined in 
(\ref{cocycle}) to the previous Fredholm modules 
$(\AO,\Hha ,\hat{\pi}_{\omega}, F)$; it
can already be found in \cite{Co85}. 
This is a result specific to the algebra $\AO$ and depends upon 
the dimension $D=2$ of the physical space (or of the Brillouin zone).

\begin{theo}[Second Connes formula]
\label{prop-Connesformula} 
For $A_0,A_1,A_2 \in \AO$, we have the following formula: 

\begin{equation}
\label{Connesformula}
\int_{\Omega} d\pro 
 \TR_S(
  \hat{A}_{0,\omega}
   d\hat{A}_{1,\omega}
    d\hat{A}_{2,\omega}
     )=
     \TV_2(A_0,A_1,A_2) 
\mbox{ . }
\end{equation}

\end{theo}

\noindent {\bf Proof.} We remark that the left hand side is well defined
thanks to Theorem~\ref{theo-DixmierConnes}. For indeed, if $A\in \AO$ then
$dA_{\omega}\in \Ll^{2+}\subset \Ll^3$ $\PP$-almost surely.
The trace of an integral operator on $L^2(G)$
with continuous and compactly supported integral kernel is given by the
integral of its diagonal. We can therefore evaluate the left hand side:

\begin{eqnarray}
\label{eq-intermediate}
\lefteqn{ 
 \int_{\Omega}  d\pro     
  \TR_S(
   \hat{A}_{0,\omega}
   d\hat{A}_{1,\omega}
    d\hat{A}_{2,\omega}
     )} \nonumber \\   
& = & \int_{\Omega}
	d\pro 
	 \int_{G^3} d^2x_0 \; d^2x_1\; d^2x_2 \;
      \left[ 
       -(1-
        \frac{\overline{x_0}}{|x_0|}
	 \frac{{x_1}}{|x_1|})
	  (1-
	   \frac{\overline{x_1}}{|x_1|}
	    \frac{{x_2}}{|x_2|})
	     (1-\frac{\overline{x_2}}{|x_2|}\frac{{x_0}}{|x_0|}) 
      \right] \nonumber
   \\  
& &
\\
& & 
e^{\imath\lambda
 (x_0\wedge x_1+x_1\wedge x_2+x_2\wedge x_0)} 
  A_0(T^{-x_0}\omega,x_1-x_0)
   A_1(T^{-x_1}\omega,x_2-x_1)
    A_2(T^{-x_2}\omega,x_0-x_2)
\mbox{ , } \nonumber
\end{eqnarray}

\noindent where $\lambda=q\Bb /2\hbar$. The main ingredient of the proof
is now the following lemma for which
there are two different proofs in \cite{Co85} and \cite{ASS}. We shall follow
\cite{ASS}, but present a discrete version of the proof; 
for the continuous case $G=\RR^2$ we refer to \cite{ASS}.

\begin{lemma}
\label{lem-triangle}
Let $a,b \in G$ which we write as $a=a_1+\imath a_2$, $b=b_1+\imath
b_2$. Then we have:

$$
-2\pi \imath \;a\wedge b=\int_{s\in G}\;
(1-\frac{\overline{s}}{|s|}\frac{s-a}{|s-a|})
(1-\frac{\overline{s-a}}{|s-a|}\frac{s-b}{|s-b|})
(1-\frac{\overline{s-b}}{|s-b|}\frac{s}{|s|})
\mbox{ . }
$$

\end{lemma}

\noindent {\bf Proof.} In the discrete case $G=\ZZ^2$, the integral
appearing in the lemma is in fact a sum, let us denote it by $C(a,b)$. Then
$C(a,b)=-C(b,a)$ and $\ov{C(a,b)}=-C(a,b)$. Hence $C(a,b)$ is purely imaginary.
Now we define:

$$
e(s,t)= 
	(\frac{s}{|s|}\frac{{\ov t}}{|t|}-\frac{t}{|t|}\frac{{\ov s}}{|s|})
		=-e(t,s)
			=-\ov{ e(s,t)}
\mbox{ . }
$$

\noindent A direct calculation leads to

$$
C(a,b) = -\sum_{s\in \ZZ^2} (e(s-a,s-b) + e(s-b,s) +e(s,s-a))\mbox{ . }
$$

\noindent We introduce $C_N(a,b)$ as the same sum in which $s$ is restricted to
be smaller or equal to
$N\in\NN$. The finite difference operators are defined with help of the
translation operators $T_1,T_2$ on functions on $\ZZ^2$ as:

$$
\Delta_{j}=T_{j}-1 \qquad
\tilde{\Delta}_{j}=1-T_{j}^{-1}
\hspace{0.8cm} j=1,2\mbox{ . }
$$

\noindent We consider the finite differences of of $C_N(a,b)$:

\begin{eqnarray*}
\lefteqn{(\Delta_{a_1}\Delta_{b_2}-\Delta_{b_1}
	\Delta_{a_2})\;C_N(a,b)  }  \\
\vspace{0.2cm}
& = & -\sum_{|s|\leq N}
	\tilde{\Delta}_{s_1}\frac{s-a}{|s-a|}
		\tilde{\Delta}_{s_2}\frac{\ov{ s-b}}{|s-b|} -
			\tilde{\Delta}_{s_1}\frac{{ s-b}}{|s-b|}
				\tilde{\Delta}_{s_2}\frac{\ov {s-a}}{|s-a|}
					\;+\;(1\leftrightarrow 2)
\mbox{ . }
\end{eqnarray*}

\noindent A discrete analogue of Stokes' theorem allows us to transform the sum
over the square into a sum over the border of the square. As $N \to \infty$, 
this sum converges to the Riemann integral 
$\int_0^{2\pi} e^{\imath\phi}d e^{-\imath\phi}$. 
The term $(1\leftrightarrow 2)$ gives the same contribution and we obtain:

$$
(\Delta_{a_1}\Delta_{b_2}-\Delta_{b_1}\Delta_{a_2})\;C(a,b)
\;=\;-4\pi \imath\mbox{ . }
$$

\noindent As this is true for every $a,b \in \ZZ^2$, $C(a,b)$ is of the form

$$
C(a,b)=\alpha + \beta (a,b) -2\pi \imath \;a\wedge b\mbox{ , }
$$

\noindent where $\alpha$ is a constant and $\beta$ is linear in the 
vector $(a,b)$. As $C(0,0)=0$, we have $\alpha=0$.
Because $C(a,b)$ and $a\wedge b$ are both odd under permutation of
$a$ and $b$, $\beta$ must be odd as well so that it vanishes.
\hfill $\Box$

\vspace{0.2cm}

\noindent {\bf Proof of Theorem~\ref{prop-Connesformula} (end).} 
We set $s=x_0,\; s_0=x_0-x_1, \; s_1=x_1-x_2$ in equation 
(\ref{eq-intermediate}) and use the
invariance of the measure ${\bf P}$ 
in order to replace $T^{-s}\omega$ by $\omega$. Applying
Lemma~\ref{lem-triangle} we get

\begin{eqnarray*}
\lefteqn{ \int_{\Omega}
		d\pro     \TR_S(\hat{A}_{0_{\omega}}d\hat{A}_{1_{\omega}}d
				\hat{A}_{2_{\omega}})
	 }
\\
& = &	   -\int_{\Omega}
		d\pro     \int_{
}
		d^2s_0d^2s_1	e^{\imath\lambda s_0\wedge s_1} 
			2\pi \imath \;s_0\wedge s_1
		 A_0(\omega,-s_0)A_1(T^{s_0}\omega,-s_1)
			A_2(T^{s_0+s_1}\omega,s_0+s_1)
\mbox{ . }
\end{eqnarray*}

\noindent The right hand side is precisely the formula for
$\TV_2(A_0,A_1,A_2)$. 
\hfill $\Box$


\vspace{.2cm}

\subsection{Chern character and Fredholm index}
\label{sec-Index}

The main 
interest of Connes' theory of non-commutative Fredholm modules \cite{Co85} is
the following in our context: for a given 3-summable 
projection $P$ in an algebra $\Aa$, the expression 
$ \sigma(P,P,P)=\frac{1}{2}\TR_S(\hat{P}d\hat{P}d\hat{P})$ can be
related to the index of a Fredholm operator. In order to make this
article self-contained, we will reproduce here the main steps 
relevant for us.

First we need the following formula due to Fedosov \cite{Fed}. A proof can be
found in the appendix of \cite{Co85}.

\begin{proposi}[Fedosov's formula]
Let $F$ be a bounded operators on a Hilbert space $\Hh$.
We suppose that $(1-F^*F)\in{\cal L}^p(\Hh)$ 
and $(1-FF^*)\in{\cal L}^p(\Hh)$ for some $p\in[1,\infty)$.
Then $F$ is a Fredholm operator and for every integer $n\geq p$ 
its index satisfies 

$$
\mbox{\rm Ind}(F)= 
 \TR ((1-F^*F)^n) - 
  \TR((1-FF^*)^n)
\mbox{ . }
$$
\end{proposi}

\begin{proposi}
\label{pro-ConInd}
Let $(\Dd,\Hha,\hat{\pi},F)$ be a Fredholm module and $P\in\Dd$ be
a 3-summable projection. Then $F^{+-}_P=PF^{+-}|_{P\Hh^-}$ is a 
Fredholm operator and 

$$
\mbox{\rm Ind}(F^{+-}_P)=
 \TR_S(\hat{P}d\hat{P}d\hat{P}) 
\mbox{ , }\qquad 
\hat{P}=
 \left(
  \begin{array}{cc}
      {\pi (P)} &     0     \\ 
          0     & {\pi (P)} \\
   \end{array}
 \right)
\mbox{ . }
$$

\end{proposi} 

\noindent {\bf Proof.} Suppose $\Hh^+=\Hh^-$ for simplicity. Then 

\begin{eqnarray*}
-\hat{P}[F,\hat{P}]^2\hat{P} 
 &=& \hat{P} - \hat{P}F\hat{P}F\hat{P} \\
 &=& \left(
       \begin{array}{cc}
         {(1-F^{+-}_PF^{-+}_P)|_{P\Hh^+}} &        0  \\
		             0            & {(1-F^{-+}_PF^{+-}_P)|_{P\Hh^-}} \\
       \end{array}
     \right) 
\mbox{ . }
\end{eqnarray*}

\noindent By hypothesis $[F,\hat{P}] \in {\cal L}^3(\Hha)$. 
H\"older's inequalities imply $(1-F^{+-}_PF^{-+}_P)|_{P\Hh^+}
\in {\cal L}^2(\Hh^+)$ and $(1-F^{-+}_PF^{+-}_P)|_{P\Hh^-}
\in {\cal L}^2(\Hh^-)$. By Fedosov's formula we get

\begin{eqnarray*}
\mbox{\rm Ind}(F^{+-}_P) 
& = & 	
	\TR_{P\Hh^-}((1-F^{-+}_PF^{+-}_P)^2)
		-\TR_{P\Hh^+}((1-F^{+-}_PF^{-+}_P)^2)
\\
& = & 	
	-\TR_{\Hha}(\hat{G}(\hat{P} - \hat{P}F\hat{P}F\hat{P})^2) 
\mbox{ . }
\end{eqnarray*}

\noindent We can check that this is equal to 
$\TR_S(\hat{P}d\hat{P}d\hat{P})$ by using the following
algebraic identities

$$
\begin{array}{cclcrcl}
 [F,\hat{P}] & = &\hat{P}[F,\hat{P}]+[F,\hat{P}]\hat{P}\mbox{ , } & &
\hat{P}[F,\hat{P}]^2 & = & [F,\hat{P}]^2\hat{P}\mbox{ , } \\
 &  &  &  &  &  & \\
F[F,\hat{P}]^{2n+1} & = &  -[F,\hat{P}]^{2n+1}F\mbox{ , }  & &
F\hat{G} & = &  -\hat{G}F \mbox{ . } 
\end{array}
$$

\noindent The proof can easily be completed. \hfill $\Box$

\vspace{.2cm}

We shall now extend this result to stochastic operators.

\begin{theo}[\cite{Be85}]
\label{theo-indexloc} 
Let $P\in \Ww$ be a projection belonging to the non-commutative Sobolev 
space $\Ss$. Then for ${\bf P}$-almost every $\omega\in\Omega$, $P$ is
$2+$-summable and 

$$
\CH(P)=\mbox{\rm Ind}(P_{\omega}u|_{P_{\omega}(\Hh^-)}) 
\mbox{ , } 
$$

\noindent where $P_{\omega}=\pi_{\omega}(P)$ and $u=\frac{X}{|X|}$. In
particular, $\CH(P)$ is an integer.
\end{theo}

\noindent {\bf Proof.} Using Theorem~\ref{theo-DixmierConnes}, 
Theorem~\ref{prop-Connesformula} and Proposition~\ref{pro-ConInd},
it just remains to show that the index is 
$\PP$-almost surely independent of $\omega$. Using the ergodicity of $\PP$,
it is enough to show that the index is translation invariant. For indeed,
translating $P_{\omega}$ by $a\in G$ just changes $\omega$ into $T^{-a}
\omega$. On the other hand, translating $u$ by $a$ changes it into
$u+O(1/|X|)$. Thus $P_{\omega}u|_{P_{\omega}(\Hh^-)}$ is changed into
$P_{T^{-a}\omega}u|_{P_{T^{-a}\omega}(\Hh^-)}$ modulo a compact operator.
Since a compact perturbation does not change the index of a Fredholm operator, 
the result is achieved. \hfill $\Box$

\vspace{0.2cm}

\noindent {\em Remark.} The above theorem is true independently of the choice of the 
probability measure $\PP$. However, changing $\PP$ is equivalent to changing the
disorder. Therefore we cannot expect the Sobolev condition to hold independently
of $\PP$. In particular, if $H$ is a given bounded selfadjoint operator, 
the spectrum of its representative $\pi_{\omega}(H)$ is $\PP$-almost surely 
constant \cite{KuSo}, but changing $\PP$ may change it. Therefore, $\PP$ has a
physical content. We will see in the Section~\ref{sec-HarperLoc} some of the 
consequences of changing the disorder. \hfill $\Box$

\vspace{.2cm}

\subsection{Quantization, Fredholm and relative index}
\label{sec-QuaRelIndex}

In this section, we will discuss the links between the Laughlin argument as
presented in Section~\ref{sec-IQHcond} 
and our approach. The essential
ingredient for that will be the relative index of two projections as defined by
Avron, Seiler and Simon \cite{ASS}. It turns out that the singular gauge 
transformation of Laughlin corresponds to the unitary operator $u=X/|X|$. 
Thus the charge transported after changing the flux by one quantum is then
exactly given by the index we computed in Theorem~\ref{theo-indexloc}. The main
improvement upon Laughlin's argument is that we control completely the effect of
the disorder now, since this index is a topological invariant. Let us remark 
that the topology we are talking about is the one of the Brillouin zone and
not of the sample as can be erroneously derived from a superficial understanding
of Laughlin's argument. 

More precisely, following Section~\ref{sec-IQHcond}, let the varying
flux be $\phi(t)=\hbar t/e\tau$ with $\tau$ so large as to
produce an adiabatic change. Then at time $t=\tau$ 
the new Hamiltonian in (\ref{polarcoordinates}) is given by 

\begin{equation}
\label{eq-unitaryequi}
H_\Bb (\tau ) =
	uH_\Bb (0)u^* 
\mbox{ , }
\hspace{1cm}
	u=\frac{X}{|X|}
\mbox{ , }
\end{equation}

\noindent namely the phase of the wave function changes by 
$e^{\imath \theta}$ where $\theta$ is the polar angle of the position $x$. 
Formally, one gets an equivalent formula at each intermediate times, 
but the domains change with time. The eq.~(\ref{eq-unitaryequi}) implies
for the corresponding Fermi projections:

\begin{equation}
\label{eq-unifermi}
P_F(\tau)=uP_F(0)u^*
\mbox{ , }
\hspace{1cm}
  P_F(t)=\chi_{\leq E_F}(H_\Bb (t))
\mbox{ . }
\end{equation}

\noindent The charge transported to infinity after this adiabatic change
is the number of states in $P_F(\tau)$ that are not in $P_F(0)$. Since 
both projections are infinite dimensional, we must be careful in computing 
this number. It is the purpose of the following definition \cite{ASS} to take
care of this difficulty. Namely, given two projections $P$ and $Q$ on 
a Hilbert space we set

$$
\mbox{\rm Index} (P,Q) = 
	\mbox{\rm dim}(\mbox{\rm Ker}(P-Q-1)) -
		\mbox{\rm dim}(\mbox{\rm Ker}(Q-P-1))
\mbox{ , }
$$

\noindent whenever the right side is well defined. Then (see \cite{ASS}
for a proof)

\begin{proposi}
Let $P$ and $Q$ be two projections on Hilbert space. 
If $(P-Q) \in \Ll^p$, then for every integer $m$ such that $2m+1\geq p$, 
the relative index can be calculated as

$$
\mbox{\rm Index} (P,Q) = \TR((P-Q)^{2m+1})
\mbox{ . }
$$
\end{proposi}

\noindent We then apply this formalism to the $2+$-summable
Fermi projections $P_{F,\omega}$ and $uP_{F,\omega} u^*$. It is an immediate 
consequence of the proof of Proposition~\ref{pro-ConInd} that:

\begin{equation}
\label{eq-indexrel}
\mbox{\rm Index}(P_{F,\omega},uP_{F,\omega} u^*)=
 -\TR_S(\hat{P}_{F,\omega} d\hat{P}_{F,\omega} d\hat{P}_{F,\omega}) 
\mbox{ , }
\end{equation}

\noindent which is $\PP$-almost surely equal to $-\CH (P_F)$. The suitable $p=2m+1$
is $p=3$ here. We remark that it is the smallest possible one giving a non-zero
relative index because, if $P_F-uP_Fu^*$ were trace class, the relative index 
would vanish. Hence the Chern character of the Fermi projection can be identified 
up to sign, with the charged transported at infinity during the adiabatic
flux change. Let us then call this index the {\em charge deficiency index}.

Owing to the stability of the Fredholm index, if one pierces the flux tube at
some other place than the origin and then uses another unitary operator than
$u=e^{\imath\theta}$, the Fredholm index will not change. Moreover, adding some
disorder potential to the Landau Hamiltonian will not change the index as well
as long as the Fermi projection belongs to the Sobolev space (we will discuss
this condition more precisely in the next section).
 
\vspace{.5cm}

\section{Localization and non-commutative Sobolev space}
\label{chap-loc}

In this chapter, we relate the Sobolev condition on Fermi projections to the 
Anderson localization. We will give several mathematical tools to describe
rigorously localization in terms of our formalism. 
The main results are
Theorems~\ref{theo-locpoint} and \ref{theo-continuity} below. As a consequence,
the remaining results in the main Theorem~\ref{theo-main} follow. 

In Section~\ref{sec-loccrit} we review the
well-known Anderson-Pastur criterion for localization. Then we give a new 
definition of the localization
length and discuss its relation to other notions of localization. Notice,
however, that a good part of Section~\ref{sec-Sobolev} has already been published
in \cite{Be92}. Our non-commutative localization length allows us to formulate a
mathematically precise sufficient condition for the existence of the plateaux.
All  the  tools we introduce fit well within the non-commutative 
framework developed so far. For technical simplicity,
however, we will restrict ourselves to the lattice case namely for $G=\ZZ^D$ (tight
binding representation) in this chapter. Even though we believe that most of
these results hold for the continuum case as well, the proofs are more
difficult and will be postponed to a future work \cite{SCB}.

Our first definition of the localization length starts by demanding that
the mean square distance that a particle moves from a given point in an
infinite amount of time be finite. We will give a precise definition of
the word ``mean'' we use here. In Theorem~\ref{theo-locpoint}, we show
that provided this condition holds, the localization length can be defined as a
$L^{2}(\RR ,d\Nn )$-function of the energy, where $\Nn$ is the DOS of the
Hamiltonian under study. Furthermore we give another definition of the
localization length, based upon the Sobolev norm of the eigenprojections,
and show in Proposition~\ref{pro-loclength2} that it is equivalent to the
first one.
Then Theorem~\ref{theo-locpoint} shows that if
this localization length is finite in some energy interval, the spectrum of
the Hamiltonian in this interval is pure point for almost all disorder
configurations.
Theorem~\ref{theo-continuity} shows that under the same condition,
the spectral projection $P_{E}$ on energies
lower than or equal to $E$ is a continuous function of $E$ with respect to
the the Sobolev norm. This technical result, together with the results of the
previous chapter, implies that the Chern number of $P_{E}$ is
constant on that interval. This is the reason why we get the plateaux of
the Hall conductance. 

Let us remark that our localization length can be computed as
a disorder average of a product between two Green functions, or
equivalently by means of a current-current correlation. In particular, let
$m$ be the positive measure on $\RR^{2}$ defined by \cite{Pa93}

$$ \int_{\RR^{2}}
    dm(E,E') \;f(E)g(E')=
     \TV (\naV H f(H) \naV H g(H))
\mbox{ , }
\hspace{1cm}
f,g\; \in \Cc_{0}(\RR )
\mbox{ . }
$$

\noindent Using spectral theory in $L^{2}(\Aa ,\TV )$, one can indeed
show that such a measure exists and that it can be calculated 
using Green's functions. Then the localization condition that we define below
is given by 

$$
l^2(\Delta)=
  2\int_{\Delta\times\RR}
   \frac{dm(E,E')}{(E-E')^{2}}
\mbox{ , }
$$

\noindent whenever the integral 
exists. We will not develop this point of view here,
but we emphasize that our approach is equivalent to the one used in solid
state physics. 

\vspace{.2cm}

\subsection{The Anderson-Pastur localization criterion}
\label{sec-loccrit}

Most results in this section are due to Pastur \cite{Pa,FiPa}.
The underlying physical idea can be traced back to
Anderson \cite{An}. Since this theory holds in
any dimension, we will assume that the lattice in space is $\ZZ^D$.
Let then $\Omega$ be a compact space endowed with an action of $\ZZ^D$ 
by homeomorphisms. The magnetic field $\Bb $ in $D$-dimension is an 
antisymmetric bilinear form on $\ZZ^D$ written as $\Bb a\wedge b$.
We define the \CS $\Aa=C^{\ast}(\Omega \times \ZZ^D,\Bb )$ as in 
Section~\ref{sec-ObsCal}. In this discrete case, this algebra 
has a unit. Given a $\ZZ^D$-invariant ergodic 
probability measure $\PP$ on $\Omega$, we get a trace $\TV $
on $\Aa$, which is actually normalized. We will denote by $\Ww$ the
von Neumann algebra $L^{\infty}(\Aa, \TV )$ of the corresponding GNS 
representation. Recall that the representations $\pi_{\omega}$
extend to $\Ww$ and give random operators.
Then, to avoid inessential difficulties, the Hamiltonian $H$ will
be a selfadjoint element of $\Aa$ in this section.
We will denote by $\sigma(\omega)$ the spectrum of the operator 
$\pi_{\omega}(H)=H_{\omega}$. 
As a preliminary, let us recall the  

\begin{lemma}[Wiener criterion]
\label{the-Wiener} 
Let $\mu$ be a finite complex measure on the
real line, i.e. $\mu$ is a linear combination of four finite positive measures
$\mu=\mu_1-\mu_{2}+\imath\mu_3-\imath\mu_4$.
Let $F_{\mu}(t)=\int
e^{itx}d\mu(t)$ be its Fourier
transform, then:

$$
\lim_{T\rightarrow\infty}\int^T_0\frac{dt}{T}|F_{\mu}(t)|^2
=\sum_{E\in\RR}|\mu(E)|^2
\mbox{ . }
$$

\end{lemma}

\noindent A proof for a positive measure can be found in \cite{RS}; it can be completed to
a proof of Lemma~\ref{the-Wiener} without any difficulties.

The next result is given by

\begin{proposi}
\label{prop-dimproj}
Let $P$ be a projection in $\Ww$, 
the von Neumann algebra $L^{\infty}(\Aa, \TV )$ generated
by $\Aa$ in the GNS representation of the trace per unit volume $\TV $
associated to the probability measure $\PP$. Then $\PP$-almost surely,
$\pi_{\omega}(P)$ is a projection on $\ell^2(\ZZ^D)$. Moreover its
dimension 
$\mbox{\rm dim}(\pi_{\omega}(P))=\TR(\pi_{\omega}(P))$ is either zero or infinity
$\PP$-almost surely.
\end{proposi}

\noindent {\bf Proof.} Because of the covariance of $P$, $\mbox{\rm dim} 
(\pi_{\omega}(P))$ is translation invariant.
Let us define $\Pi=\int d \pro \pi_{\omega}(P)$ as a weak integral acting on 
$\ell^2(\ZZ^D)$. By
construction, $\Pi$ commutes to the translation group. Since the measure
$\PP$ is invariant and ergodic, using the monotone convergence theorem,
we get for $\PP$-almost all $\omega \in \Omega$:

$$
\mbox{\rm dim}(\pi_{\omega}(P))=
 \int d\pro \TR(\pi_{\omega}(P))=
  \TR(\Pi)=
   \sum_{x\in\ZZ^D} <x|\Pi|x>=
    \sum_{x\in\ZZ^D} <0|\Pi|0>
\mbox{ , }
$$

\noindent and this last expression is either zero or infinity. \hfill $\Box$

\vspace{.2cm}

Let now $\Delta \subset \RR$ be a Borel subset 
of the spectral axis. Let then $P(\Delta)$ be the corresponding spectral
projection of $H$. Let $A_n(\Delta,\omega)$ be the time-averaged probability
for a particle initially at site $|n>, n \in \ZZ^D$, to stay at this same site
$|n>$ the time evolution being governed only by the restriction of the
Hamiltonian $H_{\omega}$ to the interval $\Delta$. More precisely 

$$
A_n(\Delta,\omega)=\lim_{T\rightarrow\infty}\int_0^T\frac{dt}{T}
\;|<n|\pi_{\omega}(e^{\imath Ht}P_{\Delta})|n>|^2
\mbox{ . }
$$

\noindent The covariance of the Hamiltonian leads to
$A_n(\Delta,\omega)=A_0(\Delta,T^{-n}\omega)$. After averaging over the
disorder configurations, it will therefore be sufficient to consider only 
$A(\Delta,\omega)=A_0(\Delta,\omega)$. We introduce the (disorder
average) return probability:

$$
\xi(\Delta)=
 \int d\pro \; A(\Delta,\omega)
\mbox{ . }
$$

\noindent $\xi(\Delta)$ can also be expressed as follows

$$
\xi(\Delta)=
 \lim_{T\rightarrow\infty}
  \int_0^T
   \frac{dt}{T} 
    \int
     \frac{d^D\theta}{(2\pi)^D}
      \TV(
       (e^{-\imath Ht}P_{\Delta})
        e^{\imath \vec{\theta}\cdot\naV}
         (e^{\imath Ht}P_{\Delta})
          )
          \mbox{ , }
$$

\noindent as can be checked by a direct calculation. Here 
$\naV=(\partial_1,\ldots,\partial_D)$ is the derivation on $\Aa$.
If $\xi(\Delta)>0$, we expect that there are some localized states
corresponding to energies within the interval $\Delta$, this is 
contained in Theorem~\ref{theo-pastur} below. Using Wiener's
criterion we obtain:

\begin{eqnarray}
\label{eq-pp}
\xi(\Delta) 
& = & 
	\int d \pro     \; \lim_{T\rightarrow\infty}\int_0^T\frac{dt}{T} |\int
		<0|\pi_{\omega}(dP_{E} P_{\Delta})|0> e^{\imath Et}|^2 
\nonumber
\\
& = & 
	\int d \pro \; \sum_{E\in\sigma_{pp}(\omega)\cap\Delta}
		|\psi_{\omega ,E}(0)|^4 
\mbox{ , }
\end{eqnarray}

\noindent here $P_{E}\in \Ww$ is the spectral projection of $H$ on
the interval $(-\infty ,E]$, $\psi_{\omega ,E}$ is the eigenstate 
of $H_{\omega}$ corresponding to the eigenvalue $E$ and $\sigma_{pp}(\omega)$ 
is the set of eigenvalues of $H_{\omega}$. Hence we see that $\xi(\Delta)$ 
is related to the so-called ``inverse participation ratio''.
The main result is summarized in the following 

\begin{theo}
\label{theo-pastur}
Let $\Delta \subset \RR$ be an open interval. Then the following results hold:
 
\bed
\item[ i)] Let $\Nn$ be the DOS of $H$. If $\lambda $ is a growth point of $\Nn$
then $\lambda \in \sigma (\omega)$ $\PP$-almost surely. If $\lambda $ is not 
a growth point of $\Nn$ then $\lambda$ is outside $\sigma (\omega)$ 
$\PP$-almost surely.

\item[ ii)] The number of eigenvalues of $H_{\omega}$ counted with their multiplicity
and contained in $\Delta$ is either zero or infinity $\PP$-almost surely.

\item[iii)] A given real number $\lambda$ is $\PP$-almost never an eigenvalue
of $H_{\omega }$ with finite multiplicity.

\item[iv)] The number of eigenvalues in $\Delta$ is infinite if and only if
$\xi(\Delta)>0$. 
\ed
\end{theo} 

\noindent {\bf Proof.} i) Suppose $\lambda $ is a growth point of $\Nn$.
This implies that
$<0|\pi_{\omega}(P_{\Delta})|0>>0$ on a set of positive measure 
for any neighborhood $\Delta $ of $\lambda$. Thus, for any such
neighborhood, the set of 
$\omega$'s such that $\sigma(\omega) \cap \Delta \neq \emptyset$
has probability one, because it is measurable and translation invariant.
Since 

$$
\{\omega\in\Omega|\lambda \in\sigma(\omega)\}
 =\bigcap_{j>0}
  \{
   \omega\in\Omega|
    \sigma(\omega)
     \cap
      (\lambda-\frac{1}{j},\lambda+\frac{1}{j})
       \neq\emptyset
       \}
\mbox{ , }
$$

\noindent then $\lambda \in\sigma(\omega)$ $\PP$-almost surely.

Conversely let us suppose that $\lambda $ is not a growth point of $\Nn$. Then there
is a small open interval $\Delta$ containing it such that
$<0|\pi_{\omega}(P_{\Delta})|0>=0$ on a set $\Omega_0$ of probability one.
We set $\Omega_{\infty}=\bigcap_{n\in \ZZ^D}T^{-n}\Omega_0$ to get a translation
invariant subset 
of probability one. On this subset $<n|\pi_{\omega}(P_{\Delta})|n>=0$ for all
$n\in \ZZ^D$ by covariance. Therefore on this subset, $\sigma(\omega)\cap \Delta
= \emptyset$. This implies that $\lambda $ is outside $\sigma(\omega)$ with
probability one.

ii) Let $P_{pp}(\omega)$ be the projection onto the subspace spanned by the 
eigenvectors of $H_{\omega}$. It defines a covariant, measurable family of 
bounded operators \cite{KuSo}. So by Theorem~\ref{the-Wstar} it defines an
element $P_{pp}$ of $\Ww$. Thus the number of eigenvalues counted with their
multiplicity and contained in the Borel set $\Delta$ is the dimension of 
$\pi_{\omega}(P_{pp}P_{\Delta})$, namely it is zero or infinity 
$\PP$-almost surely by Proposition~\ref{prop-dimproj}.

iii) $\lambda$ is an eigenvalue of $H_{\omega}$ of finite multiplicity
if and only if $0< \TR (\pi_{\omega}(P_{\lambda})) <\infty$. By 
Proposition~\ref{prop-dimproj} this happens with probability zero. 

iv) Let $\Omega_{\Delta}$ be the set $\{\omega \in
\Omega\;|\;\sigma_{pp}(\omega)\cap\Delta\neq\emptyset\}$. Thanks to 
eq.~(\ref{eq-pp}) this set has positive probability if and only if
$\xi(\Delta)>0$. Since this set is translation invariant and measurable
it has probability one if and only if $\xi(\Delta)>0$. By 
Proposition~\ref{prop-dimproj} again, the result is achieved. \hfill $\Box$

\bed
\item[{\em Remark 1:}] ii) says that, with probability one, no eigenvalue 
of $H_{\omega}$ with finite
multiplicity is isolated, whereas iii) shows that such eigenvalues are
fluctuating with the disorder. There are however examples of models having
non-fluctuating eigenvalues of infinite multiplicity.
\item[{\em Remark 2:}] 
The criterion iv) does not eliminate the occurrence of some continuous spectrum.
\ed 

\vspace{.2cm}

\subsection{Non-commutative localization criterion and localization length}
\label{sec-Sobolev}

We introduce a second physical idea of localization: the average mean
square displacement $\delta X(T)$. Let $X$ be the
position operator in $G$ and $X_{\omega}(t)=e^{\imath H_{\omega}t}X
e^{-\imath H_{\omega}t}$ its time evolution. Then we consider:

$$
\delta X_{\omega,n}(T)^2 =\int^T_0 \frac{dt}{T}<n|(X_{\omega}(t)-X)^2|n>
\mbox{ . }
$$

\noindent The covariance relation leads to $\delta X_{\omega,n}(T) =
\delta X_{T^{-n}\omega,0}(T)$, so that it is  sufficient to examine the
behavior of the state $|0>$. Averaging over the disorder and using again our algebra
we get 

\begin{equation}
\delta X (T)^2 =
 \int^T_0\frac{dt}{T} \TV(|\naV(e^{-\frac{\imath}{\hbar}Ht})|^2)
\mbox{ . }
\end{equation}

\noindent Boundedness of $\delta X(T)$ in
time $T$ will be an indicator for localization and behavior proportional to
$T^{\sigma}, \sigma\in (0,1]$ will be interpreted as diffusive or ballistic
quantum motion. In order to localize in energy we restrict the motion to 
energies in a
Borel subset $\Delta$ of the real line and get $\delta X_{\Delta}(T)$ 
in much the same way. On the other hand taking 
$T\rightarrow \infty$ leads us to define the $\Delta $-localization length as

\begin{equation}
\label{eq-strongloc}
l^2(\Delta)=
 \limsup_{T\rightarrow \infty }
  \int^T_0
   \frac{dt}{T} 
    \TV(
     |\naV(e^{-\frac{\imath}{\hbar}Ht}P_{\Delta})|^2
       )
\mbox{ . }
\end{equation}

\noindent Boundedness of $l^2(\Delta)$ will be our localization condition for the
spectral subset $\Delta$.

\begin{theo}
\label{theo-locpoint}
Suppose that $l^2(\Delta) < \infty$.
Then $H_{\omega}$ has pure-point spectrum in $\Delta$ for almost every
$\omega\in\Omega$.
Moreover, if ${\cal N}(E)=\TV(P_E)$ is the density of states, there is a
${\cal N}$-measurable function $l$ on $\Delta$ such that for every Borel
subset $\Delta'$ of $\Delta$:

\begin{equation}
\label{eq-loclen}
l^2 (\Delta')=\int_{\Delta'}d{\cal N}(E)\; l(E)^2
\mbox{ . }
\end{equation}

\noindent Finally, if $l^2(\Delta) < \infty$, $l^2(\Delta')$ is given  by

\begin{equation}
\label{eq-sobloc}
l^2(\Delta') = 
 \int d\pro \sum_{n\in\ZZ^2}
  |n|^2
   \sum_{E\in\sigma_{pp}(\omega)\cap \Delta'}
    |<0|\pi_{\omega}(P_{\{E\}})|n>|^2
\mbox{ . }
\end{equation}

\end{theo} 

\begin{theo}
\label{theo-continuity}
 If the localization condition $l^2(\Delta)<\infty$
is satisfied, then at every regularity point of the density of states,
the application $E\in\Delta \rightarrow P_E\in \Ss$ 
is continuous.
\end{theo}

\bed
\item[{\em Remark 1:}] Letting $\Delta$ shrink to one point $E$, the function
$l^2(E)$ represents a kind of average of the quantity
$\sum_{n\in\ZZ^D}|n|^2|\psi_{E,\omega}(n)|^2$, where $\psi_{E,\omega}$ is an
eigenvector of $H_{\omega}$ corresponding to energy $E$; here the average is
taken over the disorder and a small spectral set around $E$. 
This quantity measures
the extension of this eigenstate. For this reason the function $l$ will be
called localization length for $H$. 
Note that no exponential decay of the wave functions is needed for our
localization length to be finite. However, such behavior may be studied within
the present framework; we postpone the details to future work.

\item[{\em Remark 2:}]
The index theorem which we proved in 
Section~\ref{sec-Index} only requires that $P_E$ be in $\Ss$ in order to insure the
integrality of the corresponding Chern character (compare to results of 
Section~\ref{sec-Dixmier}).
Thus boundedness of $l^2(\Delta)$ is sufficient but not necessary to
prove the index theorem. 

\item[{\em Remark 3:}]
The localization condition $l^2(\Delta)<\infty$ implies
pure-point spectrum in $\Delta$ almost surely.
In mathematical physics, pure-point spectrum of a Hamiltonian $H$
in a certain region of the density of states
has been considered as a the criterion for localization. The eigenstates being
square integrable are thus localized. The classical RAGE-theorem  \cite{RS}
permits to make such a statement more accurate.
\ed

\noindent {\bf Proof of Theorem~\ref{theo-locpoint}}.
i) The basic argument we will use here is due to Guarneri 
\cite{Gu2}. The spectral projection on the continuous part of the spectrum is 
then given by
$\pi_{\omega}(P_{c})=1-\pi_{\omega}(P_{pp})$. Using the definition of the trace,
then applying the theorems of Fubini and monotone convergence, we find (with the
notation $|n|_{\infty}=\max_{1\leq j \leq D}|n_j|$):

$$
\delta X_{\Delta} (T)^2 
 = 	\int d \pro     \lim_{N \rightarrow
		\infty}\sum_{|n|<N}|n|^2 p_T(\omega,n) 
\mbox{ , }
$$

\noindent where

$$
p_T(\omega,n)=\int^T_0\frac{dt}{T}|<0|\pi_{\omega}(e^{\imath Ht}P_{\Delta})|n>|^2 
\mbox{ . }
$$

\noindent Then $p_T(\omega,n)$ satisfies

$$
0\leq p_T(\omega,n)\leq 1 
\mbox{ , } 
\hspace{0.7cm}
 \sum_{n\in\ZZ^D}
  p_T(\omega,n)\;=\;
   <0|\pi_{\omega}(P_{\Delta})|0> \;\;\leq \;1
\mbox{ . }
$$

\noindent Now, we use the spectral theorem for $H_{\omega}$ and the Wiener
criterion

\begin{eqnarray*}
\lim_{T \rightarrow
\infty} p_T(\omega,n) 
& = & 
	\lim_{T \rightarrow\infty} \int^T_0\frac{dt}{T}
		|\int <0|\pi_{\omega}(dP_{E}P_{\Delta})|n>e^{\imath Et}|^2 
\\
& = &  
	\sum_{E\in\sigma_{pp}(\omega)\cap \Delta}
		|<0|\pi_{\omega}(P_{\{E\}})|n>|^2 
\mbox{ . }
\end{eqnarray*}

\noindent In particular, for fixed positive integer $L$:

\begin{eqnarray}
\label{eq-pt}
\lim_{T \rightarrow
\infty} \sum_{|n|<L}p_T(\omega,n) 
\vspace*{2cm}
& = &  
	\sum_{E\in\sigma_{pp}(\omega)\cap \Delta}\;\;\;\sum_{|n|<L}
		<0|\pi_{\omega}(P_{\{E\}})|n><n|\pi_{\omega}(P_{\{E\}})|0> 
\nonumber
\\
& & 
\\
& \leq &  
	 <0|\pi_{\omega}(P_{\Delta})|0>
		-<0|\pi_{\omega}(P_{c}P_{\Delta})|0> 
\nonumber
\mbox{ . }
\end{eqnarray}

\noindent Let us now introduce the following notations:

$$
p_T(n) = \int d\pro      p_T(\omega,n)\mbox{ , }
$$

$$
 r= \int d\pro      <0|\pi_{\omega}(P_{c}P_{\Delta})|0>
\qquad s= \int d\pro      <0|\pi_{\omega}(P_{\Delta})|0>
\mbox{ . }
$$

\noindent By the dominated convergence theorem it gives

$$
\int d\pro \lim_{T \rightarrow
\infty} \sum_{|n|<L}p_T(\omega,n)
\;=\;\lim_{T \rightarrow
\infty}\sum_{|n|<L}p_T(n) \;\;\leq \;\;s-r 
\mbox{ . }
$$

\noindent Since $r\geq 0$, one can find $T_L>0$ such that, if $T\geq T_L$,
one has $\sum_{|n|<L}p_T(n)\leq s-\frac{r}{2}$.
Thus, for $T\geq T_L$:

\begin{equation}
\delta X_{\Delta} (T)^2 
 \geq 
  \int d\pro
   \sum_{|n|\geq L} |n|^2 p_T(\omega,n)\;\; \geq 
    \;\;\frac{L^2r}{2}
\mbox{ . }
\end{equation}

\noindent Taking the $\limsup$ over $T$ we get $Lr \leq l^2(\Delta)$ for all
$L$, implying that $r=0$. Thus for almost all $\omega\in
\Omega$: $<0|\pi_{\omega}(P_c P_{\Delta})|0>=0$. Using the covariance relation
and since $\ZZ^D$ is countable,
there is $\Omega'\subset\Omega$ of probability one such that 
$<n|\pi_{\omega}(P_c P_{\Delta})|n>=0$
for all $n\in\ZZ^D$. In particular $\pi_{\omega}(P_c)|n>=0$ for all  $n\in\ZZ^D$
and all $\omega\in \Omega'$, namely the continuous spectrum in $\Delta $ is empty.

\vspace{0.2cm}

ii) Given two 
Borel subsets $\Delta_1,\Delta_2\subset \Delta$, we define the
following expression:

$$
{\cal E}_{T,\omega}^{(L)}(\Delta_1,\Delta_2) =\int^T_0\frac{dt}{T}
\sum_{|n|<L}|n|^2<0|\pi_{\omega}(e^{\imath Ht}P_{\Delta_1})|n>
\overline{<0|\pi_{\omega}(e^{\imath Ht}P_{\Delta_2})|n>}
\mbox{ . }
$$

\noindent This expression gives a Borel function in $\omega$. We use the Wiener criterion:

$$
\lim_{T \rightarrow\infty}{\cal E}_{T,\omega}^{(L)}(\Delta_1,\Delta_2) =
\sum_{|n|<L}|n|^2
\sum_{E\in\sigma_{pp}(\omega)\cap \Delta_1\cap \Delta_2}
|<0|\pi_{\omega}(P_{\{E\}})|n>|^2=
{\cal E}_{\omega}^{(L)}(\Delta_1\cap \Delta_2)
\mbox{ . }
$$

\noindent From this definition of ${\cal E}_{\omega}^{(L)}(\Delta')$
for a Borel set $\Delta'\subset \Delta$ it follows that:
\bed
\item[(a)] $0\;\leq \;{\cal E}_{\omega}^{(L)}(\Delta')\;\;\leq \;\;L^2 <0|
\pi_{\omega}(P_{pp}P_{\Delta'})|n>
\;\;\leq \;L^2
$
\item[(b)] If $\Delta_1\cap \Delta_2=\emptyset$, then
${\cal E}_{\omega}^{(L)}(\Delta_1\cup\Delta_2)=
{\cal E}_{\omega}^{(L)}(\Delta_1)+
{\cal E}_{\omega}^{(L)}(\Delta_2)
$
\item[(c)]
$
{\cal E}_{\omega}^{(L)}(\Delta')\leq {\cal E}_{\omega}^{(L+1)}(\Delta')
$
\item[(d)] ${\cal E}_{\omega}^{(L)}(\Delta')$ is a Borel function of $\omega$ as
pointwise limit of Borel functions.
\item[(e)] If $(\Delta_j)_{j\in\NN}$ is a decreasing sequence of Borel subsets such
that $\bigcap_{ j\in\NN}\Delta_j=\emptyset$, then 
${\cal E}_{\omega}^{(L)}(\Delta_j)$ decreases to zero.
\ed

\noindent Averaging over the disorder, we obtain 
${\cal E}^{(L)}(\Delta')= \int d\pro {\cal E}_{\omega}^{(L)}(\Delta')$ 
which fulfills (a),(b),(c), and also (e) thanks to 
the dominated convergence theorem.
Moreover, since ${\cal E}^{(L)}(\Delta')\leq L^2$, we can use the dominated
convergence theorem, Fubini's theorem and the definition of $l^2(\Delta)$ to get

$$
{\cal E}^{(L)}(\Delta') 
\; = \;
	\lim_{T \rightarrow\infty}
		\int^T_0\frac{dt}{T}\int d \pro
			\sum_{|n|<L}|n|^2|<0|\pi_{\omega}(e^{\imath 
				Ht}P_{\Delta'})|n>|^2 
\; \leq  \; 
	l^2(\Delta') 
\mbox{ . }
$$

\noindent In much the same way, we get, thanks to (b),
${\cal E}^{(L)}(\Delta')\leq l^2(\Delta)<\infty\mbox{ , }\forall\;\;L\in\NN$. 
As ${\cal E}^{(L)}(\Delta')$
is bounded and increasing in $L$, it follows that ${\cal E}(\Delta')
\;=\;\lim_{L \rightarrow\infty}{\cal E}^{(L)}(\Delta')$ exists. $
{\cal E}$ defines a non-negative set function on the set of Borel subsets of
$\Delta$. Because of property (e), it is continuous from above and since it is
moreover finite, we can conclude its $\sigma$-additivity. Therefore ${\cal E}$
is a Radon measure. Moreover, using the monotone convergence theorem:

$$
{\cal E}(\Delta') \; =\; 
 \int d \pro
  \sum_{E\in\sigma_{pp}(\omega)\cap \Delta'}\;\;\;
   \sum_{n\in\ZZ^D}
    |n|^2
     |<0|\pi_{\omega}(P_{\{E\}})|n>|^2
\mbox{ . }
$$

\vspace{0.1cm}

(iii) Let us show now $l^2(\Delta')\leq {\cal E}(\Delta')$. Recall
that:

$$
{\cal E}(\Delta')\; =\; \int d \pro \sup_{N\in\NN} \sum_{|n|<N}|n|^2
\lim_{T \rightarrow\infty}\int^T_0\frac{dt}{T}
|<0|\pi_{\omega}(e^{\imath Ht}P_{\Delta'})|n>|^2 \;< \; \infty
\mbox{ . }
$$

\noindent We can replace the limit by the $\limsup$. Hence for  a fixed
$T_0$ large enough, the following expression is finite and
we may apply Fubini's theorem:

\begin{eqnarray*} 
\lefteqn{ \int d \pro \sup_{N\in\NN} \sum_{|n|<N}|n|^2
		\sup_{T >T_0}\int^T_0\frac{dt}{T}
			 |<0|\pi_{\omega}(e^{\imath Ht}P_{\Delta'})|n>|^2 }
\\
& \geq &
	\sup_{T >T_0}\int d \pro 
		\int^T_0\frac{dt}{T}\sup_{N\in\NN}
		\sum_{|n|<N}|n|^2|<0|\pi_{\omega}(e^{\imath Ht}P_{\Delta'})|n>|^2  
\\
& \geq &
	\limsup_{T >T_0} 
		\int^T_0\frac{dt}{T}\int d \pro \sup_{N\in\NN}
		\sum_{|n|<N}|n|^2|<0|\pi_{\omega}(e^{\imath Ht}P_{\Delta'})|n>|^2    
\\
& = &  
	l^2(\Delta')
\mbox{ . }
\end{eqnarray*}

\noindent As this is true for all $T_0$, we obtain $l^2(\Delta')\leq {\cal
E}(\Delta')$, and with (ii) their equality.

\vspace{0.2cm}

(iv) To finish the proof we use the Radon-Nikodym theorem.
It is thus sufficient to show that the measure $l^2(\Delta')=  {\cal
E}(\Delta')$ is absolutely continuous with respect to the density of states.
Let $\Delta'\subset\Delta$ be such that ${\cal N}
(\Delta')=\TV(P_{\Delta'})= 0 $. From the definition of the trace it
follows that $<0|\pi_{\omega}(P_{\Delta'})|0>= 0 $ almost surely. By
covariance and because $\ZZ^D$ is countable, this gives
$\pi_{\omega}(P_{\Delta'})|n>=0$ for all $n\in \ZZ^D$ almost surely. Then the
definition of ${\cal E}_{T,\omega}^{(L)}(\Delta',\Delta')$ implies that it
is zero for any $l,T$ and almost all $\omega$; consequently $0= {\cal
E}(\Delta')=l^2(\Delta')$.
\hfill $\Box$

\vspace{0.2cm}

Let us give another useful expression for $l^2(\Delta)$. We consider
finite partitions $\Pp$ of
$\Delta$ into disjoint Borelian subsets:

$$
 \Pp=\{\Delta_j\subset \Delta\; , \;
\Delta_j\; \mbox{Borelian}\;| \;\;j=1\ldots q, \;
\bigcup_j\Delta_j=\Delta ,\;\Delta_j\cap\Delta_k = \emptyset\}
\mbox{ . }
$$

\noindent The set $\ZA$ of such finite partitions is ordered by refinement

$$
\Pp\leq\Pp' \qquad \Leftrightarrow \qquad \forall \;\Delta' \in \Pp' \;\;
\exists
\;\Delta'' \in \Pp \;\;\;\mbox{such that} \;\;\;\Delta' \subset \Delta''
\mbox{ . }
$$

\noindent This gives an ordered net. We define:

\begin{equation}
\label{def-loclen2}
\hat{l}^2(\Delta)=
 \lim_{\Pp\in\ZA}
  \sum_{\Delta'\in\Pp}
   \TV  (
    |\naV P_{\Delta'}|^2
     )
\mbox{ , }
\end{equation}

\noindent where the limit is understood to be the one under refinements of partitions.

\begin{proposi}
$\hat{l}^2(\Delta)$ is well defined in $\overline{\RR}=\RR\cup\{\infty\}$.
Moreover:

$$
\hat{l}^2(\Delta)=
 \sup_{\Pp\in\ZA}
  \sum_{\Delta'\in\Pp}
   \TV  (
    |\naV P_{\Delta'}|^2
     )
\mbox{ . }
$$

\end{proposi}

\noindent {\bf Proof.} It is sufficient to show that refining the partition
results
in the increase of the quantity $\sum_{\Delta'\in\Pp}\TV 
(|\naV P_{\Delta'}|^2)$. Take
$\Delta=\Delta_1\cup\Delta_2,\;\;\Delta_1\cap\Delta_2=\emptyset$,
and suppose $\TV(|\naV P_{\Delta}|^2)<\infty$. We have:

$$
\TV(|\naV P_{\Delta}|^2)=\TV(|\naV P_{\Delta_1}|^2 + |\naV
P_{\Delta_2}|^2 +\naV P_{\Delta_1}\cdot 
\naV P_{\Delta_2}+\naV
P_{\Delta_2}\cdot \naV
P_{\Delta_1})
\mbox{ . }
$$

\noindent Now, either one of the Sobolev norms of $P_{\Delta_1}$ and $P_{\Delta_2}$
is infinite and in this case the inequality 
$\TV  (|\naV P_{\Delta_1\cup\Delta_2}|^2)\;\;\leq\;\;
\TV  (|\naV P_{\Delta_1}|^2)\;+\;
\TV  (|\naV P_{\Delta_2}|^2)$ is trivially satisfied, or
the components of $\naV P_{\Delta_1}$ and $\naV P_{\Delta_2}$ are in 
$L^2(\Aa,\TV)$. The H\"older inequality implies
$\naV P_{\Delta_1}\cdot \naV P_{\Delta_2}\in L^1(\Aa,\TV)$. We may
therefore treat each term separately.

$$
\TV(\naV P_{\Delta_1}\cdot \naV P_{\Delta_2}) =
\TV(P_{\Delta_1}\naV 
P_{\Delta_1}(1-P_{\Delta_1})\cdot \naV P_{\Delta_2} )
+\TV((1-P_{\Delta_1})\naV P_{\Delta_1}P_{\Delta_1}
\cdot \naV P_{\Delta_2})
\mbox{ . }
$$

\noindent We apply the cyclicity of $\TV$ and the formul{\ae}

$$
P_{\Delta_1}\leq 1 -P_{\Delta_2} \mbox{ , }\qquad
P_{\Delta_2}\leq 1 -P_{\Delta_1} \mbox{ , }\qquad
\naV P_{\Delta_2}P_{\Delta_1}=-P_{\Delta_2}\naV P_{\Delta_1}
\mbox{ , }
$$

\noindent to get:

\begin{eqnarray*}
\TV(\naV P_{\Delta_1}\cdot \naV P_{\Delta_2}) 
& = &
	\TV(P_{\Delta_1}\naV P_{\Delta_1}\cdot \naV P_{\Delta_2})
	+\TV(\naV P_{\Delta_1}\cdot (-\naV P_{\Delta_1} P_{\Delta_2})) 
\\
& = &
	-\TV(|\naV P_{\Delta_2} P_{\Delta_1}|^2) 
		-\TV(|\naV P_{\Delta_1} P_{\Delta_2}|^2) \;\;\;\leq \;\;0
\mbox{ . }
\end{eqnarray*}

\noindent The same calculation for the other term then implies the 
result. \hfill $\Box$

\vspace{0.2cm}

Because of equation (\ref{eq-sobloc}), one expects $\hat{l}^2$ to be equal to $l^2$
and this is what we shall prove in the sequel.
We first need the following technical lemma:

\begin{lemma}
\label{lem-measures}
 Let $\mu,\nu$ be two finite positive measures on a Borel subset
$\Delta$ of the real line and $\mu_{pp},\nu_{pp}$ their pure-point parts. For a
finite
partition $\Pp$ of $\Delta$ we set:

$$
{\cal N}(\Delta,\Pp)=\sum_{\Delta'\in \Pp}\mu(\Delta')\nu(\Delta')
\mbox{ . }
$$

\noindent Then:

$$
\lim_{\Pp\in\ZA}{\cal N}(\Delta,\Pp)=
\inf_{\Pp\in\ZA}{\cal N}(\Delta,\Pp)
=\sum_{E\in\Delta}\mu_{pp}(\{E\})\nu_{pp}(\{E\})
\mbox{ . }
$$

\end{lemma}

\noindent {\bf Proof.} First, we show that ${\cal N}(\Delta,\Pp)$ decreases as the
partition is refined. We introduce ${\cal F}(\Pp)=\bigcup_{\Delta'\in\Pp}
\Delta'\times\Delta'$. Suppose that $\Pp\leq\Pp'$, then
${\cal F}(\Pp')\subset{\cal F}(\Pp)$. As $\mu\otimes\nu$ is a finite positive
measure on a Borel subset of $\RR^2$, we have:

$$
{\cal N}(\Delta,\Pp)
=\mu\otimes\nu({\cal F}(\Pp))\geq\mu\otimes\nu({\cal F}(\Pp'))
={\cal N}(\Delta,\Pp')
\mbox{ . }
$$

\noindent Let us now consider the Lebesgue decomposition of the measures $\mu$ and $\nu$
in continuous and pure-point parts:

$$
\mu\otimes\nu=\mu_{pp}\otimes\nu_{pp}+\mu_{c}\otimes\nu_{pp}+
\mu_{pp}\otimes\nu_{c}+\mu_{c}\otimes\nu_{c}
\mbox{ . }
$$

\noindent As the measures are finite, the continuous parts satisfy:

$$
\forall \;\epsilon>0 \;\; \exists\; \delta>0\;\;\;\mbox{such that}\;\;\; 
\mu_{c}(\Delta')<\epsilon,\;\;\nu_{c}(\Delta')<\epsilon
\;\;\;\;\forall \;\Delta' \;\;\;\mbox{with}\;\;\; \mbox{diam}(\Delta')<\delta
\mbox{ . }
$$

\noindent Here, the diameter is defined by:
$\mbox{diam}(\Delta')=\sup_{x,y\in\Delta'}|x-y|$. Now we choose and fix a
sequence of $\Pp_n$ of finite partitions which satisfies:

$$
\lim_{n\rightarrow\infty}\left(
\max_{\Delta'\in\Pp_n}\mbox{diam}(\Delta')\right) = 0
\mbox{ . }
$$

\noindent For such a sequence $\Pp_n$ the following holds:

$$
\bigcap_{n=1}^{\infty}\left(\bigcup_{\Delta'\in\Pp_n}
(\Delta'\times\Delta')\right)
=\{(x,x)|x\in\Delta\}= \mbox{Diag}(\Delta\times\Delta)
\mbox{ . }
$$

\noindent In the limit, the contribution containing a continuous part
vanishes for indeed for instance

$$
\lim_{n\rightarrow\infty}\mu_{c}\otimes\nu_{pp}
\left(\bigcup_{\Delta'\in\Pp_n}
(\Delta'\times\Delta')\right)
\;\; \leq \;\;  
	\lim_{n\rightarrow\infty} \epsilon(n)\sum_{\Delta' \in\Pp_n}
		\nu_{pp}(\Delta') 
\;\; \leq \;\; 
	\lim_{n\rightarrow\infty} \epsilon(n)\nu(\Delta) \;\;= \;\;0
\mbox{ . }
$$

\noindent Therefore, using the $\sigma$-additivity of the finite measure 
$\mu\otimes\nu$
(which is equivalent to continuity from above), we find:

\begin{eqnarray*}
\lim_{n\rightarrow\infty}\sum_{\Delta' \in\Pp_n}\mu(\Delta')
\nu(\Delta') 
& = & 
	\lim_{n\rightarrow\infty}
		\sum_{\Delta' \in\Pp_n}\mu_{pp}(\Delta')
			\nu_{pp}(\Delta') 
\; = \; 
	\mu_{pp}\otimes\nu_{pp}\left(
		\bigcap_{n=1}^{\infty}\bigcup_{\Delta'\in\Pp_n}
			(\Delta'\times\Delta')\right) 
\\
& & 
\\
& = & 
	\mu_{pp}\otimes\nu_{pp}(\mbox{Diag}(\Delta\times\Delta)
\;\; = \; 
	\sum_{E\in\Delta}\mu_{pp}(\{E\})\otimes\nu_{pp}(\{E\})
\mbox{ . }
\end{eqnarray*}

\noindent This is true for every such refining sequences of partitions, 
leading to the result. \hfill $\Box$

\begin{proposi}
\label{pro-loclength2}
If $l^2(\Delta)<\infty$, then 
$l^2(\Delta')=\hat{l}^2(\Delta')$ for every Borelian subset
$\Delta'\subset \Delta$.
\end{proposi} 

\noindent {\bf Proof.} As $l^2(\Delta)<\infty$, 
Theorem~\ref{theo-locpoint} shows that

$$
{\cal E}^{(L)}(\Delta)= \int d\pro
\sum_{|n|<L} |n|^2 \sum_{E\in \sigma_{pp}(\omega)\cap\Delta}
|<0|\pi_{\omega}(P_{\{E\}})|n>|^2
\mbox{ . }
$$

\noindent We decompose the complex measure
$\mu(\Delta')=<0|\pi_{\omega}(P_{\Delta'})|n>$ by polarisation into four
positive measures:

$$
\mu(\Delta')=\mu_1(\Delta')-\mu_2(\Delta')
+\imath \mu_3(\Delta')-\imath \mu_4(\Delta')
\mbox{ . }
$$

\noindent With these notations, we apply Lemma~\ref{lem-measures} 
to each terms to get

\begin{eqnarray*}
{\cal E}^{(L)}(\Delta) 
& = & 
	\int d\pro
		\sum_{|n|<L} |n|^2 \sum_{E\in \sigma_{pp}(\omega)\cap\Delta}\left(
			\sum_{k=1}^4 \mu_k\otimes\mu_k -2\mu_1\otimes\mu_2
				-2\mu_3\otimes\mu_4\right)(\{E\times E\}) 
\\
 & = & 
	\int d\pro
		\sum_{|n|<L} |n|^2 \lim_{\Pp \in {\cal Z}(\Delta')}
			\sum_{\Delta''\in\Pp}|<0|\pi_{\omega}(P_{\Delta''})|n>|^2
\mbox{ . }
\end{eqnarray*}

\noindent The dominated convergence theorem now gives:

$$
{{\cal E}}^{(L)}(\Delta) = \lim_{\Pp \in {\cal Z}(\Delta')}
\sum_{\Delta''\in\Pp} \int d\pro\sum_{|n|<L} |n|^2
\;|<0|\pi_{\omega}(P_{\Delta''})|n>|^2 =\hat{{\cal E}}^{(L)}(\Delta)
\mbox{ , }
$$

\noindent  by definition of $\hat{{\cal E}}^{(L)}(\Delta)$. Obviously we have:

$$
\hat{{\cal E}}^{(L)}(\Delta) 
\;\; \leq \;\;
	\lim_{\Pp \in {\cal Z}(\Delta')}
		\sum_{\Delta''\in\Pp} \int d \pro\sum_{n\in\ZZ} |n|^2
			\;|<0|\pi_{\omega}(P_{\Delta''})|n>|^2 \;\;
			 = \;\; \hat{l}^2(\Delta') 
\mbox{ . }
$$

\noindent Moreover, $\hat{{\cal E}}^{(L)}(\Delta')$ is increasing in $L$.
Since ${\cal E}^{(L)}(\Delta) $ is 
bounded by $l^2(\Delta)$, so is $\hat{{\cal E}}^{(L)}(\Delta) $. 
Its limit therefore exists. Actually, it converges to 
$\hat{l}^2(\Delta')$. For indeed, with the dominated convergence theorem, we find:

\begin{eqnarray*}
\hat{l}^2(\Delta')  
& = &
	\sup_{\Pp \in {\cal Z}(\Delta')}
		\sum_{\Delta''\in\Pp} \int d\pro\lim_{L\rightarrow\infty}\sum_{|n|<L} |n|^2
			|<0|\pi_{\omega}(P_{\Delta''})|n>|^2 
\\
 & = &
	\sup_{\Pp \in {\cal Z}(\Delta')}\;\sup_{L\in\NN}
		\sum_{\Delta''\in\Pp} \int d\pro\sum_{|n|<L} |n|^2
			|<0|\pi_{\omega}(P_{\Delta''})|n>|^2  
\\
 & = & 
	\lim_{L\rightarrow \infty}\hat{{\cal E}}^{(L)}(\Delta')
\mbox{ . }
\end{eqnarray*}

\noindent This finishes the proof. \hfill $\Box$

\vspace{0.2cm}

\noindent {\bf Proof of Theorem~\ref{theo-continuity}}.
Fix $E,E'\in\Delta,\;E\leq E'$, then:

\begin{eqnarray*}
 \parallel P_{E'}-P_{E}\parallel_{_{\Ss}}^2 
& = &
	 \parallel P_{[E,E']}\parallel_{_{\Ss}}^2 \;\; \leq \;\;
		\hat{l}^2([E,E'])+ \Nn (E')-\Nn (E)
 \\
& = & 
	\int^{E'}_E (l(E'')^2 + 1)d{\cal N}(E'')
\mbox{ . }
\end{eqnarray*}

\noindent Now, in the limit $E'\rightarrow E$, this is zero 
if $E$ is a regularity point of ${\cal
N}$. \hfill $\Box$

\vspace{.2cm}

\subsection{Localization in physical models}
\label{sec-physicalmodel}

In this section we give an example of a physical model for which
the localization condition $l^2(\Delta)<\infty$ is satisfied both for weak
disorder at the band edges and for high disorder all over the spectrum. For the
mathematical treatment we will, once again, restrict ourselves to the discrete
case. Our line of arguments will use results of Aizenman and
Molchanov \cite{Ai,AM}. They give a simple proof of mathematical
results proved earlier \cite{FrSp,Ko,SW,KuSo,DSL}. We will conclude with
some remarks about the continuous case.

As a preliminary, let us remark that if the spectrum has a finite gap
$\Delta$, then the condition $l^2(\Delta)<\infty$ is satisfied for the
simple reason that $P_{\Delta}=0$. In dimension two, the Chern character
corresponding to every energy band is therefore an integer and the Hall
conductivity an integer multiple of $e^2/h$. However, this integer may
be zero.

The analysis by Aizenman and Molchanov consists of two steps: first, one shows
exponential decay of low moments of Green's function for concrete classes
of models and in specified regions $\Delta$ of the spectrum; then, this decay
is used \cite{Ai} to show exponential decay in $|n-m|$ of the quantity $\int
d\pro \sup_{t\geq 0}<n|\pi(e^{\imath Ht}P_{\Delta})|m>$. This, in turn, will allow us
to show that $l^2(\Delta)<\infty$ is satisfied.

Here, we fix our attention to the $D$-dimensional Anderson model with constant
magnetic field. The Hamiltonian acting on $\ell^2(\ZZ^D)$ is given by:

$$
H_{\omega,\lambda}=H_0+\lambda V_{\omega} \qquad \lambda \in \RR
\mbox{ , }
$$

\noindent where $H_0$ is the $D$-dimensional analog of 
Hamiltonian given in equation (\ref{eq-Har1}).
 $V_{\omega}$ is the disorder
potential: at every site of the lattice it takes a random value in the interval
$[-1,1]$; the probability distribution is supposed to be uniform and the sites
are independent. In \cite{AM,Ai} much more general situations are considered:
the kinetic part $H_0$ may have non-zero elements farther off the diagonal as
long as they decay exponentially in distance from the diagonal; the random
potential may also have gaussian or Cauchy distribution; moreover, 
correlations between the sites are allowed in a sense to be made precise. For us,
however, the 
Anderson model captures the essential of the localization phenomenon and it has
the advantage that the hull $(\Omega, T, {\bf P} )$ is easily constructed as
topological product of intervals $[-1,1]$; by Tychonow's theorem, $\Omega$ is
compact; the action $T$ on $\Omega$ is given by the translations in physical
space and the product measure of the uniform probability on $[-1,1]$ is a
$T$-invariant and ergodic measure ${\bf P}$ on $\Omega$.

We will now describe precisely in which situations Aizenman and Molchanov prove
exponential decay of low moments of Green's functions for the Anderson model:

\begin{equation}
\label{decay}
\int d \pro \;|<n|\pi_{\omega}(\frac{1}{H-E})|m>|^s \;\;\leq\;\;
De^{-c|n-m|} \qquad s\in (0,1)
\mbox{ . }
\end{equation}

The first situation is that of strong disorder: for every $s\in (0,1)$ there
exists an $\lambda_c=\lambda_c(D,s)$ such that for $\lambda\geq\lambda_c$ the
bound holds on all of the spectral axis.

The second situation considers weak disorder, that is small $\lambda$. Recall
that the free Hamiltonian $H_0$ has energy bands of continuous spectrum. In the 
interesting case for us of dimension two, the energy bands and their dependence on the
magnetic field are given by Hofstadter's butterfly; $H_0$  then gives Harper's
equation.
For the Anderson model the bound (\ref{decay}) is obtained for energies $E$
situated at band edges:

$$
E\in R(\lambda)= \bigcup_{s\in(0,1)} \{ E\in\RR -  \sigma(H_0) \;|\; \frac{1}{2}\lambda
g_s(E) <1 \}
\mbox{ , }
$$

\noindent where

$$
g_s(E) = \sup_{x\in\ZZ^D} \left(\sum_{y\in\ZZ^D} |<n|\pi_{\omega}(\frac{1}{H-E})|m>|^s
\right)^{-s}
\mbox{ . }
$$

\noindent (For the Anderson model, the constant $\kappa(s)$ appearing in \cite{Ai} is
equal to $\frac{1}{2}$ independent of $s$.)
By a Combes-Thomas argument, one gets \cite{Ai} the asymptotic behavior of
$g_s(E)$ in $\xi=\mbox{dist}(E,\sigma(H_0))$:

$$
g_s(E)=\left\{\begin{array}{cc}{s^{-\frac{D}{s}}\xi^{-(1+\frac{D}{s})}
\hspace{2cm}} {\xi\rightarrow 0} \\
{\xi^{-1}(1+\Oo(\frac{1}{\xi})) \hspace{1.8cm} }{\xi\rightarrow \infty} 
\end{array}\right.
\mbox{ . }
$$

\noindent Therefore, because $\sigma(H_{\lambda})\subset
\{E\;|\; E\in\{\sigma(H_o)+\nu\};\nu\leq\lambda\}$, the intersection of $R(\lambda)$
with $\sigma(H_{\lambda})$ is non-empty for $\lambda$ sufficiently big. In other
words, the bound (\ref{decay}), is satisfied for some energies belonging to the
spectrum and situated at the band edges. This is probably the case as soon as
$\lambda\neq 0$; dimension two is critical \cite{An}.

Let us now come to the second step of Aizenman's analysis. If the bound
(\ref{decay}) is satisfied for an interval $\Delta$ of the spectral axis, he
proves that the unitary evolution operator filtered with the spectral
projection $P_{\Delta}$ satisfies:

\begin{equation}
\label{estimate}
\int d \pro \sup_{t\geq 0} |<n|\pi_{\omega}(P_{\Delta}e^{\imath Ht})|m>| \;\;\leq
C\;\;
e^{-D|n-m|} \qquad C,D\in \RR_+
\mbox{ . }
\end{equation}

\noindent Let us now use this result to show that the non-commutative localization
condition is satisfied. Because $\parallel \pi_{\omega}(P_{\Delta}e^{\imath Ht})
\parallel_{_{\Ll(\ell^2(\ZZ^D)}} \leq 1$, we have:

\begin{eqnarray*}
\TV (|\partial(P_{\Delta}e^{\imath Ht})|^2) 
& = &
	\sum_{n\in \ZZ^D} |n|^2\int 
		d \pro |<0|\pi_{\omega}(P_{\Delta}e^{\imath Ht})|n>|^2
\\
& \leq & 
	\sum_{n\in \ZZ^D} |n|^2\int 
		d \pro |<0|\pi_{\omega}(P_{\Delta}e^{\imath Ht})|n>|
\\
& \leq & 
	\sum_{n\in \ZZ^D} |n|^2 C e^{-D|n|} \;\; \leq \;\; C' \;\;<\;\;\infty 
\mbox{ . }
\end{eqnarray*}

\noindent Now, taking the time-average we see that   $l^2(\Delta)<\infty$. 
This is, of
course, only true if either $\lambda
>\lambda_c$ or $\Delta \subset R(\lambda)$.

Let us now comment on the  case of continuous physical space. First of all, if
the Fermi energy $E_F$ lies in a finite gap, then it can be shown that $P_F \in
\Ss$ \cite{NB90}. For strong disorder, one expects a finite localization length,
but near the set  $\{\hbar \omega_c(n+\frac{1}{2}) | \omega_c=\frac{e\Bb }{m}, \; n\in
\ZZ\}$ the localization length probably diverges.
This would be the analog of the discrete case.
\vspace{.5cm}

\section{Applications and complements}
\label{chap-phys}

\vspace{.2cm}

\subsection{Low-lying states do not contribute to the IQHE}
\label{sec-Lowlying}

In this section, we address the following question: consider the Landau
Hamiltonian $H_L$ and add a periodic potential $V_p$ of varying strength

$$
H\;=\;H_L + \sigma V_p \mbox{ , } \qquad 0\leq V_p\leq1 \mbox{ . }
$$

How do the Chern numbers evolve as the coupling parameter $\sigma$ is increased?
In particular, what happens if $\sigma \gg \hbar \omega_c$

Halperin et al \cite{TH} made corresponding numerical calculations. They took a
finite size sample and computed the Chern numbers for various values of $\sigma$
by the method described in Section~\ref{sec-HarperLoc}. For small $\sigma$, the weak
periodic potential approximation is valid and the Chern numbers are those given
by the Diophantine equation (\ref{eq-Dioph}). For intermediate values of
$\sigma$, the energy bands cross each other in a complicated way and it is
difficult to follow the Chern numbers. However, for $\sigma\gg\hbar \omega_c$,
Halperin et al observed the following striking result: all low energy bands of
the spectrum carry zero Hall current. This result was put on a rigorous basis by
Nakamura and one of the authors \cite{NB90}.

\begin{theo}
Consider the Hamiltonian $H=H_L+\sigma V$ where $V$ is a potential satisfying
the conditions below. Then for $\sigma$ sufficiently large the spectrum has gaps
between his low energy bands (energy smaller than $\epsilon \hbar \omega_c$,
where $\epsilon$ appears in the conditions below). If the Fermi energy lies in
one of these gaps, the Hall conductivity vanishes.
\end{theo}

\noindent Hypothesis on the potential $V$:

i) $\inf_{x\in\RR^2}V(x)=0\mbox{ , } 
\sup_{x\in\RR^2}V(x)\leq C<\infty\mbox{ . } $

ii) There is a countable set of $\RR^2\mbox{ , } \{x_n,n=1,2,\ldots\}$ such that
\\
\hspace*{1cm}
$|x_n-x_m|\geq d$ if $n\neq m$ with $d>0$.

iii) There are $\epsilon > 0$ and $\Vv \in C^2(B_{\epsilon}$ where
$B_{\epsilon}=\{x\in\RR^2 | |x|<  \epsilon\}$ such that $d>2\epsilon$ and
\\
\hspace*{1cm}
$V(x_n+x)=\Vv(x)$ for $x\in B_{\epsilon}$ and for all $n$

iv) $0$ is the unique non-degenerate minimum of $\Vv\in B_{\epsilon}$

v) If $|x_n-x|\geq \epsilon$ for all $n$, then $V(x)\geq \delta >0$.

Note that this potential is not strictly periodic, although the bottoms of the
wells need to be identical. The framework of the proof is that of 
Non-Commutative Geometry completed by semi-classical analysis. Let us just 
describe the strategy here. The existence of the gaps follows from semiclassical
analysis: the energy levels of a single, isolated quantum well are enlarged by
the tunneling effect; the band width can be estimated by $e^{-a\sqrt{\sigma}}
\mbox{ , } a> 0$. In order to show that the Chern character 
is zero, one proceeds as
follows. First consider the situation of quantum wells separated by infinitely
high barriers. The projection $P^j_E$ on the energy level of quantum well number
$j$ is one-dimensional. Its Chern character therefore vanishes. Now we consider
the projection $P_{\infty}=\oplus_j P^j_E$, where the sum runs over all wells.
Because of the additivity of the Chern character (compare
Lemma~\ref{lem-additivity}), the
Chern character of $P_{\infty}$ is zero as well. Finally, semiclassical analysis
allows us to show that $P_{\infty}$ is Murray-von Neumann equivalent to the
projection $P_E$ onto the energy band which arises as the barriers are reduced
to finite height. In view of Lemma~\ref{lem-cherninvariance}, this finishes the proof. It is
likely that the result also holds if the projection is not in the gaps. 

\vspace{.2cm}

\subsection{Where and how does the localization length diverge?}
\label{sec-pointsp2}

In this section, we review the most important results on the behavior of the
localization length at the center of the broadened Landau level. Discussion of
theoretical ideas is followed by a brief presentation of numerical techniques and
results. As before, we consider a system of independent fermions in a disorder
potential described by a one-particle Hamiltonian. 

In numerical calculations and scaling theory, a
 rather strong notion of localization is often used, that of a Lyapunov exponent
$\lambda(E)$. If one supposes exponential decay of the correlation function of
the wave function $\psi_{E}$ corresponding to the energy $E$, $\lambda(E)$ is
defined by:

$$
<\ov{ \psi_{E}(r)}\;\psi_{E}(r')>\;\; \propto\;\;
e^{-\lambda(E){|r-r'|}} 
\mbox{ . }
$$

\noindent Here the average is taken over disorder configurations. Physicists often call
the inverse of $\lambda(E)$ {\sl the localization length}. However, except for
in the one-dimensional case, there is yet no clear mathematical formulation.

In 1979, the gang of four \cite{AALX} used scaling theory (ideas due to Thouless
\cite{Thou}) and renormalization group calculations to show that, at absence of
magnetic field, two is the critical dimension for localization in the following
sense: for dimension smaller than two, all states are exponentially localized at
any strength of disorder, that is, for all wavefunction there are positive
Lyapunov exponents. For dimension greater than two, there exist extended states
for low disorder. At dimension two, all states are localized except for states
corresponding to isolated critical energies at which the localization length
diverges. In the case of perpendicular magnetic field in two dimension, the same
result holds as shows a development of higher order \cite{Hilcani}.

In the QHE, the Hall conductivity does not vanish. This led Halperin to postulate
the existence of extended states \cite{Halp}. A corollary of the Theorem
\ref{theo-main} is the following:

\begin{coro} \label{divergence} \cite{Ku}
Between the plateaux, there has to be a spectral interval $\Delta$ such that
the localization condition $\l^2(\Delta)<\infty$ is violated.
\end{coro}

Remark that the interval may be very small. It is even likely that there is, in
fact, a single energy $E$ such that $P_{\leq E}$ is not in the Sobolev space $\Ss$.
This implies, of course, $\l^2(\Delta)=\infty$ for every interval containing
$E$. Corresponding numerical results are presented at the end of this section.

We now come to the question how the localization length diverges. If one
considers a disorder potential varying on a length scale much bigger than the
magnetic length, the motion can semi-classically (in the magnetic field) be well
approximated by a motion along equipotentials of the disorder potential.
Percolation theory now indicates that there is one critical energy $E_c$ to which
correspond extended states. Moreover, at this energy 
the localization length diverges as

\begin{equation}
\label{locdiv}
\frac{1}{\lambda(E)} \propto |E-E_c|^{-\nu}
\mbox{ . }
\end{equation}

Trugman evaluated $\nu=\frac{4}{3}$ \cite{Tr}, but more thorough examination of
the effect of tunneling between the wells and mountains led Mil'nikov and
Sokolov to predict
$\nu=\frac{7}{3}$ \cite{MS}. In spite of the simplicity of the theoretical
approach, the agreement with numerical results obtained by three different
methods is very good. These latter are the Thouless number method,
the Chern number method  and the third one is a scaling
theory approach; the original ideas behind are due to \cite{ThEd},
\cite{RB} and \cite{PiSa,Kramers} respectively.

The Thouless number method is the most elementary because only  energies
and no eigenvectors need to be calculated. In the us interesting case of strong
magnetic
field, calculations were made by Ando \cite{An2}.
The Chern number method \cite{HuoB} is described in some more detail in the next
section. Let us concentrate on the scaling theory approach. One considers a
cylinder of (small) circumference $M$ and extended to infinity in one direction
(denoted by the variable $z$). The magnetic field is perpendicular to the
cylinder surface. Chosing the Landau gauge $A=(0,0,\Bb z)$, the eigenstates of the
free Hamiltonian are centered on circles around the cylinder corresponding to
fixed values of $z$. This discrete set of eigenstates will serve as base for the
Hilbert space; in this base one also expresses Hamiltonian perturbed by
disorder. 
 Now for fixed circumference and fixed energy, one
calculates numerically the disorder averaged Lyapunov exponent $L_M$ of a wave
function by the transfer matrix method. Finite size scaling theory \cite{PiSa}
then allows to estimate the behavior of the localization length near the
critical energy; in particular, the exponent $\nu$ in equation (\ref{locdiv})
can be calculated. In presence of magnetic field, this method was first used in
\cite{Kramers2,SKmcK}; more recent and more 
extended numerical studies \cite{HuKr,Sar}
confirm the theoretical prediction $\nu=\frac{7}{3}$. It seems that the exponent
is independent of the disorder distribution and the Landau level index.

What is the behavior of the wave function at the critical energy $E_c$ itself?
Pichard and Sarma \cite{PiSa} suggest the following in order to calculate the
exponent $x$ for the powerlaw decay of the wave function $\psi_{E_c}$:

$$
|\psi_{E_c}(r)\psi_{E_c}(r')| \;\;\propto \;\;|r-r'|^{-x}
\mbox{ . }
$$

One calculates as before the Lyapunov exponent $L_M$ for different $M$ at fixed
energy $E_c$. With the hypothesis of conformal invariance, $x$ can then be shown
to be equal to the slope of the linear relation between $M$ and $1/L_M$. Chalker
and Coddington \cite{CC} estimate $x\approx 0.27$. We remark that this implies
that the weak localization condition $P_{\leq E_c}\in \Ss$ 
is violated. Therefore, at
critical energies (that is at the centers of the broadened Landau levels) the
Chern number may change.

\vspace{.2cm}

\subsection{Chern numbers and localization in Harper's equation}
\label{sec-HarperLoc}

The Chern characters or Chern numbers of the energy bands of Harper's equation
corresponding to rational flux have already been computed by
 Thouless, Kohmoto, Nightingale and den Nijs \cite{TKN}; we gave some insight
in Section~\ref{sec-ChKu}. Ando \cite{An1,An2} and more recently Tan 
\cite{Yon} have made numerical studies on the influence of an added disorder
potential on these Chern numbers. We describe some of the results because they
constitute a good illustration of theoretical ideas.

Let us first recall that Harper's equation (\ref{eq-Har2})
is obtained in two different limiting cases \cite{AnAO,Yon}.
One consists in adding a weak periodic potential $V_p$ to the Landau Hamiltonian
$H_L$, for example, a two-dimensional sinusoide; 
projecting $H_L+V_p$ on a subspace corresponding to one Landau level then
leads to Harper's equation. On the other hand, if one adds a strong periodic
potential (which means that its well is much deeper than
$\hbar\omega_c$), the low energy region of the spectrum is well described by a
tight binding model (equation \ref{eq-Har1}). As shown in
Section~\ref{sec-ChKu}, Bloch's representation also leads to Harper's
equation. 

The spectrum of Harper's equation is well known and its dependence on the
magnetic flux per unit area $\phi$ is represented in Hofstadter's butterfly
\cite{Hof}. The parameter $\alpha$ appearing in equation (\ref{eq-Har2}) is
proportional to $\phi$ in the tight-binding case and proportional to the inverse
of $\phi$ 
in the case of weak periodic potential. Let us now suppose the rationality of
$\alpha=\frac{q}{p}$, where $p$ and $q$ are relatively prime integers. The
spectrum then consists of $p$ energy bands of extended states. The Chern number
of the $r^{\mbox{th}}$ band can be calculated with the Diophantine equation in
the following way \cite{TKN,St}. There exist unique integers $s_r,t_r$ (except
for even $p$ and $r=p/2$) such that
$|s_r|\leq \frac{p}{2}$ and satisfying:

\begin{equation}
\label{eq-Dioph}
s_rq+t_rp=r \qquad r=0,\ldots,p
\mbox{ . }
\end{equation}

The Chern number of the  $r^{\mbox{th}}$ band and therefore its contribution
in units of $\frac{e^2}{h}$ to the Hall conductance is given by $t_r-t_{r-1}$ in
the weak periodic potential case and by $s_r-s_{r-1}$ for the tight-binding
model. Remark that as $t_0=0$ and $t_p=1$, the sum of all $p$ Chern numbers in the
weak periodic potential limit is equal to one. This reflects the fact that the
bands result from the splitting of one Landau level with Chern number $1$. On
the other hand, $s_0=s_p=0$, such that the sum of the Chern numbers in the tight
binding approximation is zero.

Now a disorder potential will be added. This could be, for example, an
Anderson-type perturbation (compare Section~\ref{sec-physicalmodel}), 
but in the numerical
simulations a densely distributed potential with two values $V_d,-V_d$ was used.
The fractal structure of Hofstadter's butterfly disappears as soon as disorder
is turned on. Because many gaps close, it becomes possible to label the
remaining energy bands by a finite number of rationals for any (maybe
non-rational) parameter $\alpha$. What are the Chern numbers of the bands and
how do they evolve as a function of the disorder strength?

Before starting the discussion, let us comment on the physical relevancy of the
two approximations. In the usual QHE samples presented in
Section~\ref{sec-QHE}, neither seems to describe the reality: the
disorder potential is much stronger than the weak periodic potential due
to the ions within the surface. On the other hand, because the mobility is very
high, the tight-binding approximation is not appropriate. However, lately
so-called anti-dot 
superlatices seem to open the possibility of
the experimental realization of parts of
Hofstadter's spectrum (of course, only in a very approximative way).
In these systems the  ideas we present here could be
tested.

For all details on how the numerical results were obtained, we refer to the
original works \cite{An1,Yon}, but let us describe the principal steps.
Of course, the calculations are made in finite size samples and therefore
disorder is just treated in an increased unit cell, causing thus a splitting of
the Harper bands corresponding to a fixed rational flux into subbands. 
Chern numbers for these subbands can be calculated by the contour
integrals given in equation (\ref{eq-Chern2}). Remark
that because we are in dimension two, these subbands have corresponding Hilbert
sub-spaces generically well separated except for in isolated points. 
If this happens, the Chern number is simply calculated for a two dimensional
fiber bundle (this is also the case for the closed central gaps of the Harper
equation with even denominator).
 In this way, one gets a
distribution of Chern numbers over the subbands of every Harper band.  Now, changing
the disorder configuration will give a different distribution of the Chern
numbers over the subbands. This reflects the fact that changing the disorder
configuration is equivalent to chosing another probability measure on the space
of disorder configurations; we already commented that this will change the Chern
numbers (see Section~\ref{sec-Index}). However, the sum of all Chern numbers of the
subbands is constant and equal to the Chern number of the Harper band, as long as
the disorder is weak. As it becomes stronger, the Harper bands overlap
 energetically; the Hofstadter spectrum loses more of its structure as
indicated above. But the distribution of Chern numbers may still be calculated.

\begin{figure}
\label{fig-phasedia1}
\includegraphics{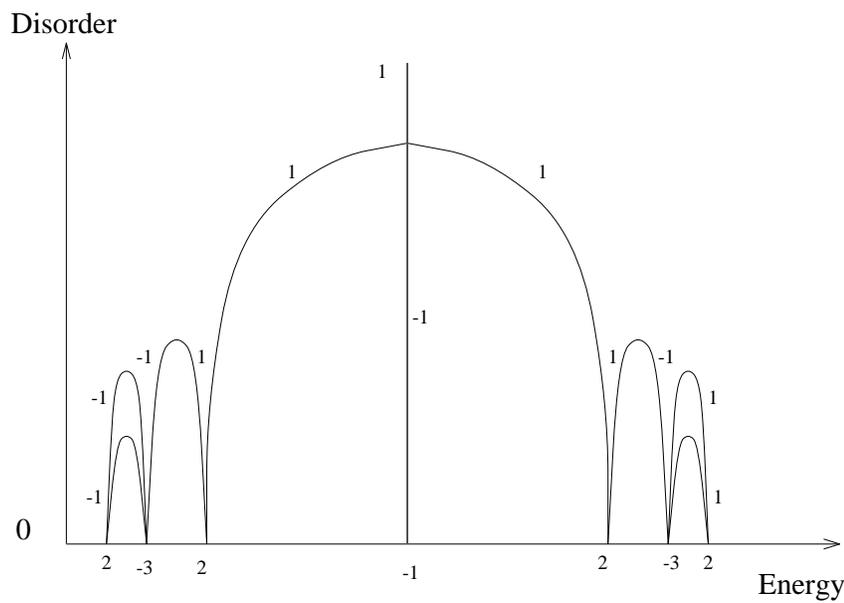}
\vglue 7.5truecm
\caption{{\sl Suggested phase diagram for the weak
potential approximation}}
\end{figure}

\begin{figure}
\label{fig-phasedia2}
\includegraphics{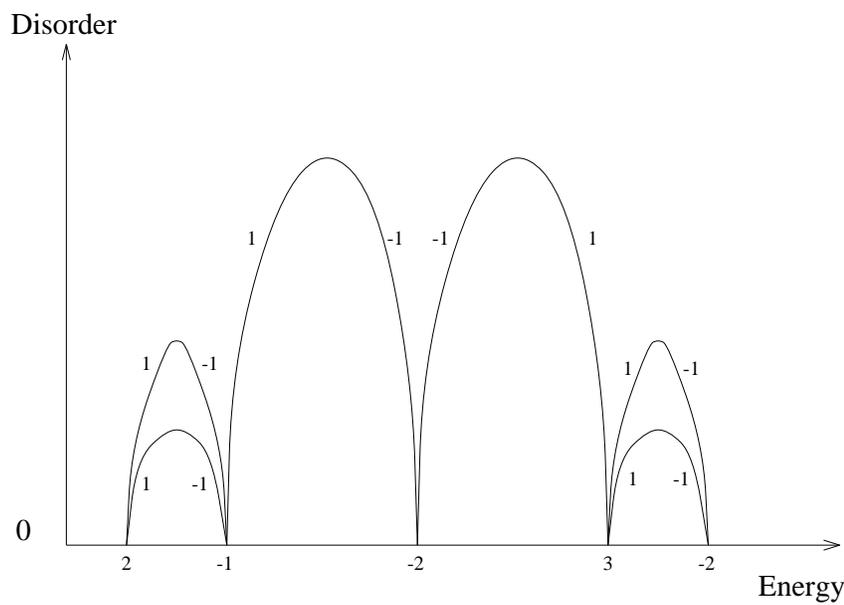}
\vglue 7.5truecm
\caption{{\sl Suggested phase diagram for the tight-binding
approximation}}
\end{figure}

Now, for every strength of disorder, averages of the Chern numbers over the
disorder configurations
 can be calculated for every subband. The
numerical results then indicate that, on average, the Hall current is
essentially carried by states corresponding to one central band situated
somewhere at the center of the Harper band. This belief is supported by the
calculation of the localization length by the Thouless number method \cite{ThEd}:
it diverges at about the energy corresponding to the Hall current carrying
subband \cite{Yon}. In the limit of an infinite sample, one
therefore expects the Hall current to be carried by one critical energy, just
as in the case without periodic potential. This is, in fact, not very
surprising because the semiclassical argument presented in
Section~\ref{sec-pointsp2} applies to the present situation as well. 

In summary, the numerical calculations suggest the phase diagram \cite{Yon}
shown in an idealized form in Figures~\ref{fig-phasedia1}and \ref{fig-phasedia2}. 
Starting at rational
flux without disorder, the evolution of the current carrying states and their
Chern numbers is given as a function of increasing strength of disorder. At some
critical values, they amalgamate; the Chern number of the resulting state is
then given by the sum of the merging ones (this sum may also be zero, Hall
currents then annihilate each other). To conclude, let us comment on the
limiting behavior for strong disorder. In the weak periodic potential
approximation, one eventually 
obtains a Chern number one as for the Landau level.
It is an
interesting question what happens to the Chern number as the quantum Hall regime
breaks down, certainly the corrections to the Kubo formula will be
important at very high disorder. In the tight
binding approximation, the system evolves to zero Hall current, this reflects
the results of Section~\ref{sec-Lowlying}.

\vspace{.5cm}

\section{Introduction to the FQHE}
\label{chap-FQHE}

\vspace{.2cm}

\subsection{Overview}
While in 1982 Thouless, Kohmoto, Nightingale and den Nijs made an important 
step towards the understanding of the IQHE \cite{TKN}, D.C.~Tsui, 
H.L.~St\"ormer and A.C.~Gossard from Bell Laboratories made the surprising 
discovery that plateaux of the Hall conductivity could be observed not only at
integer, but also at fractional multiples of $\frac{e^2}{h}$ \cite{Tsui}. Using
a high-quality $\rm GaAs-Al_xGa_{1-x}As$ heterojunction, they observed a 
plateaux at $\sigma_H / \frac{e^2}{h} = \frac{1}{3}$. The corresponding minimum 
of the direct resistance was also observed. The filling factor was varied by 
changing the magnetic field while keeping the charge-carrier density fixed. In
order to obtain the filling factor $\nu = \frac{1}{3}$, they needed a magnetic 
field strength of about 15 T. The experiment was undertaken at different values 
of temperature (0.48, 1.00, 1.65 and 4.15 K) and even at the lowest of them, the
accuracy of $\sigma_H$ on the plateau was much more modest than in the integer 
effect.

Since this first experiment, the FQHE has been observed for many values of 
$\sigma_H$. The most striking results are the following:
\begin{enumerate}
\item Let $\sigma_H / \frac{e^2}{h} = \frac{p}{q}$ with $p,q$ relatively prime
integers. Then, $q$ is odd (`odd-denominator rule'). Recently, some deviations
of this rule have been observed, but this seems to correspond to non
spin-polarized electrons. In the sequence, we will not consider this case.
\item The observation of the fractional effect requires very clean samples. It
is easily destroyed by impurities causing inelastic scattering. Moreover,
the stability of the plateaux with respect to temperature and impurity effects
is very dissimilar for different fractions. In general, fractions with small 
denominator seem to be more stable. Thus, the plateaux appear in a well-defined 
order when lowering the temperature. This gives rise to series of fractions, 
also called `hierarchy of states'.
\item In all devices, the accuracy of the Hall conductivity on the plateaux is
much less for fractional than for integer values. Whereas in the IQHE we have
$\frac{\delta \sigma_H}{\sigma_H} \approx 10^{-7}$, this is about $10^{-5}$ for
the FQHE.
\end{enumerate}
Although we have given a rigorous explanation of the IQHE, the situation is much
less satisfactory for the FQHE. The QHE contains two distinct aspects: the 
existence of plateaux (which, for the IQHE, follows from localization of the
states at the Fermi edge) and the integrality (respectively fractionality)
of the Hall conductivity on them (which has a topological origin). Usually,
these two points are not well separated in the literature on the 
FQHE. In fact, there are conceptual difficulties in understanding the r\^ole of
localization in the case of interacting electrons. The mechanism which gives 
rise to the existence of plateaux is quite unclear.
The main problem is that no Kubo formula is known to calculate the Hall 
conductivity for interacting Fermions.
Instead, one usually follows another strategy: since we know that 
non-interacting electrons only exhibit an IQHE, it is clear that the FQHE might
appear when electron-electron interaction can no longer be neglected. It is 
generally believed that at certain values of the filling factor, the ground state
becomes incompressible owing to the repulsive two-particle interaction. Thus one 
tries to find explicit ground states which exhibit a fractional Hall 
conductivity.
Then arguments are given to explain that this value remains constant while the
filling factor is slightly changed.
The most promising states that have been found show a remarkable property:
they exhibit particle-like excitations with some charge $q \not= - e$. The 
states are only defined for specific values of the filling factor, but it seems 
that the existence of quasi-particles is stable against perturbations if 
$\frac{q}{e}$ is a fractional number. Apart from numerical evidence, there is
no serious explanation of this property. The stability of quasi-particles is
due to an energy gap separating the ground state from the excitations. It is
conceptually  not clear  why the existence of such a gap is related to the
fractionality of the quasi-particle charge. No rigorous results are known here.

\vspace{0.2cm}

\subsection{Laughlin's ansatz for the $\nu = \frac{1}{m}$ groundstate}

Up to now, most attempts to explain the FQHE - in particular the classical 
hierarchy of Haldane and Halperin \cite{Hald1,Halp2}
as well as Jain's composite-fermions approach \cite{Jain} - 
are based on Laughlin's wave functions which describe a finite number of
electrons in a finite volume, yielding a filling factor $\nu = \frac{1}{m}$ 
with $m$ an odd integer. In the following, we will review the ideas which
led Laughlin to his proposal. Following Haldane, we will then construct a
class of Hamiltonians for which Laughlin's states are exact ground states.

Before doing this, we want to clarify the r\^ole of these states and try to 
justify our interest in them. As we have already mentioned, there does not
exist a satisfying theory of the FQHE at the moment. In particular, the
universality of the effect and its insensitivity to the exact form of the 
Hamiltonian is not quite clear. The main advantage of our approach to the IQHE
is that we do not need to know the exact form of the one-particle potential,
but only have to impose some mild conditions on it. To explain the existence of
plateaux and to prove the integrality of $\sigma_H / \frac{e^2}{h}$,
we did not need to calculate the ground state explicitly. In our opinion,
despite the wide use of Laughlin's wave functions in the literature, the
explicit form of the ground state does not lie at the heart of the theory,
neither for the integer nor for the fractional effect.

Nevertheless, in our theory of the IQHE we were not able to give the explicit
values of the Hall conductivity just by means of general considerations. In
fact, the value given by the Kubo-Chern formula  might depend on the
choice of the probability measure $P$ on the hull of the one-particle 
Hamiltonian. To derive the `right' values, we computed explicitly the 
Chern character of the eigenprojections of the Landau Hamiltonian. A homotopy
argument then shows that these values remain constant when the disorder is
switched on, at least for small values of the potential.

The states which arise from Laughlin's wave functions in the thermodynamical 
limit might play the same r\^ole for interacting particles: They yield the 
explicit values of the conductivity for a Hamiltonian with a restricted 
two-particle interaction and without one-particle potential. The main task
which remains is to show that these values stay constant as the Hamiltonian
is deformed continuously. This is going beyond the scope of our article.
 
We will now come to the to the description of Laughlin's states. The general
Hamiltonian for $N$ electrons is given by
$$ H_N = \frac{1}{2 m_\ast} \sum_{j=1}^N 
           (\vec{p}_j + \frac{e}{c}\vec{A}(\vec{x}_j))^2 
       + \sum_{1 \le j < k \le N} U(|\vec{x}_j - \vec{x}_k |)
       + \sum_{j=1}^N V(\vec{x}_j), $$
where in our case $\vec{\nabla} \times \vec{A}(\vec{x}) = \Bb  \vec{e}_z$, 
$\Bb  = \mbox{const}$. Thus we assume that the two-particle interaction depends 
only on the distance between the particles. Under certain conditions on the 
functions $U$ and $V$, $H_N$ is essentially self-adjoint on 
$C_0^\infty (\RR^{2N})$. Furthermore, the $N$-Fermion space 
$\bigwedge^N L^2(\RR^2)$ is an invariant subspace. Choosing suitable 
boundary conditions (we will choose Dirichlet boundary conditions on the disk 
$\Lambda_R = \{ \vec{x} \in \RR^2 : |\vec{x}| \le R \}$) yields a 
self-adjoint operator $H_{N,\Lambda}$ in the $N$-fold tensorproduct 
$\bigotimes^N L^2(\Lambda)$. We will denote its restriction to 
$\bigwedge^N  L^2(\Lambda)$ by the same symbol. Since the Laughlin states should
play the same r\^ole for interacting electrons as the ground-state of the 
Landau-Hamiltonian does for non-interacting ones, we will set $V=0$. In fact, up
to now it is unclear what happens when the one-particle interaction is switched 
on, even for small values of the potential. The basic idea which leads to 
Laughlin's states is the following: for a strong magnetic field, the 
two-particle interaction can be treated perturbatively; if we start from a 
system of non-interacting electrons in the lowest Landau level, the excitations 
to higher Landau levels in the perturbation series can be neglected. Thus, for 
filling factors $\nu \le 1$, we assume the groundstate to be a linear 
combination of Slater determinants of one-particle wavefunctions belonging to 
the lowest Landau level. 

For $V = 0$, the remaining operator $H_N$ is rotationally invariant. Thus, we 
can look for joint eigenvectors of energy and total angular momentum. It is
clear that, for a repulsive two-particle interaction, the particles will escape
to infinity when not restricted to a finite volume. Classically, the trajectory
of an electron in a constant magnetic field with angular momentum $l$ and energy
$E$ is a circle around the origin with radius $R = \sqrt{\frac{l}{m_\ast E}}$.
Thus, for $E = \frac{\hbar \omega_c}{2}$, restricting the particles to a disk 
with radius $R$ is essentially equivalent to restricting the angular momentum 
to values less or equal than 
$\frac{m_\ast \hbar \omega_c R^2}{2} = \frac{R^2 \hbar}{2 {\ell_\Bb}^2}
                                     = \frac{N \hbar}{\nu} = N_\Phi \hbar$ 
(with $\ell_\Bb$ the magnetic length). Thus, instead of calculating the ground 
state of $H_{N,\Lambda}$, we are looking for an eigenstate of $H_N$ built of 
lowest Landau level wavefunctions with maximal angular momentum $N_\Phi$. 
Laughlin, strongly influenced by the theory of liquid $\rm ^3He$, proposed the 
ansatz
$$ \psi(z_1, \ldots , z_N) = \prod_{1 \le j < k \le N} g(z_k - z_j) $$
for the $N$-particle groundstate. Here and in the following, we set the magnetic 
length to unit and we use complex coordinates $z = x + \imath y$ in the plane.
The lowest Landau level is spanned by the 
functions $\phi_m \propto z^m e^{- \frac{1}{2} |z|^2}$, thus
$$ \psi(z_1, \ldots , z_N) = \prod_{1 \le j < k \le N} f(z_k - z_j) 
                             e^{-\frac{1}{2} \sum_{l=1}^N |z_l|^2} $$
with an analytic function $f$. Since $\psi$ should be anti-symmetric under
particle exchange, we have $f(-z) = - f(z)$. Furthermore, as we are looking for
eigenstates of total angular momentum $M$ and the operator of angular momentum
for one particle is given by 
$z \partial_z - \overline{z} \partial_{\overline{z}}$, 
the product in the above formula has to be a polynomial of degree $M$ in each 
variable $z_j$. From this, it follows that $f(z) = z^m$ with an odd integer 
$m$. We will denote the resulting normalized wave function by $\psi_{N,m}$. 
The total angular momentum is $M = \frac{N(N-1)}{2} m$ and the particles are 
essentially restricted to a disk with radius $R \approx \sqrt{2m(N-1)}$ 
(in units of the magnetic length) or, equivalently, $\nu \approx \frac{1}{m}$.

Before we investigate the question for which Hamiltonians
Laughlin's ansatz yields an exact groundstate, we want to make some remarks on 
the thermodynamical limit. The QHE (the integer as well as the fractional) is 
assumed to be exact in the case of infinite volume and zero temperature, thus 
necessarily we have to involve statistical mechanics. Laughlin's states are 
states for a finite number of particles in a finite volume, i.e. we are working 
in the canonical ensemble. The Gibbs states are
$$ \langle a \rangle_{\beta,N,\Lambda} 
   = \frac{\mbox{tr}(e^{-\beta H_{N,\Lambda}} a)}{
       \mbox{tr}(e^{-\beta H_{N,\Lambda}})}, $$
where $a$ is a bounded linear operator in $\bigwedge^N L^2(\Lambda)$. At zero 
temperature, they reduce to the vector states
$$ \langle a \rangle_{\infty,N,\Lambda} 
   = (\psi_{N,\Lambda}, a \psi_{N,\Lambda}), $$
where $\psi_{N,\Lambda}$ is the (non-degenerate) groundstate of $H_{N,\Lambda}$.
To investigate the limit $N \to \infty$, $\Lambda \to \RR^2$ with fixed 
particle density $\frac{N}{|\Lambda|}$, we have to include vector states with 
an arbitrary number of particles, i.e. we have to work in Fock space. 
The algebra of observables should be a suitable subalgebra of bounded linear 
operators in this space. We will choose the canonical anti-commutation 
relations (CAR) algebra. We recall the definition: Let
$$ {\cal F}_-(L^2(\RR^2)) := \bigoplus_{N=0}^{\infty} 
                                    \bigwedge{}^N L^2(\RR^2) $$
denote the Fermi-Fock space. For $\phi \in L^2(\RR^2)$ and 
$\psi \in \bigoplus^N L^2(\RR^2)$, we shall now define 
$a^\ast(\phi) \psi := \sqrt{N+1} \phi \otimes \psi$. Let $P_-^N$ denote the 
projection onto $\bigwedge^N L^2(\RR^2)$ in 
$\bigotimes^N L^2(\RR^2)$, i.e.
$$ P_-^N \psi_1 \otimes \ldots \otimes \psi_N 
   = \frac{1}{N!} \sum_{\sigma \in S_N} 
       (-)^\sigma \psi_{\sigma(1)} \otimes \ldots \otimes \psi_{\sigma(N)}, $$
where the sum runs over all permutations of $(1, \ldots, N)$. Let 
$a^\ast_- (\phi) = P_-^{N+1} a^\ast (\phi) P_-^N$ on 
$\bigwedge^N L^2(\RR^2)$, hence, for $\psi \in \bigwedge^N L^2(\RR^2)$
$$ a^\ast_- (\phi) \psi (z_1, \ldots , z_{N+1}) 
   = \frac{1}{\sqrt{N+1}} \sum_{j=1}^{N+1} 
       (-)^{j+1} \phi(z_j) \psi(z_1,\ldots,\hat{z_j},\ldots,z_{N+1}), $$
where $\hat{z_j}$ denotes that the $j$th variable has to be omitted. This yields
a bounded linear operator $a_-^\ast(\phi)$ in ${\cal F}_-(L^2(\RR^2))$ with
$\|a_-^\ast (\phi) \| = \| \phi \|$. The adjoint operator is given by
$$ a_-(\phi) \psi (z_1, \ldots , z_{N-1}) 
   = \sqrt{N} \int_{\RR^2} d^2z \, 
       \overline{\phi(z)} \psi(z,z_1,\ldots,z_{N-1}) $$
for $\psi \in \bigwedge^N L^2(\RR^2)$. These operators satisfy the CAR
$$ \{a_-(\phi),a_-(\psi) \} = \{a_-^\ast(\phi),a_-^\ast(\psi) \} = 0, \quad 
   \{a_-(\phi),a_-^\ast(\psi)\} = (\phi,\psi). $$
The CAR algebra over $L^2(\RR^2)$, which we will denote by ${\cal A}$, is 
the $C^\ast$-algebra of bounded linear operators in 
${\cal F}_-(L^2(\RR^2))$ generated by the identity and the set of operators
$\{a_-(\phi) : \phi \in L^2(\RR^2) \}$. 

We regard a state as a linear mapping $\langle \cdot \rangle$ from this algebra
into the complex numbers such that 
$\langle A^\ast A \rangle \ge 0 \quad \forall A \in {\cal A}$ 
and $\langle 1 \rangle = 1$. Given a sequence $(\psi_N)_{N \in \NN}$ of 
vectors $\psi_N \in \bigwedge^N L^2(\RR^2)$, we can ask for the limit of 
the corresponding vector states. Here, the limit has to be understood in the 
weak-*-topology, i.e. 
$\langle \cdot \rangle = \lim_{N \to \infty} \langle \cdot \rangle_N$ 
if and only if 
$\langle A \rangle = \lim_{N \to \infty} \langle A \rangle_N 
\quad \forall A \in {\cal A}$.
It is clear that the limit is not a vector state.

As an example, we will treat the case $m = 1$. Then, Laughlin's wave function is
a single Slater determinant (namely Vandermonde's determinant). The normalized
wave functions are
$$ \psi_{N,1} (z_1, \ldots , z_N) = 
   \frac{1}{\sqrt{\pi^N \sqrt{2}^{N(N+1)} 1! \ldots N!}} 
   \prod_{1 \le j < k \le N} (z_k - z_j) 
   e^{-\frac{1}{4}\sum_{l=1}^N |z_l|^2}. $$
We will denote the limit state by $\langle \cdot \rangle_{\infty,1}$. It is 
completely determined once the correlation functions 
$\langle 
 \prod_{j=1}^m a_-^\ast (\phi_j) \prod_{k=1}^n a_- (\psi_k) 
 \rangle_{\infty,1}$ 
are known. Obviously, these vanish for $m\not= n$. We will calculate the 
two-point function explicitly: We have 
$\langle a_-^\ast (\phi) a_- (\psi) \rangle_{\infty,1} = (\psi, T \phi)$, 
where the integral kernel of $T$ is given by
\begin{eqnarray*} 
  (z',T z) &=& \lim_{N \to \infty} N \int d^2z_2 \ldots d^2z_N \,
               \overline{\psi_{N,1} (z,z_2, \ldots, z_N)} 
               \psi_{N,1}(z',z_2,\ldots,z_N) = \\
           &=& \lim_{N \to \infty} \frac{1}{2 \pi} 
               e^{-\frac{1}{4}(|z|^2 + |z'|^2)} 
               \sum_{j= 0}^{N-1} \frac{(\frac{\overline{z}z'}{2})^j}{j!} 
               = \frac{1}{2 \pi} 
               e^{-\frac{1}{4} |z - z'|^2 + \frac{i}{2} z \wedge z'}. 
\end{eqnarray*}
Thus, the operator $T$ is just the projection onto the lowest Landau level. As
a consequence, we get the correct filling factor $\nu = 1$. Furthermore, it can
be shown that $\langle \cdot \rangle_{\infty,1}$ is a quasi-free state, i.e. 
the $2n$-point functions are certain sums over products of two-point functions,
for example
$\langle
    a_-^\ast(\phi_1) a_-^\ast(\phi_2) a_-(\psi_1) a_-(\psi_2) 
 \rangle_{\infty,1}
  =   \langle a_-^\ast(\phi_1) a_-(\psi_1) \rangle_{\infty,1}
      \langle a_-^\ast(\phi_2) a_-(\psi_2) \rangle_{\infty,1} 
    - \langle a_-^\ast(\phi_1) a_-(\psi_2) \rangle_{\infty,1} 
      \langle a_-^\ast(\phi_2) a_-(\psi_1) \rangle_{\infty,1}$; 
see \cite{BraRob} for an exact definition. But this is just the limit Gibbs 
state (for temperature $T=0$ and filling factor $\nu = 1$) for non-interacting 
electrons in a constant magnetic field. Thus in the thermodynamical limit we 
get the exact groundstate.

This can also be seen from the work of Haldane \cite{Hald2} which we will now
review. Let $\Pi$ denote the projection onto the lowest Landau level
in the one-particle space. We 
will investigate the Hamiltonian $H_{N,eff} := \Pi^{\otimes N} H_N
\Pi^{\otimes N}$. The kinetic part just equals $\frac{N \hbar \omega}{2}$. 
In order to calculate the integral kernel of 
$\Pi \otimes \Pi U(|z_j - z_k|) \Pi \otimes \Pi$, we perform the unitary 
transformation
$$ {\cal U}: L^2(\RR^2) \bigotimes L^2(\RR^2) 
             \rightarrow L^2(\RR^2) \bigotimes L^2(\RR^2),$$ 
$$ ({\cal U} \psi) (Z_{jk},z_{jk}) 
   = \psi(z_j = Z_{jk}+z_{jk}, z_k = Z_{jk}-z_{jk}). $$
Then,
$$ {\cal U} (\Pi \otimes \Pi) U(|z_j - z_k|) (\Pi \otimes \Pi) {\cal U}^{-1} 
   = \Pi \otimes (\Pi U(|z_{jk}|) \Pi) . $$
Since $U(|z_{jk}|)$ is rotationally invariant, the second factor in the 
tensorproduct is diagonalized by the simultaneous eigenstates of the Landau
Hamiltonian and angular momentum. In this basis, 
$$ (\phi_m, U(|z_{jk}|) \phi_{m'}) = U_m \delta_{m,m'} $$
with
$$ U_m = \int_0^\infty \frac{dr}{2^{2m+1} m!} 
           e^{- \frac{1}{4} (\frac{r}{\ell_\Bb})^2} 
           (\frac{r}{\ell_\Bb})^{2m+1} U(r), $$
where we have reintroduced the magnetic length $\ell_\Bb$. Thus, in some sense, the 
projection onto the lowest Landau level yields a `quantization of 
inter-particle separation'. For example, for the Coulomb interaction 
$U(r) = \frac{e^2}{4 \pi r}$ we have
$$ U_m = \frac{e^2}{4 \pi} \frac{\Gamma(m + \frac{1}{2})}{2 m!}. $$
For $m \to \infty$, this falls off as $\frac{1}{\sqrt{m}}$.
It is remarkable that the coefficients $U_m$ are independent of $\ell_\Bb$ exactly
for the Coulomb interaction.

As a result, we state: The set of functions
$$ \psi(Z,z) \propto Z^M z^m 
                     e^{- \frac{1}{4}(|z|^2 + |Z|^2)}
             \propto (z_j + z_k)^M (z_j - z_k)^m 
                     e^{- \frac{1}{4}(|z_j|^2 + |z_k|^2)}
$$
with $M,m \in \{0,1,2,\ldots\}$ forms a basis of eigenfunctions of 
$\Pi \otimes \Pi U(|z_j - z_k|) \Pi \otimes \Pi$. Each eigenvalue $U_m$ is 
infinitely degenerate (according to an arbitrary angular momentum of the 
center-of mass motion).

The general two-particle wave function in the lowest Landau level with angular
momentum maximal equal to $N_\Phi$ is given by
$$ \psi(z_1,z_2) 
   = P_{N_\Phi}(z_1,z_2) e^{- \frac{1}{4}(|z_1|^2 + |z_2|^2)} $$
where $P_{N_\Phi}$ is a polynomial of degree $N_\Phi$ in each of its variables.
These states form a $N_\Phi \choose 2$-dimensional subspace, but this subspace
is not invariant under the two-particle interaction. Let us assume that the 
sequence $U_m$ decreases monotonically. Then the desired ground state would be
a state with relative angular momentum $N_\Phi$, or, since we are dealing with
Fermions and therefore the relative angular momentum has to be odd, $N_\Phi -1$.
In the latter case, we have a degenerate ground state since we can choose $M=0$ 
or $M=1$. Let us remark that these functions are not equal to Laughlin's ansatz
for $N=2$.

In the case of more than two particles, it is in general not possible to give 
the groundstate. The general $N$-particle wave function in the lowest Landau 
level with maximal angular momentum equal to $N_\Phi$ is given by
$$ \psi(z_1,\ldots,z_N) 
   = P_{N_\Phi} (z_1,\ldots,z_N) e^{- \frac{1}{4} \sum_{l=1}^N |z_l|^2}, $$
where again $P_{N_\Phi}$ is a polynomial of degree $N_\Phi$ in each of its
variables. These states form a $N_\Phi + 1 \choose N$-dimensional subspace.
Let us require that only pairs with a relative angular momentum of at least 
$m_0$ appear in the wave function. This is possible if and only if 
$m_0 \le \frac{N_\Phi}{N-1} \to \frac{1}{\nu}$, and the general form of the 
wave function is given by
$$ \psi(z_1,\ldots,z_N) 
   = P_{N_\Phi - m_0(N-1)} \prod_{1 \le j < k \le N} (z_k - z_j)^{m_0}  
   e^{- \frac{1}{4} \sum_{l=1}^N |z_l|^2}. $$
Let us now consider the truncated Hamiltonian which results from setting 
$U_m = 0$ for $m \ge m_0$. Then, the above $\psi$ is an eigenstate with energy 
$0$. Furthermore, if we assume $U_m = 0$ for $m < m_0$, it is a ground state. 
In general, for $\nu < \frac{1}{m_0}$, the ground state is degenerate due to the
polynomial pre-factor $P_{N_\Phi - m_0(N-1)}$.  Let us assume that $m_0$ is odd.
Then, to get a Fermionic state, this pre-factor has to be symmetric under 
particle exchange. Thus, if we assume $N_\Phi - m_0(N-1) = 0$ which is 
equivalent to $\frac{1}{m_0} = \nu - \frac{1}{N_\Phi} \to \nu$, we get a unique 
ground state which is equal to Laughlin's state $\psi_{N,m}$ with $m = m_0$. For
$m_0$ even, $P_{N_\Phi - m_0(N-1)}$ has to be antisymmetric. There is (up to a 
constant factor) an unique possibility if the polynomial is of degree one, 
namely $P_1(z_1,\ldots,z_N) = \prod_{1 \le j < k \le N} (z_k - z_j)$. In this 
case, we get $\psi_{N,m}$ as a unique ground state for $\nu = \frac{1}{m}$ and 
$m = m_0 + 1$.

Let us recapitulate the result: if we project the Hamiltonian onto the lowest 
Landau level and assume that all coefficients $U_m$ vanish for $m$ greater or
equal to some $m_0$ and $U_m > 0$ otherwise, then the unique (infinite volume)
groundstate with filling factor $\nu = \frac{1}{m_0}$ for $m_0$ odd 
respectively $\nu = \frac{1}{m_0 + 1}$ for $m_0$ even is the limit state which
arises from Laughlin's wave functions $\psi_{N,m_0}$ or $\psi_{N,m_0+1}$ 
respectively.

In fact, we have not proved this rigorously. First, it remains to show that
Laughlin's wave functions indeed yield the correct filling factor in the 
thermodynamic limit. This has been shown by Laughlin by means of a somewhat
heuristic argument \cite{Laug2}. Then, it has to be shown that the restriction
of angular momentum indeed yields the same thermodynamic limit as a classical 
boundary condition. This should not be too hard to prove.

\vspace{0.2cm}

\subsection{The elementary theory of the $\nu=\frac{1}{m}$-FQHE}

The elementary theory of the $\nu = \frac{1}{m}$-FQHE is solely based on 
Laughlin's states. Thus we assume in the following that the correct ground state
is indeed given by their thermodynamical limit. We then have to explain that
\begin{enumerate} 
\item the Hall conductivity for the state given by $\psi_{N,m}$ equals 
$\frac{e^2}{mh}$,
\item plateaux of the Hall conductivity arise when the one-particle potential
is switched on. 
\end{enumerate}
We remark that the first point is just the classical result 
$\sigma_H / \frac{e^2}{h} = \nu$. Indeed, for $V=0$, the classical result should
hold for any value of the filling factor.

To calculate $\sigma_H$, Laughlin \cite{Laug2} proposed a Gedanken-experiment 
similar to that we have already described in Section~\ref{sec-IQHcond}. We 
consider the  $N$-particle wavefunction $\psi_{N,m}$. We introduce a infinitely 
thin solenoid at the origin, perpendicular to the disk in which the particles 
are confined. Then we force adiabatically one flux quantum 
$\Phi_0 = \frac{hc}{e}$ through the flux tube. We assume that $\psi_{N,m}$ is 
the non-degenerate groundstate for some Hamiltonian and that there is a gap 
between the groundstate energy and the rest of the spectrum. Let us mention 
that this `gap-condition' is a sufficient condition for our considerations, but 
in general need not be necessary for $\sigma_H/ \frac{e^2}{h}$ to be equal to 
$\frac{1}{m}$. After the flux quantum has been added, the Hamiltonian has not 
changed up to a gauge transformation. Since we can apply the adiabatic theorem, 
we know that the system is still in an eigenstate of the Hamiltonian.
The single-particle wave functions $z^m e^{-\frac{1}{4}|z|^2}$ evolve, up
to a phase factor and normalization, to $z^{m+1} e^{-\frac{1}{4}|z|^2}$ during
this operation. The particles which were essentially confined in a disk with 
radius $R=\sqrt{2 N_\Phi}$ at the beginning are now restricted to the radius 
$R'=\sqrt{2 (N_\Phi + 1)}$. These two disks differ by an area $2 \pi$, and 
since the charge density equals uniformly $\frac{e}{2 \pi m}$, a charge 
$\frac{e}{m}$ has left the original disk while adding the flux quantum. Similar 
to the IQHE, this `gauge argument' does not prove directly the fractionality of 
the Hall conductivity, but shows that the classical result 
$\sigma_H / \frac{e^2}{h} = \nu$ is valid. Fractionality of 
$\sigma_H / \frac{e^2}{h}$ follows from the fractionality of the filling factor.

Let us discuss the excited state which appears in Laughlin's Gedanken-experiment
in the case of Haldane's truncated Hamiltonian. After the flux quantum has been
added, the single-particle angular momentum is restricted to values less than 
or equal to $N_\Phi + 1$. Since we are looking for an excited state, the 
relative angular momentum for each pair is still at least $m$. Thus,
$$ \psi(z_1,\ldots,z_N) = P_1(z_1,\ldots,z_N) \psi_{N,m} (z_1,\ldots,z_N). $$
In general, $P_1(z_1,\ldots,z_N) \propto \prod_{j=1}^N (z_j - u_j)$ with 
arbitrary $u_j$. But as $P_1$ has to be symmetric under particle exchange, all
$u_j$ have to be equal. This yields the excited states already proposed by
Laughlin \cite{Laug2}:
$$ S_{z_0} \psi_{N,m} (z_1,\ldots,z_N) 
   = \prod_{j=1}^N (z_j - z_0) \psi_{N,m} (z_1,\ldots,z_N). $$
The adjoint operator is given by
$$ S_{z_0}^\ast \psi_{N,m} (z_1,\ldots,z_N) 
   = e^{- \frac{1}{4} \sum_{l=1}^N |z_l|^2} 
     \prod_{j=1}^N (2 \partial_{z_j} - \overline{z_0})
     \prod_{1 \le k < l  \le N} (z_l - z_k)^m. $$
The operators $S_{z_0}$ and $S_{z_0}^\ast$ are called `quasiparticle' 
respectively `quasihole' creation operators for the following reasons: 
Denote by $\Phi_{N,m,z}$ the $N$-particle state 
$e^{-\frac{1}{4m} |z|^2} S_z \psi_{N,m}$. Then, one can show that
$$ (\Phi_{N,m,z},\Phi_{N,m,z'}) 
   \propto e^{-\frac{1}{4m}(|z|^2 + |z'|^2 - 2\overline{z}z')}. $$
Thus $\Phi_{z,m}$ looks like the wavefunction of a `quasi-particle' with 
charge $\frac{e}{m}$ in the lowest Landau level. More formally: Let 
$\psi \in L_2(\RR^2)$ (a quasi-particle wave function). Define the 
$N$-electron state
$$ Q_{N,m} \psi 
   := \frac{1}{\sqrt{2 \pi m}} \int d^2\eta \, 
      \overline{\psi(\eta)} e^{-\frac{1}{4m} |\eta|^2} S_{\eta} \psi_{N,m}.$$ 
Then, for 
$\psi_z(\eta) 
 = \frac{1}{\sqrt{2 \pi m}} 
   e^{-\frac{1}{4m}(|z|^2 + |\eta|^2 - 2 \overline{z}{\eta})}$, 
we have $Q_{N,m} \psi_z = \Phi_{N,m,z}$ and moreover
$(\psi_z,\psi_{z'}) = (Q_{N,m} \psi_z, Q_{N,m} \psi_{z'})$. 

Of course, the state $\Phi_{z_,m}$ is in general not an exact eigenstate of
the truncated $N$-particle Hamiltonian. But as $N \to \infty$, its overlap with
an eigenstate becomes larger and larger. Let us assume, for simplicity, that it 
is  an eigenstate. Since the quasi-particle state $\psi_z$ is an eigenstate of 
the Hamiltonian 
$H_q = (\frac{1}{\imath} \vec{\nabla}_\eta + \frac{e}{m} \vec{A}(\eta))^2$, we have 
$$(\Phi_{N,m,z}, H_{N,eff} \Phi_{N,m,z}) \propto (\psi_z, H_q \psi_z), $$
where, for simplicity, we have denoted also the truncated Hamiltonian by 
$H_{N,eff}$. One is led to define a quasi-particle potential $V_q$ by the 
equation
$$ (\Phi_{N,m,z}, \sum_{j=1}^N V(z_j) \Phi_{N,m,z}) 
   \propto (\psi_z,V_q \psi_z) 
      =    \frac{1}{2 \pi m} \int d^2\eta \, 
             e^{-\frac{1}{2m}|z - \eta|^2} V_q (\eta). $$
Thus, we have defined an effective Hamiltonian which governs the quasi-particle
motion. This yields a qualitative explanation for the existence of plateaux: If
the quasi-particles which are created by an electric field are localized due to
randomness of the potential $V_q$, they do not contribute to the Hall
conductivity. The FQHE looks like the IQHE for quasi-particles.

But this only concerns the mechanism which causes the existence of plateaux. The
fractionality of $\sigma_H/\frac{e^2}{h}$ is due only to the fact that the 
charge density of Laughlin's states is exactly $\frac{e}{2 \pi m}$. As we have
shown, these states are indeed the non-degenerate ground states for some 
Hamiltonian, but it is not clear at all why the quasi-particle charge should be
invariant under slight deformations of the ground state. Thus, the topological 
origin of the FQHE remains undiscovered.

\vspace{0.2cm}

\subsection{The r\^ole of gauge invariance and incompressibility}

We have tried to emphasize that the main problem of the common approach to the
FQHE is the lack of a Kubo formula to calculate the Hall conductivity. To 
calculate it for Laughlin's states, we used gauge-invariance and assumed the
existence of a gap separating the ground state energy from the rest of the
spectrum. These two assumptions - gauge-invariance and incompressibility - also
form requirements for a completely different approach which is based on a 
work of Halperin \cite{Halp} and has been developed further e.g. by Wen
\cite{Wen} and Fr\"ohlich et al. \cite{FrKe,FrZe,Fr,FrSt1,FrSt2,FrSt3}. 
We will particularly refer to the work of the latter group.

Let us first look at incompressibility from a classical point of view. How does
the system respond to external ``small'' fields $\vec{\Ee}$ and $\vec{\Bb }$ (where 
the latter does {\em not} include the strong magnetic field which causes the 
quantum Hall state)? In $(2+1)$-dimensional space-time (we will only consider 
cases where space is a subset $\Lambda \subset \RR^2$ - in particular a 
disk around the origin - with the usual `flat' metric and space-time will then 
be $\RR \times \Lambda$), we only have two components of the electric and 
one component of the magnetic field. In covariant notation, we have the 
electromagnetic-field tensor 
$F = \frac{1}{2} F_{\mu \nu} dx^\mu \wedge dx^\nu$
with $x^0 = ct$ and
$$ (F_{\mu \nu})_{\mu,\nu \in \{0,1,2\}} 
   = \left( 
       \begin{array}{ccc}    0 & \Ee_x & \Ee_y  \\ 
                          -\Ee_x &  0  & -\Bb _z \\ 
                          -\Ee_y & \Bb _z & 0 
       \end{array} 
     \right). $$
The homogeneous Maxwell equations $dF = 0$ reduce to Faraday's induction law
$$ \frac{1}{c} \partial_t \Bb_z + \vec{\nabla} \wedge \vec{\Ee} = 0. $$
Let $\vec{j}$ denote the electric current density. Then we have the Ohm-Hall
law (respectively the definition of the conductivity tensor $\sigma$) 
$\vec{j} = \sigma \vec{\Ee}$. Furthermore, $\sigma = \rho^{-1}$, where $\rho$ is 
the symmetric resistivity tensor. Incompressibility means that the diagonal 
components of $\rho$ vanish and hence
$$ \sigma = \left( 
              \begin{array}{cc}     0     & \sigma_H \\ 
                                -\sigma_H &     0 
              \end{array} 
            \right). $$
Furthermore, we have the continuity equation (i.e. charge conservation)
$$ \frac{1}{c} \partial_t j^0 + \vec{\nabla} \vec{j} = 0, $$
where $c j^0$ denotes charge density minus the uniform charge density of the 
unperturbed quantum Hall state. From these three basic equations, we get 
$\partial_t j^0 = \sigma_H \partial_t \Bb$. Since, with our above definition of 
$j^0$, we have $j^0 = 0$ for $\Bb = 0$, the integration of this equation yields 
$j^0 = \sigma_H \Bb$.

We introduce the $2$-form $J$ dual to $(j^0,\vec{j})$:
$$ J = \frac{1}{2} J_{\mu \nu} dx^\mu \wedge dx^\nu, \quad
  (J_{\mu \nu})_{\mu,\nu \in \{0,1,2\} } 
  = \left( 
      \begin{array}{ccc}    0  &  j_y & - j_x \\ 
                         - j_y &   0  &   j^0  \\ 
                           j_x & -j^0 &    0 
      \end{array}
    \right). $$
Then, the continuity equation reads $dJ = 0$ and we can combine the Ohm-Hall 
law and the connection between $j^0$ and $\Bb_z$ into one equation: 
$J = - \sigma_H F$. Since we work in simply connected space-time, we have
$$ J = db, \quad F = da $$
with $1$-forms $a$ and $b$. Hence, $db = -\sigma_H da$. This is the 
Euler-Lagrange equation for some action. Whether we regard it as an equation for
$a$ or for $b$ is just a matter of taste, corresponding to the choice that we
can either create a current by applying an external field or, equivalently, can
create a Hall voltage by forcing some current. For our purpose, we will choose
the latter possibility. Thus, our `dynamical variable' is the gauge potential
whereas we keep the potential for the current density fixed. A possible choice
for the action is:
$$ S_{\Lambda,b}(a) 
   = \frac{\sigma_H}{2} \int_{\Lambda \times \RR} a \wedge da 
   + \int_{\Lambda \times \RR} b \wedge da 
   + b.t., $$
where, since we work in a finite volume, we have to include some (unknown) 
boundary term `$b.t.$'.

Can these phenomenological equations be derived from a microscopical 
description? And is there a connection between the quantization of the classical
theory (by means of path integrals) and our original quantum mechanical problem?
Formally, the Hamiltonian $H_\Lambda = \bigoplus_{N=0}^\infty H_{N,\Lambda}$
(with $H_{N,\Lambda}$ as in the previous sections) can be described by means of
`fields' $\Psi$ and $\Psi^\ast$ which satisfy at each time $t$ the CAR
$$   \{ \Psi(\vec{x},t), \Psi(\vec{y},t) \} 
   = \{\Psi^\ast(\vec{x},t), \Psi^\ast(\vec{y},t) \} 
   = 0, \quad 
     \{\Psi(\vec{x},t),\Psi^\ast(\vec{y},t)\} = \delta (\vec{x} - \vec{y}). $$
The fields act on the Fermi-Fock space ${\cal F}_- (L_2(\RR^2))$. The 
particle-number operator is given by 
$N_\Lambda = \int_\Lambda \Psi^\ast(\vec{x},t) \Psi(\vec{x},t)$. 
We will work in the grand-canonical ensemble. We have
\begin{eqnarray*} 
   H_\Lambda(t) - \mu N_\Lambda 
   &=& - \frac{\hbar^2}{2m_\ast} \int_{\Lambda} d^2x \, 
       [\vec{\nabla} - \frac{\imath e}{\hbar c} \vec{a}_{tot}(\vec{x},t)]
       \Psi^\ast(\vec{x},t) 
       [\vec{\nabla} + \frac{\imath e}{\hbar c} \vec{a}_{tot}(\vec{x},t)] 
       \Psi(\vec{x},t) + \\ 
   &+& \int_{\Lambda} d^2x \, 
       \Psi^\ast(\vec{x},t) [V(\vec{x}) - \mu +e a^0 ] \Psi(\vec{x},t) + \\ 
   &+& \frac{1}{2} \int_{\Lambda \times \Lambda} d^2x d^2y \, 
       \Psi^\ast(\vec{x},t) \Psi^\ast(\vec{y},t) 
       U(|\vec{x} -  \vec{y}|) 
       \Psi(\vec{y},t) \Psi(\vec{x},t). 
\end{eqnarray*}
Here, $\vec{a}_{tot}$ represents the potential for the total magnetic field 
(i.e. the potential $\vec{A}$ for the constant field $\Bb \vec{e}_z$ plus some 
- in general time-dependent - source term $\vec{a}$) and $a^0$ also is a source 
term.

We then have the Heisenberg equation of motion
$$ \partial_t \Psi(\vec{x},t) 
   = \frac{\imath}{\hbar} [H_\Lambda (t) - \mu N_\Lambda, \Psi(\vec{x},t)]. $$
But this is the Euler-Lagrange equation derived from the action
$$ S_\Lambda(\Psi^\ast, \Psi) 
   = \int_\RR dt 
       \left\{ 
         \int_\Lambda d^2x \, 
         \Psi^\ast(\vec{x},t) \imath \hbar \partial_t \Psi(\vec{x},t) 
         - H_\Lambda (t) + \mu N_\Lambda 
       \right\}. $$
The partition function (at zero temperature and chemical potential $\mu$) is
given by the path integral
$$ Z_\Lambda 
   = \int {\cal D}\Psi^\ast {\cal D}\Psi 
     e^{\frac{\imath}{\hbar} S_\Lambda (\Psi^\ast,\Psi)}. $$
We define the effective action as 
$S_\Lambda^{eff} := \frac{\hbar}{\imath} \ln Z_\Lambda$. Regarded as a functional of 
the gauge potential $(a_{tot})_\mu dx^\mu = a^0 dx^0 - \vec{a}_{tot} d\vec{x}$, 
it is the generating functional for connected time-ordered Green functions of 
the current-density operators
$$  j^0   = \Psi^\ast \Psi, \quad 
  \vec{j} =  -\frac{\imath \hbar}{2m_\ast c} 
            \{ 
              [(\vec{\nabla} - \frac{\imath e}{\hbar c} \vec{a}_{tot}) \Psi^\ast]\Psi
            - \Psi^\ast (\vec{\nabla} + \frac{\imath e}{\hbar c} \vec{a}_{tot}) \Psi 
            \}. $$
We have
$$ \langle T j^{\mu_1}(x_1) \ldots j^{\mu_n}(x_n) \rangle^c 
   = \imath^{(n+1)} \frac{\delta}{\delta a_{\mu_1} (x_1)} 
               \ldots 
               \frac{\delta}{\delta a_{\mu_n} (x_n)} 
     S_\Lambda^{eff}(a^0,\vec{a}). $$
We now perform a scale transformation, which means the following: We enlarge 
the domain $\Lambda$ by multiplying the space-time $\RR \times \Lambda$ 
with a scalar $\lambda > 1$. We introduce new coordinates 
$\xi = (\xi^0,\vec{\xi}) = \lambda^{-1} (x^0,\vec{x})$
(thus $\vec{\xi} \in \Lambda$) and define the rescaled sources 
$a_\mu^\lambda (\lambda \xi):= \lambda^{-1} a_\mu (\xi)$,
such that $a_\mu^\lambda (\lambda \xi) d(\lambda \xi) = a_\mu (\xi) d\xi$. 
The `scaling-limit' $\lambda \to \infty$ corresponds to a thermodynamical, but 
also adiabatic, limit. We assume some `strong-clustering property', namely we 
require  that for $n \le 3$ the distributions
$$ \phi^{\mu_1 \ldots \mu_n} (\xi_1,\ldots,\xi_n) 
   := \frac{\imath^{n+1}}{n!} \lim_{\lambda \to \infty} \lambda^{2n} 
      \langle 
         T j^{\mu_1}(\lambda \xi_1) \ldots j^{\mu_n}(\lambda \xi_n) 
      \rangle^c $$
are {\em local} distributions, i.e. 
$$ \mbox{supp} \, \phi^{\mu_1 \ldots \mu_n} 
   = \{ (\xi_1 ,\ldots, \xi_n) \in (\RR \times \Lambda)^n : 
        \xi_1 = \ldots = \xi_n \}. 
$$
Then, assuming some differentiability condition on $S_\Lambda^{eff}$, we have up
to second order (with $A = - \vec{A} d\vec{x}$):
\begin{eqnarray*} 
  S^{eff}_{\lambda \Lambda} (A + a^\lambda) 
  &=&   S^{eff}_{\lambda \Lambda} (A) 
      + \int_{\lambda (\RR \times \Lambda)} d^3x \, 
        \left. 
          \frac{\delta}{\delta a_\mu(x)} 
        \right|_{a=A} 
        S^{eff}_{\lambda \Lambda}(a) \, a^\lambda_\mu(x) + \\ 
  &+&   \int_{[\lambda (\RR \times \Lambda)]^2} d^3x d^3y \, 
        \left. 
          \frac{\delta^2}{\delta a_\mu(x) \delta a_\nu(y)} 
        \right|_{a=A} 
        S^{eff}_{\lambda \Lambda}(a) \, a^\lambda_\mu(x) a^\lambda_\nu(y).
\end{eqnarray*}
Now, $\left. 
        \frac{\delta}{\delta a_\mu(x)} 
      \right|_{a=A} S^{eff}_{\lambda \Lambda}(a) 
  = - \left. 
        \langle j^\mu(x) \rangle^c 
       \right|_{a=A}$. 
We define
$j^\mu_c(\xi) 
 := \lim_{\lambda \to \infty} \lambda^2 
 \left. 
   \langle j^\mu(\lambda \xi) \rangle^c 
 \right|_{a=A}$ 
and $J_c$ as the 2-form dual to $j_c$. 

Furthermore, we have
$$ \frac{1}{2} \left. 
                 \frac{\delta^2}{\delta a_\mu(x) \delta a_\nu(y)} 
               \right|_{a=A} 
                  S^{eff}_{\lambda \Lambda}(a) 
   = - \frac{\imath}{2} \left. 
                     \langle j_\mu(x) j^\nu(y) \rangle^c 
                   \right|_{a=A} 
                     \stackrel{\lambda \to \infty}{\longrightarrow} 
                     \phi^{\mu \nu} (\lambda^{-1}x,\lambda^{-1}y). $$
By the above clustering property, we can split $\phi^{\mu \nu}$ in the manner
$$ \phi^{\mu \nu} (\xi,\eta) 
   = \alpha \varepsilon^{\mu \nu \rho} (\partial_\rho \delta)(\xi - \eta) 
   + R^{\mu \nu} (\xi,\eta), $$ 
where $\alpha$ is a constant and $R^{\mu \nu}$ consists of second or higher 
derivatives of the $\delta$-function.

It can be shown \cite{FrSt1} that neither the part with $R^{\mu \nu}$ in the 
second-order term nor any of the terms of higher order contribute to the 
scaling limit up to some boundary term. In the language of field theory, they 
are irrelevant, whereas the first-order term is relevant and the remaining part 
of second order (the `Chern-Simons term') is marginal. Hence, a simple 
substitution in the remaining integrals leads to
$$ S^{eff}_{\lambda \Lambda} (A + a) - S^{eff}_{\lambda \Lambda} (A) 
   \stackrel{\lambda \to \infty}{\longrightarrow} 
     \int_{\Lambda \times \RR} J_c \wedge a 
   + \alpha \int_{\Lambda \times \RR} a \wedge da 
   + b.t. $$
This is exactly our phenomenological action; thus the clustering conditions 
indeed correspond to incompressibility. By comparison, we expect that 
$\alpha = \frac{\sigma_H}{2}$. Let us check this explicitly. We reintroduce the 
non-rescaled variables:
$$ S^{eff}_{\lambda \Lambda} (A + a) 
   \stackrel{\lambda \to \infty}{\longrightarrow} 
     S^{eff}_{\lambda \Lambda}(A) 
   - \int_{\lambda(\Lambda \times \RR)} d^3x \, 
     \left. 
       \langle j^\mu(x) \rangle^c 
     \right|_A 
        a^\lambda_\mu(x)
   + \alpha \int_{\lambda(\Lambda \times \RR)} d^3x \, 
     \varepsilon^{\mu \nu \rho} a^\lambda_mu(x) 
     \partial_\nu a^\lambda_\rho(x). $$
By calculating the functional derivative, we obtain
$$ \left. 
     \langle j^\mu(x) \rangle^c 
   \right|_{A+a^\lambda} 
 = \left. 
     \langle j^\mu(x) \rangle^c 
   \right|_A 
 - 2 \varepsilon^{\mu \nu \rho} \partial_\nu a^\lambda_\rho(x). $$
For example, let us take as source the potential for a constant electric field 
in $y$-direction: $a_0^\lambda (x^0,\vec{x}) = - \Ee x_2$, 
$\vec{a}^\lambda = \vec{0}$. Then 
$\left. \langle j^1(x) \rangle^c \right|_{A+a^\lambda} 
= \left. \langle j^1(x) \rangle^c \right|_{A} - 2 \alpha \Ee$. 
Thus indeed we have $\alpha = \frac{\sigma_H}{2}$.
 
The boundary term in the effective action is in general unknown, but we have 
the requirement that the whole action is gauge-invariant, i.e. 
$S_\Lambda^{eff}(a + d\chi) = S_\Lambda^{eff}(a)$ for arbitrary smooth 
functions $\chi$. From this, it follows that
$$ b.t. (a + d\chi) - b.t. (a) 
   = \frac{\sigma_H}{2} \int_{\partial \Lambda \times \RR} 
   d\chi \wedge a. $$
We introduce light-cone coordinates 
$u_\pm = \frac{1}{\sqrt{2}} (vt \pm \frac{\theta}{2 \pi}L)$ on 
$\partial \Lambda \times \RR$, where $v$ is some (a priori arbitrary) 
velocity, $\theta$ is the angle of polar coordinates in the plane and $L$ is the
circumference of the disc $\Lambda$. Then the above functional equation for the 
boundary term has the general solution
$$ b.t.(a) =  \int_{\partial \Lambda \times \RR} d^2 u \, 
              (a_+ a_- - 2 a_+ \frac{\partial_-^2}{\Box} a_-) + W(a) 
           =: \Delta(a) + W(a) $$
where $a|_{\partial \Lambda} = a_+ du_+ + a_- du_-$, 
$\partial_\pm = \frac{\partial}{\partial u_\pm}$, 
$\Box = 2 \partial_+ \partial_-$ and $W$ is some arbitrary gauge-invariant 
function.

$\Delta$ is known as the generating functional for time-ordered connected Green 
functions of chiral $U(1)$-currents. We just state the result and refer the 
reader to \cite{GoOl,BuMaTo}: Assume that there are current-density operators
in two-dimensional space-time whose time-ordered connected Green functions are 
given by the functional derivatives of $\Delta$. These operators, then are sums 
of derivatives of $N$ massless scalar fields $\phi_1, \ldots, \phi_N$ for
some $N \in \{1,2,\ldots\}$. The dynamics of these fields is determined by the 
action
\begin{eqnarray*} 
  \lefteqn{S_a(\phi_1,\ldots,\phi_n) 
     =  \frac{1}{2} \int_{\partial \Lambda} d^2u \, 
        \partial_+ \phi^t K \partial_- \phi -} \\  
    &-& \int_{\partial \Lambda} d^2u \, 
        [a^t_- \partial_+ \phi - (\partial_- \phi - K^{-1}a_-)^t a_+] 
        + \frac{\kappa}{2} \int_{\partial \Lambda} d^2u \, a_-a_+, 
\end{eqnarray*}
where $K$ is a positive-definite $N \times N$-matrix, 
$\phi^t=(\phi_1, \ldots, \phi_n)$, $a^t = (a, \ldots, a)$ and
$a_\pm^t = (a_\pm, \ldots, a_\pm)$, 
$\kappa = \sum_{j,k=1}^N (K^{-1})_{jk}$ and we have to impose the chirality 
constraint
$$ \partial_- \phi - K^{-1} a_- = 0. $$
Indeed it can be checked that the effective action of this theory, which is 
defined analogously to $S_\Lambda^{eff}$, is given by $\frac{\kappa}{2} \Delta$. 
Thus after performing the `Abelian Bosonization', we see that the total 
effective action is gauge-invariant if and only if $\sigma_H/\frac{e^2}{h}
 = \kappa$. 

Up to now, we have (apart from positive-definiteness) no requirements on the 
matrix $K$ and thus $\sigma_H/\frac{e^2}{h}$ could be any real number. What are
the requirements for a rational value of the Hall conductivity? It turns out
that we have to impose some {\em further} condition. We will sketch the idea 
and omit the details. For simplicity, we will restrict ourselves to the case 
$n=1$. Thus, we have a single massless scalar field. We introduce the variables 
$\hat{a}_\pm := \frac{e}{\hbar c} a_\pm$ (which have the dimension of an 
inverse length) and $\hat{K} := \frac{e^2}{h} K$ (which is dimensionless) and 
we normalize the field $\phi$ such that the action is given by
$$ S_{\hat{a}} (\phi) 
   = \frac{1}{2 \pi} \int_{\RR \times \partial \Lambda} d^2u \, 
     \left\{ 
       \frac{\hat{K}}{2} (\partial_+ \phi) (\partial_- \phi) 
       - [\hat{a}_- \partial_+ \phi - (\partial_- \phi  
          - \frac{1}{\hat{K}} \hat{a}_-) \hat{a}_+] 
       + \frac{1}{2 \hat{K}} \hat{a}_- \hat{a}_+ 
     \right\} $$
We can break down $\phi$ into its chiral components:
$$ \phi(u_+,u_-) = \phi_L(u_+) + \phi_R(u_-). $$
`$L$' and `$R$' correspond to the sign of the charge of the (quasi-)particles
described by $\phi$. Let us treat the part $\phi_L$. We have the current-density
operator $J_L(u_+) \propto \partial_+ \phi_L (u_+)$, where at the moment it is
not quite clear how to define the constant of proportionality. We will choose 
it such that the corresponding charge operator is given by
$$ Q_L = - \frac{e}{2 \pi} \oint d\theta \, 
           (\partial_+ \phi_L - \frac{1}{\hat{K}} \hat{a}_+). $$
The form of $Q_L$ is clear for $\hat{a} = 0$ and the general case follows from
gauge invariance. Of course, this determines $Q_L$ only up to some constant 
factor. We will come back to this point which seems to be crucial to us. We 
define `vertex operators'
$$ V_n(u_+) = : e^{\imath n \phi_L(u_+)} :, $$
where `$:~:$' denotes normal ordering. These operators obey the important 
relations
$$ V_m(u_+) V_n(v_+) = e^{\pm \imath \pi \frac{mn}{\hat{K}}} V_n(v_+)V_m(u_+), $$
$$ [Q_L,V_n(u_+)] = - \frac{ne}{\hat{K}} V_n(u_+). $$
Thus, $V_n(u_+)$ creates a quasiparticle of `charge' 
$Q_L = - \frac{ne}{\hat{K}}$. If we create two such quasiparticles and 
interchange them, we get a phase factor $e^{\pm \imath \pi \frac{n^2}{\hat{K}}}$, 
i.e. these quasiparticles obey {\em fractional statistics} in general. 
If $\hat{K} = n$, then we get a phase factor $e^{\pm \imath \pi \hat{K}}$. We now 
come to the announced additional assumption: 
Assume that the quasiparticles with `charge' $-e$ created by vertex operators 
are electrons and therefore obey Fermi statistics. Then we get the celebrated 
result
$$ \hat{K} = 2l + 1 \Leftrightarrow \sigma_H 
           = \frac{1}{2l + 1} \frac{e^2}{h}, \quad l \in \ZZ. $$
Let us just mention that in the general case, i.e. with more than one field,
$\sigma_H / \frac{e^2}{h}$ is still a rational number, namely
$$ \sigma_H = \sum_{j,k=1}^N (K^{-1})_{jk} \frac{e^2}{h}, \quad
   K_{jk} \in \ZZ, \quad 
   K_{jj} \in 2\ZZ + 1. $$
Furthermore, if we also take quasiparticles with positive charge into account,
$\sigma_H / \frac{e^2}{h}$ is the difference between two such rational numbers.

We end this section with two remarks: First, it seems to us that the last 
assumption is more or less ad hoc. Why should we exclude quasiparticles with 
fractional statistics and charge $-e$? Even if we assume that the electron is 
among the possible excitations, it is not clear that we can create this 
excitation by just applying one vertex operator.

Second, there is still the question of normalization of the charge operator. 
Assume that we change its normalization by multiplying it with a factor $\gamma
$. Then, the vortex operator $V_n(u_+)$ creates a quasiparticle with `charge'  
$-\frac{\gamma n e}{\hat{K}}$. Interchanging two such quasiparticles with 
`charge' $-e$ would then yield a phase factor 
$e^{\pm \imath \pi \frac{\hat{K}}{\gamma^2}}$. Thus, by varying 
$\gamma$, $\hat{K}$ could be any real number! So 
what is the reason for the above normalization of $Q_L$? From its definition 
and Stokes' theorem, we see immediately that adding one flux quantum 
$\phi_0 = \frac{hc}{e}$ creates a `charge' $-\frac{e}{\hat{K}}$ at the boundary.
On the other hand, we know from Laughlin's gauge argument that a charge 
$-e \nu$ leaves the bulk when the flux quantum is added (and, in this framework,
then circles along the boundary). But this means that $\hat{K} = \nu$, or 
$\sigma_H = \nu \frac{e^2}{h}$, which is the {\em classical} result! Our feeling
is that we find the same problem as in our discussion of Laughlin's argument in 
Section~\ref{sec-IQHcond}. Without localization, the classical result 
$\sigma_H = \nu \frac{e^2}{h}$ should hold for any $\nu$. When will the value of
$\sigma_H$ be robust against turning on some disorder? Presumably, this is 
exactly the case if $\nu$ belongs to the set of fractions which yield an
incompressible state. But no serious arguments are known for this hypothesis, 
apart from the fact that it fits well with numerical and experimental results. 
Thus to explain the existence of plateaux in the FQHE is still an open problem, 
but we believe that the reviewed results form an important step to its solution.

\newpage


\end{document}